\documentclass[12pt]{iopart}
\usepackage[utf8]{inputenc}
\usepackage{latexsym}
\usepackage{graphicx}
\usepackage{epsfig}
\usepackage{amsopn}
\usepackage{amsfonts}
\usepackage{amssymb}
\usepackage{subfigure}
\usepackage{braket}
\usepackage{epstopdf}
\usepackage{braket}
\usepackage{multirow}

\usepackage[bookmarks,
            breaklinks,
            pdfstartview=FitH, 
            pdftitle={},
            pdfauthor={},
            pdfkeywords={}
            ]{hyperref}
\usepackage{cite}

\begin{document}
\title{Prospects of reaching the quantum regime in Li-Yb$^+$
mixtures}

\author{H~A~F\"urst$^1$, N~V~Ewald$^1$, T~Secker$^2$, J~Joger$^1$, T~Feldker$^1$
and R~Gerritsma$^1$}
\newcommand{\affA}{}
\newcommand{\affZ}{Eindhoven University of Technology, Post Office Box 513, 5600
MB Eindhoven, The Netherlands}

\address{$^1$Institute of Physics, University of Amsterdam, 1098 XH Amsterdam,
The Netherlands} \address{$^2$Eindhoven University of Technology, Post Office
Box 513, 5600 MB Eindhoven, The Netherlands}
\ead{r.gerritsma@uva.nl}
\begin{abstract}
We perform numerical simulations of trapped $^{171}$Yb$^+$ ions that are buffer
  gas cooled by a cold cloud of $^6$Li atoms. This species combination has been
  suggested to be the most promising for reaching the quantum regime of
  interacting atoms and ions in a Paul trap.
  Treating the atoms and ions classically, we compute that the collision energy
  indeed reaches below the quantum limit for a perfect linear Paul trap. We analyze the
  effect of imperfections in the ion trap that cause excess micromotion. We find
  that the suppression of excess micromotion required to reach the quantum limit
  should be within experimental reach. Indeed, although the requirements are
  strong, they are not excessive and lie within reported values in the
  literature. We analyze the detection and suppression of excess micromotion in
  our experimental setup. Using the obtained experimental parameters in our
  simulation, we calculate collision energies that are a factor 2-11 larger than
  the quantum limit, indicating that improvements in micromotion detection
  and compensation are needed there. We also analyze the buffer-gas cooling of linear and
  two-dimensional ion crystals. We find that the energy stored in the eigenmodes
  of ion motion may reach 10-100~$\mu$K after buffer-gas cooling under realistic
  experimental circumstances. Interestingly, not all eigenmodes are buffer-gas
  cooled to the same energy. Our results show that with modest improvements of
  our experiment, studying atom-ion mixtures in the quantum regime is in reach,
  allowing for buffer-gas cooling of the trapped ion quantum platform and to
  study the occurrence of atom-ion Feshbach resonances.
\end{abstract}

\submitto{\NJP}

\maketitle
\tableofcontents

\section{Introduction}
In recent years, a novel field in atomic physics has developed in which
ultracold atomic clouds are mixed with trapped ions~\cite{Smith:2005,
Grier:2009, Zipkes:2010, Schmid:2010,
Zipkes:2010b, Hall:2011, Hall:2012, Rellergert:2011, Sullivan:2012,
Ratschbacher:2012, Ravi:2012, Ratschbacher:2013, Harter:2013, Hall:2013,
Haze:2015, Meir:2016, Saito:2017, Tomza:2017cold}.  These efforts aim at
sympathetic cooling~\cite{Krych:2010,Krych:2013,Meir:2016} of ions by atoms, and
have potential applications in probing quantum many-body
systems~\cite{Kollath:2007}, quantum
computation~\cite{Doerk:2010,Secker:2016} and quantum
simulation~\cite{Bissbort:2013}. Furthermore, Feshbach resonances are predicted
to exist in atom-ion
mixtures~\cite{Idziaszek:2009,Idziaszek:2011,Tomza:2015,Gacesa:2017,Furst:2018}. Such
resonances play a pivotal role in neutral atom systems for the purpose of tuning
the interactions between the atoms~\cite{Julienne:2010} and find applications in
studies of quantum many-body physics~\cite{Bloch:2012}. However, up until now no
atom-ion Feshbach resonances have been observed which is likely because the
required ultracold temperatures have not been reached in these systems.

A crucial step towards realising the applications described above is to reach
the quantum (or $s$-wave) regime for atom-ion mixtures. It turned out that the
Paul or radio-frequency (rf) trap commonly employed for trapping the ions limits
the attainable temperatures in atom-ion mixtures, and the $s$-wave regime has so
far not been reached in this system. This limitation stems from the oscillating
electric fields employed in the rf trap, which causes the ion to perform a rapid
micromotion. During an atom-ion collision energy may be transferred from the
time-dependent trapping field into the atom-ion
system~\cite{major1968exchange,DeVoe:2009,
Zipkes:2011,Cetina:2012,
Chen:2014,Ewald:2015,Weckesser:2015,Weckesser:2016,Meir:2016,Rouse:2017,Secker:2017}.
In fact, runaway heating may occur when the atom is heavier than the ion. Cetina
{\it et al.}~\cite{Cetina:2012} calculated that the lowest temperatures may be
achieved for atom-ion combinations with large ion to atom mass ratios. They
theorize that Yb$^+$-Li, which has the largest mass ratio of any atom-ion
combination allowing straightforward laser cooling, may enter the quantum regime
after improving control over the trapping voltages to slightly beyond
state-of-the-art to compensate excess micromotion.

In this article, we calculate that the $s$-wave regime of Yb$^+$-Li should be in
reach with current technology and considering all known sources of excess
micromotion in the ion. We perform classical simulations of $^{171}$Yb$^+$ ions
in a Paul trap that are buffer-gas cooled by cold $^6$Li atoms using realistic
experimental parameters that we obtain from our experimental setup and from
parameters reported in the literature.  We further investigate the prospects of
collisional cooling of single ions and crystals of ions into the motional ground
state using a cloud of ultracold Li, taking into account experimental
imperfections. We give a limit on the remaining number of motional quanta that
can be expected and compute the cooling rate.  Motivated by the prospects of
a ultracold atom-ion system to form a solid-state emulator~\cite{Bissbort:2013}
we study the classical cooling dynamics for multiple trapped ions forming
a Coulomb-crystal within the cloud of atoms and show that the cooling dynamics is
very similar to that of a single trapped ion.

This article is organized as follows: First, we give the theoretical background
of ion trapping and micromotion as well as the model for simulating buffer-gas
cooling in section~\ref{sec:Theory}. In section~\ref{sec:Exp}, we describe the
experimental parameters and limitations in our experimental setup. We use these
parameters in the calculations of section~\ref{sec:Results1ion}, where we study
the thermalization of a single trapped ion experiencing each type of
micromotion. In sections~\ref{sect:crystals1D} and~\ref{sec:Results_crystals1D}
we describe the buffer-gas cooling of linear ion crystals, while
section~\ref{subsec:planarsub} describes the results for two-dimensional ion
crystals. Finally, we draw conclusions in section~\ref{sec:Conclusions}.

\section{Simulating buffer-gas cooled ions}\label{sec:Theory}
\subsection{Ion trapping in a linear quadrupole trap}\label{sect:traptheo}
The potential of a Paul trap as a function of ion position $\vec{r}$ can be
written as:
\begin{equation}\label{eqn:paulpotential}
  \Phi(\vec{r},t) = \frac{u_{\rm{dc}}}{2} \sum_{i=1}^3 \alpha_i r_i^2
  + \frac{u_{\rm{rf}}}{2}\cos\left(\Omega_{\rm {rf}}
  t\right)\sum_{i=1}^3\alpha_i' r_i^2
\end{equation}
with the positive, geometry- and voltage-dependent prefactors $u_{\rm{dc}}$ and $u_{\rm{rf}}$
and trap drive frequency $\Omega_{\rm {rf}}$.  To describe a linear Paul trap as
it is used in our experiment we have~\cite{Leibfried:2003}
\begin{equation}
  \alpha_1 = \alpha_2 = - \frac{1}{2} = -\frac{\alpha_3}{2} \quad {\rm{and}} \quad
  \alpha_1' = - \alpha_2' = 1\,,\quad \alpha_3'=0\,.
\end{equation}
For this choice, the confinement along the $3$-axis is supplied by
a time-independent harmonic trapping potential $\propto u_{\rm{dc}}$, whereas the radial confinement
is supplied by the oscillating field $\propto u_{\rm{rf}}$. Note that in reality the $\alpha_{1,2}$
coefficients are chosen to slightly differ from each other to lift the
degeneracy in the resulting radial trap frequencies.  The electric field is
given by
\begin{eqnarray}
  \nonumber
  \vec{E}\left(\vec{r},t\right)
  &= -\vec{\nabla}\Phi\left(\vec{r},t\right)
  \\
  &= - u_{\rm{dc}} \left(r_3 \hat{e}_3 - \frac{1}{2}\left(r_1 \hat{e}_1 + r_2
  \hat{e}_2\right)\right) - u_{\rm{rf}}\cos\left(\Omega_{\rm{rf}}t\right) \left(r_1 \hat{e}_1 - r_2 \hat{e}_2\right)\,,
\end{eqnarray}
with the unit vectors $\hat{e}_i$ in the $i$-th direction. With that, the
equation of motion for a single ion with mass $m_{\rm{ion}}$ and positive
charge $+e$ can be written as the Mathieu equation~\cite{Berkeland:1998}
\begin{equation}
  \ddot r_i + \left(a_i + 2q_i \cos\left(\Omega_{\rm{rf}}
  t\right)\right)\frac{\Omega_{\rm{rf}}}{4}r_i = 0\,,\quad
  i\in\{1,2,3\}\equiv\{x,y,x\}\,,
\end{equation}
with the parameters
\begin{equation}
  a_1 = a_2 = - \frac{1}{2} a_3 = -\frac{2 e u_{\rm{dc}}}{m_{\rm{ion}}
  \Omega_{\rm{rf}}^2}\,,\quad q_1 = - q_2 = \frac{2
  e u_{\rm{rf}}}{m_{\rm{ion}}\Omega_{\rm{rf}}^2}\,,\quad q_3 = 0\,,
\end{equation}
which are the stability parameters of the Paul trap~\cite{Leibfried:2003}.
Usually, Paul traps are operated at a region where $|a_i|$, $q_i^2 \ll 1$, which
can be achieved by properly choosing a suitable combination of
$\Omega_{\rm{rf}}$ and the static and rf electrode voltages $\propto
u_{\rm{rf}}, u_{\rm{dc}}$.  An approximate solution in first order in $q_i$ can
then be obtained by
\begin{equation}
  r_i (t)\approx
  r_i^{(1)}\cos\left(\omega_i t + \phi_i\right)
  \left(1+\frac{q_i}{2}\cos\left(\Omega_{\rm{rf}} t\right)\right)\,,
\end{equation}
where the phase $\phi_i$ and amplitude $r_i^{\left(1\right)}$
are determined by the initial condition at $t=0$. The
motion consists of a low frequency part, oscillating with the secular frequency
$\omega_i \approx \frac{1}{2}\Omega_{\rm{rf}}\sqrt{a_i + \frac{1}{2}q_i^2}$, thus
requiring $a_i + \frac{1}{2}q_i^2 > 0$ for a stable solution. In the
two radial directions, the rf field drives the so-called micromotion that
oscillates in phase with the rf drive and whose amplitude depends on the secular
motion amplitude and $q_i$-parameters. Note that in a real ion trap
imperfections in the electrode alignment can lead to a small rf field
component also in the axial direction, effectively setting $q_z \neq 0$.
By averaging over the secular oscillation period $T_{i} = \frac{2 \pi}
{\omega_{i}}$, one can obtain the average
kinetic energy in each coordinate,
\begin{equation}
  \label{eqn:ekinmode}
  \bar{E}_{\rm{kin},i}  = \frac{1}{2}
  m_{\rm{ion}} \langle \dot{r}_{i}\left(t\right)^2 \rangle_{T_{i}}
  \approx\frac{1}{4}m_i r_i^{\left(1\right)\,2} \left(\omega_i^2
  + \frac{1}{8}q_i^2\Omega_{\rm{rf}}^2\right)\,,
\end{equation}
where the assumption $\Omega_{\rm{rf}} \gg \omega_i$ was used.

\subsection{Excess micromotion}\label{sec:typesofemm}
Besides the intrinsic micromotion of the ion caused by the radiofrequency drive,
stray charges on the trap electrodes, imperfections of the trap assembly and
electrical connection as well as finite-size effects can lead to various types of so-called
excess micromotion~\cite{Berkeland:1998} that affects the average kinetic energy
of the ion and prevents reaching ultracold temperatures. Below, we will briefly
describe the three different kinds of excess micromotion that occur in a linear
Paul-trap, and in section~\ref{sec:Exp} we will describe how these can be
detected and compensated in our experiment.

Stray electric fields $E_{\rm{rad}}$ in the radial direction may push the ions
away from the rf null, where they experience the presence of the radiofrequency
field even without any secular energy. This type of excess micromotion we will
call radial micromotion.  The modified Mathieu-equation of the system including
$\vec{E}_{\rm{rad}}$ reads~\cite{Berkeland:1998}
\begin{equation}
  \ddot{r_i} + \left(a_i + 2 q_i \cos\left(\Omega_{\rm{rf}} \right)\right)
  \frac{\Omega_{\rm{rf}}}{4} r_i = \frac{e E_{\rm{rad},i}}{m_{\rm{ion}}}\,,
\end{equation}
with $i\in\{x,y\}\equiv\{1,2\}$. To lowest order in $q_i$, the solution is given by
\begin{equation}\label{eqn:position}
  r_i (t)\approx
  \left(r_i^{(0)} + r_i^{(1)}\cos\left(\omega_i t + \phi_i\right)\right)
  \left(1+\frac{q_i}{2}\cos\left(\Omega_{\rm{rf}} t\right)\right)\,,
\end{equation}
with the equilibrium position of the secular motion being shifted by $r_i^{(0)}
\approx e E_{\rm{rad},i}/(m_{\rm{ion}}\omega_i^2)$. For both radial directions
this additional shift
leads to an energy
\begin{equation}\label{eqn:enemm}
  E_{\rm{emm},i} = \frac{1}{16} m_{\rm{ion}}\left(q_i r_i^{(0)} \Omega_{\rm{rf}}\right)^2
  = \frac{4}{m_{\rm{ion}}}\left(\frac{q_i e E_{\rm{rad},i}\Omega_{\rm{rf}}}{8 \omega_i^2}\right)^2\,,
\end{equation}
in first order. Typically this micromotion can be
compensated by applying an external static electric field to cancel the stray
field at the position of the ion. Note that a stray field component in axial
direction only changes the ion's axial equilibrium position, not the kinetic
energy of the system.

Axial excess micromotion is mainly caused by the finite size of the trap leading
to a radiofrequency pickup on the dc end caps. This pickup leads to an
additional, position-independent, oscillating field with amplitude $E_{\rm{ax}}$
in axial direction that modifies the axial Mathieu-equation to
\begin{equation}
  \ddot{r}_z + a_z r_z =
  \frac{e E_{\rm{ax}} \cos\left(\Omega_{\rm{rf}} t\right)}{m_{\rm{ion}}}\,,
\end{equation}
leading to the analytic solution of a driven harmonic oscillator,
\begin{equation}
  r_z(t) = \frac{e E_{\rm{ax}}
  \cos\left(\Omega_{\rm{rf}}t\right)}{m_{\rm{ion}}\omega_z^2-\Omega_{\rm{rf}}^2}+r_z^{(1)}
  \cos\left(\omega_z t+\phi_z \right)\,,
\end{equation}
thus increasing the average kinetic energy by the term
\begin{equation}\label{eqn:enamm}
  E_{\rm{emm},z} =\frac{\left(e E_{\rm{ax}} \Omega_{\rm{rf}}\right)^2}{4
  m_{\rm{ion}}\left(\Omega_{\rm{rf}}^2-\omega_z^2\right)^2}\,.
\end{equation}
While it is hard to minimize this pickup by trap design, it can be reduced by
appropriate low-pass filters connected to the end cap electrodes or
injecting an rf field with opposite phase at one of the end cap
electrodes~\cite{MeirThesis:2016}.

Phase- or quadrature micromotion~\cite{Meir:2018} is caused by a phase
difference $\delta\phi_{\rm{rf}}$ between the radiofrequency voltages on the
opposing rf-electrodes, e.g.\ in $x$-direction.  The phase micromotion can be
approximately described by an additional homogeneous oscillating field in the
direction of the electrodes~\cite{Berkeland:1998}, $\vec{E}_{\rm{ph}} \approx
\frac{1}{4e} q_x m_{\rm{ion}} \delta \phi_{\rm{rf}} \Omega_{\rm{rf}}^2
R_{\rm{trap}} \sin\left(\Omega_{\rm{rf}} t\right)\hat{e}_x$, where
$R_{\rm{trap}}$ is half the distance between the two rf-electrodes.  The field
leads to the modified Mathieu-equation
\begin{equation}
  \ddot{r}_x + \left(a_x + 2 q_x \cos\left(\Omega_{\rm{rf}} t\right)\right)
  = \frac{1}{4} q_x R_{\rm{trap}} \delta\phi_{\rm{rf}} \Omega_{\rm{rf}}^2
  \sin\left(\Omega_{\rm{rf}}\right)\,.
\end{equation}
The solution in first order approximation then reads
\begin{eqnarray}
  r_x(t) &= r_x^{\left(1\right)} \cos\left(\omega_x t + \phi_x\right)\left(1+\frac{1}{2}q_x
  \cos\left(\Omega_{\rm{rf}}\right)\right)\\
  \nonumber
  &-\frac{1}{4}q_x
  R_{\rm{trap}}\delta\phi_{\rm{rf}}\sin\left(\Omega_{\rm{rf}}t\right)\,,
\end{eqnarray}
leading to an additional term in the average kinetic energy in the x-direction
of
\begin{equation}\label{eqn:enpmm}
  E_{\rm{phmm}} = \frac{1}{64}m_{\rm{ion}}\left(q_x
  R_{\rm{trap}}\delta\phi_{\rm{rf}}\Omega_{\rm{rf}}\right)^2\,.
\end{equation}
Compensation of the quadrature micromotion is possible but technically
challenging, for example by using two coherent rf drives with an adjustable
phase difference between their respective outputs.

\subsection{Modeling atom-ion collisions}
We model the atom-ion interaction by the long range attractive $r^{-4}$ induced
dipole-monopole potential~\cite{Langevin:1905} and an additional repulsive
$r^{-6}$ term at short ranges to simulate a hard core potential,
\begin{equation}
  V_{\rm{a-i}}(r)  = C_4\left(-\frac{1}{2 r_{\rm{a-i}}^4}+
  \frac{C_6}{r_{\rm{a-i}}^6}\right)\,,\quad\label{eqn:aipot}
  r_{\rm{a-i}}  = ||\vec{r}_{\rm{a}}-\vec{r}_{\rm{i}}||\,,
\end{equation}
where $C_6$ is given as a fraction of $C_4$, leading to a zero
crossing of the potential at a distance of $r_{\rm{hc}} = \sqrt{2 C_6}$.
The attractive
$r^{-4}$ potential leads either to glancing collisions where mainly the
momentum direction of the partners slightly change, or to Langevin
collisions where atom and ion are spiraling into each other, enabling for
a large energy and momentum transfer. Langevin collisions
occur when the impact parameter $b$ is less than the
Langevin range $b_c=(2 C_4/E_{\rm{col}})^{1/4}$~\cite{Langevin:1905}. Notably, the Langevin
collision rate $\Gamma_{\rm{L}} = 2\pi
\rho_{\rm{a}} \sqrt{C_4/\mu}$ is only dependent on the atomic density
$\rho_{\rm{a}}$ and the $C_4$ potential as well as the reduced mass $\mu$ of the
two body system but not the collision energy $E_{\rm{col}}$.

To numerically simulate the classical dynamics, a single atom is introduced on
a sphere with constant diameter $r_{\rm{0}}$ centered at the equilibrium
position of the ion before each collision. The diameter of the sphere has
to be large enough to prevent sudden changes in the potential
energy of the ion as well as leaving enough room for the ion orbit
due to micromotion and secular motion. On the other hand, the radius should
not be too large to prevent unnecessary long propagation times. To fairly sample
the flow of atoms, the atom launching coordinates are sampled from a uniform
distribution on the sphere surface at the beginning of each collision event.
To obtain a starting position, two points $p$ and $q$ are randomly picked from the
interval $[0,1]$. The azimuthal angle $\phi$ is then given by $\phi = 2 \pi
\cdot p$ and the polar angle $\theta = \arccos\left(2
q - 1\right)$~\cite{weisstein2002sphere}, from which the Cartesian
coordinates are derived,
\begin{equation}
  r_{\rm{a,1}} = r_{\rm{0}} \cos\left(\phi\right) \sin\left(\theta\right)\,\\
  r_{\rm{a,2}} = r_{\rm{0}} \sin\left(\phi\right) \sin\left(\theta\right)\,\\
  r_{\rm{a,3}} = r_{\rm{0}} \cos\left(\theta\right)\,.
  \label{eqn:sphericalcoord}
\end{equation}
The initial velocity $\vec{v}_{\rm{a}}$ of the atoms is then
sampled from the probability distribution $P_{\Phi}(\vec{v}_{\rm{a}},T_{\rm{a}})$
of the flux of thermal atoms
\begin{equation}\label{eqn:fluxatoms}
\Phi(\vec{v}_{\rm{a}})=\rho_{\rm{a}} 4\pi r_0^2
\hat{e}_r\cdot\vec{v}_{\rm{a}}\,,
\end{equation}
at a given temperature $T_{\rm{a}}$ and density $\rho_{\rm{a}}$ through the sphere,
\begin{eqnarray}
  P_{\Phi}(\vec{v}_{\rm{a}},T_{\rm{a}})\mathrm d^3 v_{\rm{a}}
  & = \frac{\Phi(\vec{v}_{\rm{a}})}{\rho_{\rm{a}} 4 \pi r_0^2} \frac{m_{\rm{a}}^2}{2\pi
  \left(k_{\rm{B}}T_{\rm{a}}\right)^2}e^{-\frac{m_{\rm{a}}\vec{v}_{\rm{a}}^2}{2
  k_{\rm{B}}T_{\rm{a}}}}\mathrm d^3v_{\rm{a}}\\
  & = \frac{m_{\rm{a}}^2}{2\pi
  \left(k_{\rm{B}}T_{\rm{a}}\right)^2}
   v_{{\rm{a}},r} e^{-\frac{m_{\rm{a}}v_{{\rm{a}},r}^2}{2
  k_{\rm{B}}T_{\rm{a}}}}
  \mathrm d v_{{\rm{a}},r}
  e^{-\frac{m_{\rm{a}}v_{\rm{a},\phi}^2}{2 k_{\rm{B}}T_{\rm{a}}}}
  \mathrm d v_{\rm{a},\phi}
  e^{-\frac{m_{\rm{a}}v_{\rm{a},\theta}^2}{2 k_{\rm{B}}T_{\rm{a}}}}
  \mathrm d v_{\rm{a},\theta}\,,
\label{eqn:vatomsflux}
\end{eqnarray}
meaning that the velocity components $v_{\rm{a},\phi}$ and $v_{\rm{a},\theta}$
tangential to the sphere surface are picked from one-dimensional Gaussian
distributions with a standard deviation of $\sigma = \sqrt{k_{\rm{B}}
T_{\rm{a}}/m_{\rm{a}}}$ each, whereas the perpendicular velocity $v_{{\rm{a}},r}$
is picked from a Weibull-distribution with shape parameter $k=2$ and scale
parameter $\lambda = \sqrt{2 k_{\rm{B}}T_{\rm{a}}/ m_{\rm{a}}}$.  Only atoms
flying towards the center of the sphere will have a chance to collide with the
ions, therefore it is enforced that $v_{{\rm{a}},r} < 0$.  The Cartesian
components of the velocity are then obtained by a coordinate transformation of
the spherical components.

After the atom is introduced, the atom-ion system is propagated forward in time
by an adaptive step-size Runge-Kutta algorithm of fourth
order~\cite{press2007numerical} maintaining a desired relative accuracy in each
coordinate $p_{\rm{tol}}$ in each coordinate as explained in~\ref{subs:reality}.
This allows for a fast
propagation when atom and ion are far away from each other and an accurate
propagation when the interaction is strong.  To define the end of a collision
event, a second sphere of radius $r_1$, slightly bigger than $r_0$ is
introduced.  Once the atom leaves this second sphere, all ion's coordinates at
this time are intermediately stored and the energy of the ion is determined.
For this, the ion motion is propagated further for a fixed amount of time
$t_{\rm{kin}} \gg 2\pi/\omega_i$ using $N_{\rm{kin}}$ fixed time steps of duration $\Delta
t_{\rm{kin}}$, sufficiently small to resolve micromotion. During this
additional propagation all ion trajectories are stored. From the velocities
$\vec{v}_{{\rm{i}},n}$ at each point in time the average kinetic energy
\begin{equation}
  \bar{E}_{\rm{kin}} = \frac{1}{N_{\rm{kin}}}\frac{1}{2}
  m_{\rm{ion}}\sum_{k=1}^{N_{\rm{kin}}} \sum_{n=1}^{N_{\rm{ions}}}
  \vec{v}_{{\rm{i}},n}\left(t_k\right)^2 = \frac{3 N_{\rm{ions}}}{2} k_{\rm{B}}
  T_{\rm{kin}},
  \label{eqn:ekinavg}
\end{equation}
is computed, which can be used to determine the collision energy but does not
contain any information about how much secular energy is stored in the
vibrational modes of the ion. The decomposition of the kinetic energy into
micromotion and vibrational energy will be discussed in
section~\ref{sect:crystals1D}.  Note that within this article we will often
mention the kinetic temperature $T_{\rm{kin}}$, although due to the included
micromotion energy, technically it is not a temperature but the average kinetic
energy in units of $3 N_{\rm{ions}} k_{\rm{B}} / 2$.

\begin{table}[htpb]
	\centering
		\begin{tabular}{cccc}
			\hline
		 \multicolumn{1}{|c|}{Parameter} & \multicolumn{1}{c|}{Value}&
      \multicolumn{1}{c|}{Comment}& \multicolumn{1}{c|}{Section} \\[0.3ex]
			\hline

       \multicolumn{1}{|c|}{\textbf{$f_z$}} &
       \multicolumn{1}{c|}{42.426\,kHz}&
       \multicolumn{1}{c|}{axial trap frequency}&  \multicolumn{1}{c|}{--} \\  \hline

       \multicolumn{1}{|c|}{\textbf{$f_{\rm{rf}}$}} &
       \multicolumn{1}{c|}{2\,MHz}&
       \multicolumn{1}{c|}{rf-drive frequency}&  \multicolumn{1}{c|}{--} \\  \hline		

       \multicolumn{1}{|c|}{\textbf{$q_x$}} &
       \multicolumn{1}{c|}{0.219}&
       \multicolumn{1}{c|}{rad. $q$-parameter}&  \multicolumn{1}{c|}{--} \\  \hline		

       \multicolumn{1}{|c|}{\textbf{$q_y$}} &
       \multicolumn{1}{c|}{$-q_x + q_z$}&
       \multicolumn{1}{c|}{rad. $q$-parameter}&  \multicolumn{1}{c|}{} \\  \hline		

       \multicolumn{1}{|c|}{\textbf{$q_z$}} &
       \multicolumn{1}{c|}{$0$}&
       \multicolumn{1}{c|}{ax. $q$-parameter}&  \multicolumn{1}{c|}{\ref{subs:qzscan}} \\  \hline		

       \multicolumn{1}{|c|}{\textbf{$T_{\rm{sec}}^{\left(0\right)}$}} &
       \multicolumn{1}{c|}{$0\,\mu$K}&
       \multicolumn{1}{c|}{initial ion temp.}&
       \multicolumn{1}{c|}{--} \\  \hline		

       \multicolumn{1}{|c|}{\textbf{$T_{\rm{a}}$}} &
       \multicolumn{1}{c|}{$2\,\mu$K}&
       \multicolumn{1}{c|}{atomic bath temp.}&
       \multicolumn{1}{c|}{\ref{subs:bathtemp}} \\  \hline		

       \multicolumn{1}{|c|}{\textbf{$r_0$}} &
       \multicolumn{1}{c|}{$0.6\,\mu$m}&
       \multicolumn{1}{c|}{atom launch sphere rad.}&
       \multicolumn{1}{c|}{\ref{subs:reality}} \\  \hline		

       \multicolumn{1}{|c|}{\textbf{$r_1$}} &
       \multicolumn{1}{c|}{$1.005\cdot r_0$}&
       \multicolumn{1}{c|}{atom escape sphere rad.}&
       \multicolumn{1}{c|}{\ref{subs:reality}} \\  \hline		

       \multicolumn{1}{|c|}{\textbf{$p_{\rm{tol}}$}} &
       \multicolumn{1}{c|}{$10^{-10}$}&
       \multicolumn{1}{c|}{relative num.\ tolerance}&
       \multicolumn{1}{c|}{\ref{subs:reality}} \\  \hline		

       \multicolumn{1}{|c|}{\textbf{$N_{\rm{fft}}$}} &
       \multicolumn{1}{c|}{$2^{14}$}&
       \multicolumn{1}{c|}{Fourier grid size}&
       \multicolumn{1}{c|}{\ref{subs:reality}} \\  \hline		

       \multicolumn{1}{|c|}{\textbf{$\Delta t_{\rm{fft}}$}} &
       \multicolumn{1}{c|}{$50\,$ns}&
       \multicolumn{1}{c|}{Fourier time resolution}&
       \multicolumn{1}{c|}{\ref{subs:reality}} \\  \hline		

       \multicolumn{1}{|c|}{\textbf{$\Delta t_{\rm{kin}}$}} &
       \multicolumn{1}{c|}{$5\,$ns}&
       \multicolumn{1}{c|}{time grid for $\bar{E}_{\rm{kin}}$}&
       \multicolumn{1}{c|}{--} \\  \hline		

       \multicolumn{1}{|c|}{\textbf{$C_4$}} &
       \multicolumn{1}{c|}{$5.607\cdot10^{-57}\,{\rm{J}}\cdot{\rm{m}}^4$}&
       \multicolumn{1}{c|}{attr.\ int.\ coeff.}&
       \multicolumn{1}{c|}{--} \\  \hline		

       \multicolumn{1}{|c|}{\textbf{$C_6$}} &
       \multicolumn{1}{c|}{$5\cdot10^{-19}\,{\rm{m}}^2\cdot C_4$}&
       \multicolumn{1}{c|}{rep.\ int.\ coeff.}&
       \multicolumn{1}{c|}{\ref{subs:reality}} \\  \hline		

       \multicolumn{1}{|c|}{\textbf{$\vec{E}_{\rm{rad}}$}} &
       \multicolumn{1}{c|}{$\{0,0,0\}\,$V/m}&
       \multicolumn{1}{c|}{dc offset field}&
       \multicolumn{1}{c|}{~\ref{subs:radex}~\&~\ref{subs:4ionsemm}~} \\  \hline		

       \multicolumn{1}{|c|}{\textbf{$E_{\rm{ax}}$}} &
       \multicolumn{1}{c|}{$0$\,V/m}&
       \multicolumn{1}{c|}{axial rf pickup ampl.}&
       \multicolumn{1}{c|}{~\ref{subs:axex}~\&~\ref{subs:4ionsemm}~} \\  \hline		

       \multicolumn{1}{|c|}{\textbf{$\delta \phi_{\rm{rf}}$}} &
       \multicolumn{1}{c|}{$0$\,mrad}&
       \multicolumn{1}{c|}{rf phase mismatch}&
       \multicolumn{1}{c|}{~\ref{subs:phasemm}~\&~\ref{subs:4ionsemm}~} \\  \hline		
		\end{tabular}
	\caption{Parameters used for the numerical simulation of the atom-ion collisions,
  unless given otherwise in the text. If varied, the last column refers to the
  respective section where it is investigated.}
	\label{tab:simparams}
\end{table}

\subsection{Collision energy and $s$-wave limit}

The $s$-wave limit of $^{171}$Yb$^+$/$^6$Li is reached at a collision energy of:
\begin{equation}
  T_{\rm{s}} = \frac{\hbar^4}{2 k_{\rm{B}}\mu^2 C_4 }=8.6\,\mu{\rm{K}}\,.
\end{equation}
\noindent  The collision energy is given by the energy in the relative atom-ion
coordinate. In the experimentally relevant situation in which the ion has a much
larger kinetic energy than the atoms $T_{\rm{kin}}\gg T_{\rm{a}}$, the collision
energy is given by:
\begin{equation}
  E_{\rm{col}}/k_{\rm{B}}= T_{\rm{col}} \approx
  \frac{3}{2}\frac{\mu}{m_{\rm{ion}}} T_{\rm{kin}}\, .
\end{equation}
Therefore, to reach the quantum regime in the limit where
$T_{\rm{a}}\rightarrow 0$, the requirement for the ion is
$T_{\rm{kin}}<168\,\mu$K~\cite{Saito:2017}. In this work, we use
$T_{\rm{a}}=2\,\mu{\rm{K}} \ll T_{\rm{s}}$ such that $T_{\rm{kin}}\gg T_{\rm{a}}$ is
fulfilled in most circumstances. At the same time, this choice still allows for
a classical treatment of the atomic bath~\cite{Krych:2013}.

\section{Micromotion detection and compensation}\label{sec:Exp}
The experimental setup is described in detail in Ref.~\cite{Joger:2017}.
The linear Paul trap is made out of four blade electrodes with a distance of
$R_{\rm{trap}}= 1.5$\,mm to the trap center. End caps with a spacing of 10\,mm are
used to confine the ion along the axial direction. Two sets of additional
electrodes can be used for compensation of stray electric fields. Oscillating
voltages at a frequency of $\Omega_{\rm{rf}} = 2\pi \times 2$\,MHz and
an amplitude of
$V_0 = 75$\,V are applied to the blades and dc voltages of V$_{\rm{dc}} \approx
15$\,V to the end caps. This results in radial and axial trap frequencies of
$\omega_{\rm{rad}}\approx 2\pi \times 150\,$kHz and $\omega_{\rm{ax}}\approx
2 \pi \times 42\,$kHz. Below, we describe how we detect and compensate
micromotion in our setup and give limits on the attainable experimental
parameters. More details on the micromotion detection and compensation can be
found in Ref.~\cite{Joger:2017}.

A radial stray field component $E_{\rm{rad},i}$ leads not only to excess
micromotion but also to a shift in equilibrium position. Within the horizontal
direction, this can be detected by tracking the ions position for different
radial trap frequencies, shifting the ions position by
$r_{\rm{h}}^{(0)}\approx\frac{e E_{\rm{rad,h}}}{m_{\rm{ion}}
\omega_{\rm{h}}^2}$, for $\omega_x\approx\omega_y=\omega_{\rm{h}}$. We measure
the shift on the ion by imaging from the top.
We extract the ion's horizontal position from averaging
over five camera images at each radial trap frequency setting and fitting
a Gaussian function. From these measurements we conclude that
$E_{\rm{emm,h}}\,<\,0.5\,$V/m under optimal circumstances.

The ion's vertical position cannot be obtained with the camera as the imaging
system and vacuum system was designed to only image the ions
from the top. Instead, we use the magnetic field dependence of the
($^2S_{1/2},F=0,m_F=0) \leftrightarrow (^2S_{1/2},F=1,m_F=1$) hyperfine
splitting in $^{171}$Yb$^+$~\cite{Olmschenk:2007} for a determination of
position shifts as a function of trap frequency. To do so, we apply a vertical
magnetic field gradient of $g_{z}=0.15$\,T/m which leads to a frequency shift of
2.1\,kHz$/\mu$m. By comparing the frequency shift at radial confinements of
$\omega_{\rm{rad}}=2\pi \times 80\,$kHz and $\omega_{\rm{rad}}=2\pi \times
230\,$kHz using microwave Ramsey spectroscopy, we measure a dc electric field of
$E_{\rm dc}=0.29(2)\,$V$\cdot$m$^{-1}$ for 1\,V applied to the compensation
electrodes. From these measurements we conclude that
$E_{\rm{emm,v}}\,<\,0.3\,$V/m at optimal compensation.

The axial micromotion is obtained by measuring the line broadening of the
4.2\,MHz wide
$^2D_{3/2} \rightarrow \,^3D[3/2]_{1/2}$ transition at 935\,nm wavelength in
Yb$^+$~\cite{Olmschenk:2007}. For this, we use a laser beam aligned along the trap
axis~\cite{Berkeland:1998,Joger:2017}. We obtain an upper bound to the amplitude
of the oscillating electric field in the trap center of $E_{\rm{ax}} \leq
15\,$V$\cdot$m$^{-1}$, limited by the observed linewidth of the
$^2D_{3/2} \rightarrow \,^3D[3/2]_{1/2}$ transition at optimal compensation. By
measuring the axial micromotion at various ion positions along the trap axis, we
obtain $q_z=0.0023$.

Aligning the beam under 45$^{\circ}$ with respect to the trap axis allows us to
also check for quadrature micromotion, but none was detected. The observed
transition linewidth results in the limit $\delta\phi_{\rm{rf}}\,<\,0.65$~mrad.
Using a transition with a narrower linewidth, e.g.\ the 22\,Hz wide
$^2S_{1/2} \rightarrow
\,^2D_{5/2}$ clock transition at 411\,nm in Yb$^+$~\cite{Taylor:1997}, could improve these limits
significantly.

\section{A single ion in the cold buffer gas}\label{sec:Results1ion}
In this section, we present the simulation results for collisions between
a single trapped ion in a Paul trap using parameters that can be
achieved with the ion trap used in our experiment. We investigate the influence
of atomic bath temperature as well as the different kinds of micromotion on the
ion's average kinetic energy for realistic parameters. For simplicity, we start
our calculations with an ion that has no energy and observe how this ion
thermalizes with the atomic bath in a similar way as described in
Refs.~\cite{Cetina:2012,Meir:2016}. Although chosen for convenience, this
situation is also of experimental relevance, as the ion may be laser-cooled
close to its ground state of motion before the atoms are
introduced~\cite{Meir:2016}.

\subsection{Influence of the atomic bath temperature}\label{subs:bathtemp}
\begin{figure}[t]
  \begin{center}
  \includegraphics[width=0.45\textwidth]{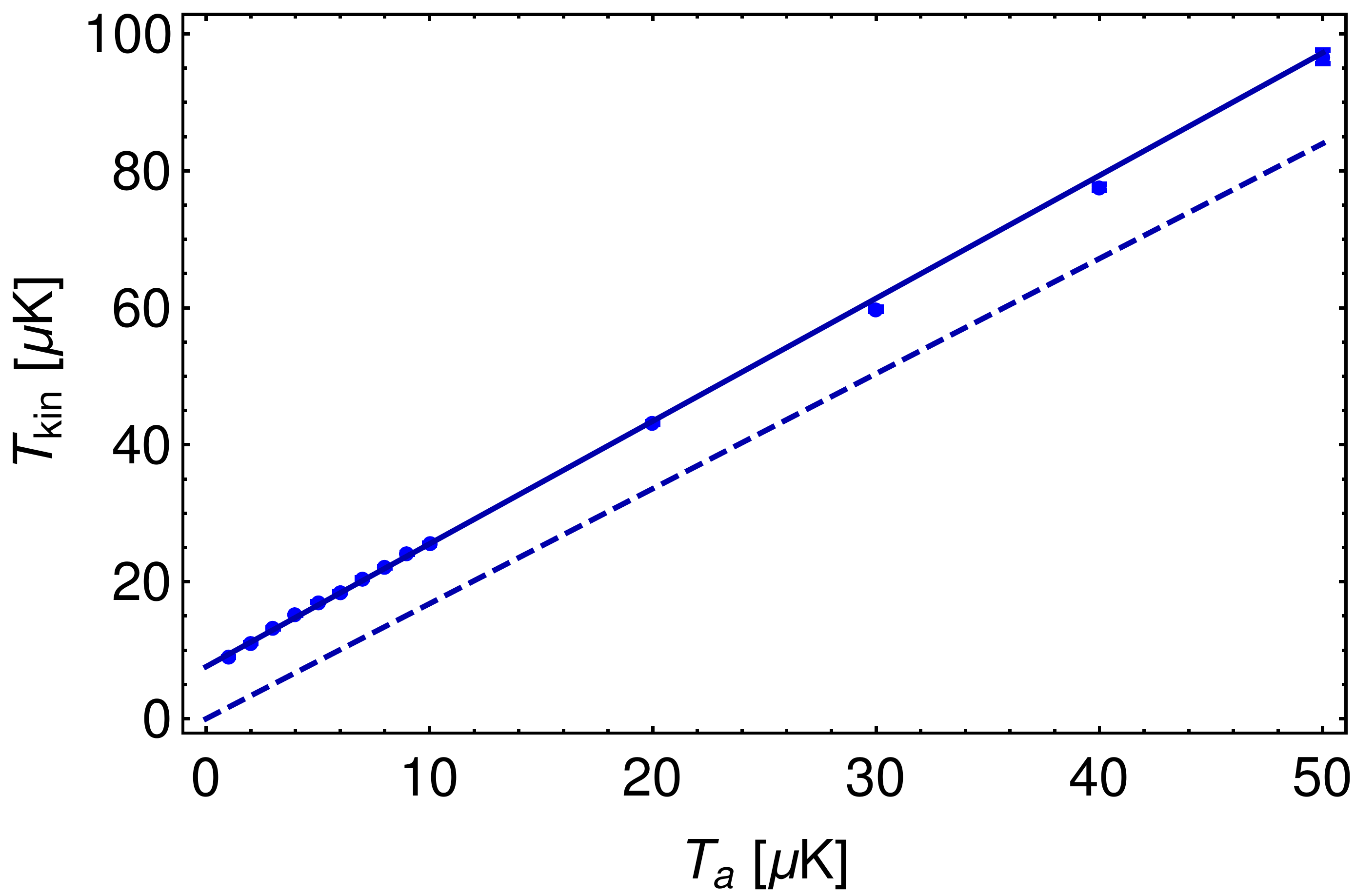}
  \hspace{0.05\textwidth}
  \includegraphics[width=0.45\textwidth]{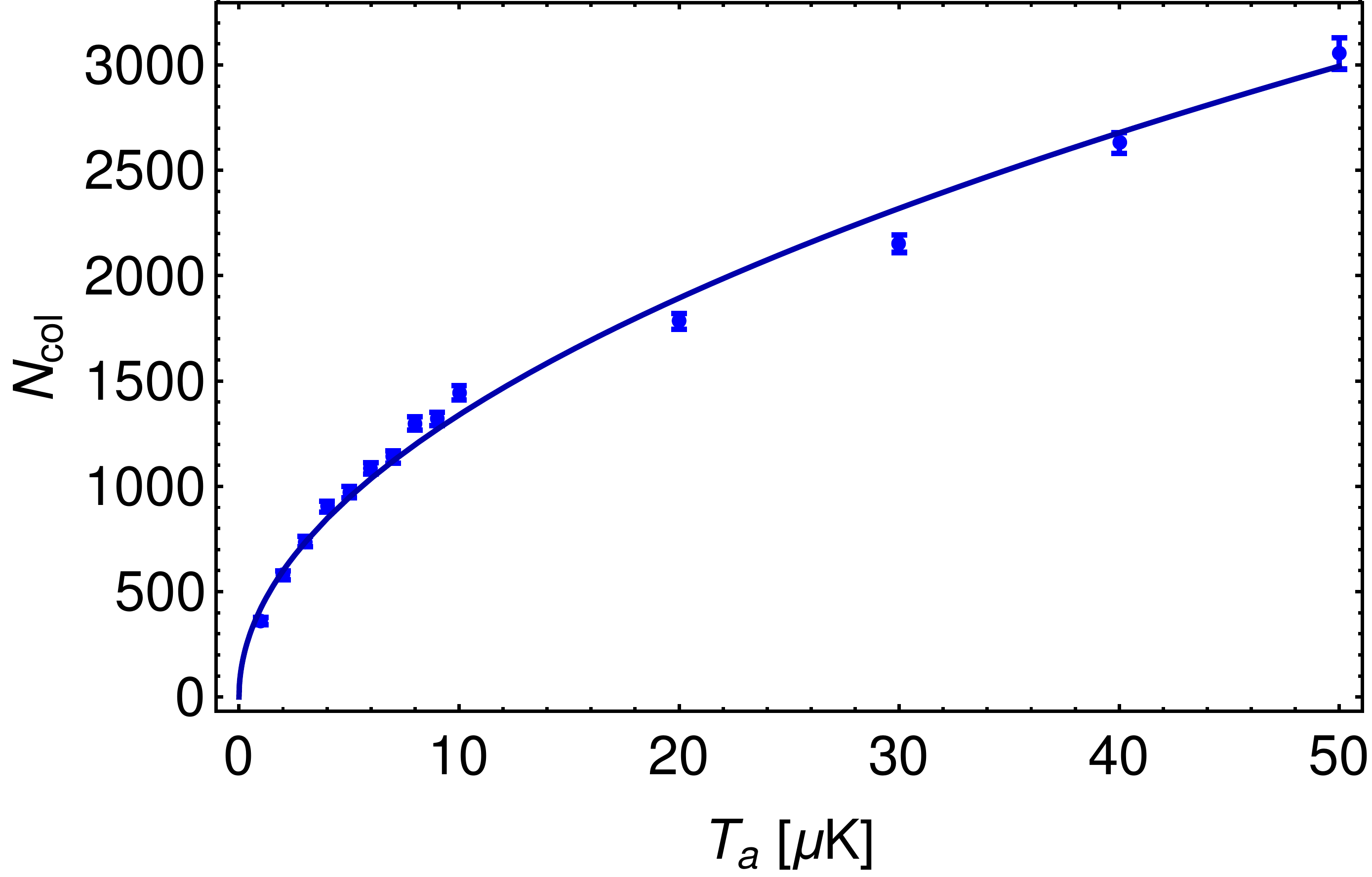}
  \end{center} \caption{Equilibrium average kinetic energy in units of
  $T_{\rm{kin}}$ of a single ion colliding with atoms (left) and number of
  collisions required to equilibrate (right) versus temperature of the atoms.
  The results were fit with a linear function (solid line, left) or a square
  root function (solid line, right) respectively. The dashed line corresponds to
  a hypothetical case without micromotion-induced heating as explained in the
  text.}
\label{fig:tatomscan}
\end{figure}

We simulated collisions for $T_{\rm{a}}$ between
$0-50\,\mu$K. The ion's averaged kinetic energy after equilibration in units of
$T_{\rm{kin}}$ and typical $1/e$ number of collisions to equilibrate $N_{\rm{col}}$
were determined by fitting an exponential function of the form
\begin{equation}\label{eq:expfit}
  T(n_{\rm{col}}) = \left(T_{\rm{kin}}-T_0\right)\left(1-e^{-\frac{n_{\rm{col}}}
  {N_{\rm{col}}}}\right)
  + T_0\,,
\end{equation}
to the results obtained by averaging at least 300 individual runs. The results
are shown in Fig.~\ref{fig:tatomscan}. The errors given in the plot correspond
to the standard errors of the fit parameters.
The average kinetic energy of the ion (left) in units of $T_{\rm{kin}}$ shows
a strictly linear dependence with a slope of 1.79(2) and offset of
$7.60(14)\,\mu$K. The dashed line shows the hypothetical case in which each
secular mode of the ion equilibrates with the temperature of the atom, according
to the approximate prediction of Eq.~\ref{eqn:ekinmode}. Its slope reads 1.68
using the trap parameters of the simulation. In particular, the deviation from
unity slope is given by the extra energy stored in the micromotion amplitude,
which is approximately $\frac{1}{2} k_{\rm{B}} T_{\rm{a}}$ extra per radial
direction~\cite{Berkeland:1998}, such that the energy of the atomic bath excites five kinetic degrees
of freedom instead of three, explaining the slope of approximately $5/3$.  The
deviation in slope of the simulated points with respect to the prediction is
expected to be caused by the approximations made to obtain the prediction (i.e.
$|a_i|$ and $q_i^2 \ll 1$, see Sec.~\ref{sect:traptheo}). The offset can be seen
as the direct influence of micromotion-induced heating, transferring energy from
the trap drive rf field into the secular motion of the ions, mediated by the
atoms.  The number of collisions required to equilibrate (right) follows
a square root function, which is to be expected, since the thermalization rate
$\Gamma_{\rm{eq}} = 1/N_{\rm{col}}$ should be directly proportional to the
fraction of events that lead to thermalization, namely Langevin collisions,
divided by the number of total events, $\Gamma_{\rm{eq}} \propto
\frac{\Gamma_{\rm{L}}}{\Phi(\vec{v}_{\rm{a}})}$, with $\Gamma_{\rm{L}}$ the
Langevin rate and $\Phi(\vec{v}_{\rm{a}})$ the flux into the sphere on which the
atoms start as defined in Eq.~\ref{eqn:fluxatoms}.  Thus, $N_{\rm{col}} \propto
\Phi(\vec{v}_{\rm{a}}) \propto \sqrt{T_{\rm{a}}}$\,.  From the fit, we obtain
a proportionality factor of $412(8)/\sqrt{\mu{\rm{K}}}$

\subsection{Influence of radial excess micromotion}\label{subs:radex}
\begin{figure}[t]
  \begin{center}
  \includegraphics[width=0.45\textwidth]{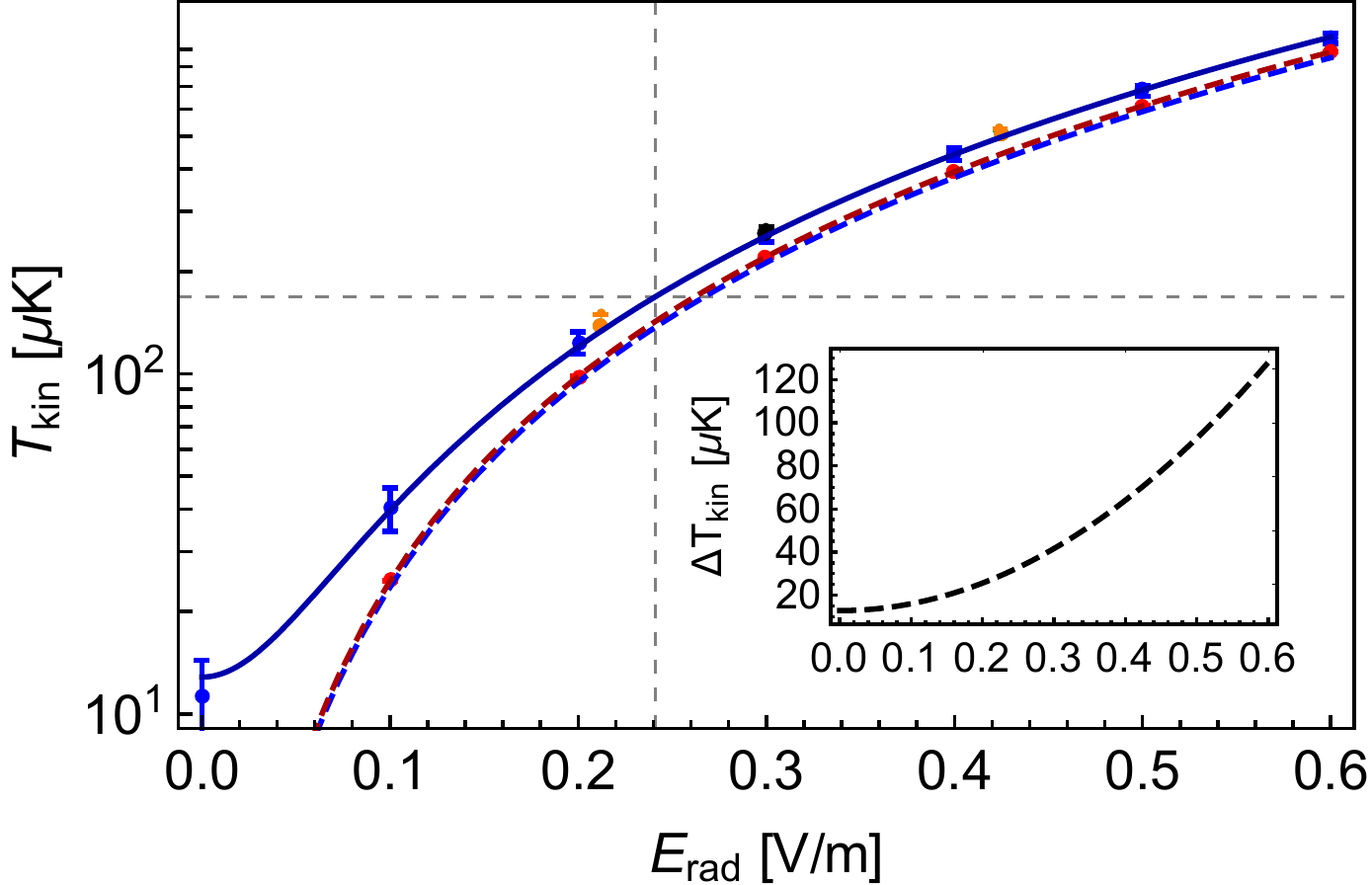}
  \hspace{0.05\textwidth}
  \includegraphics[width=0.45\textwidth]{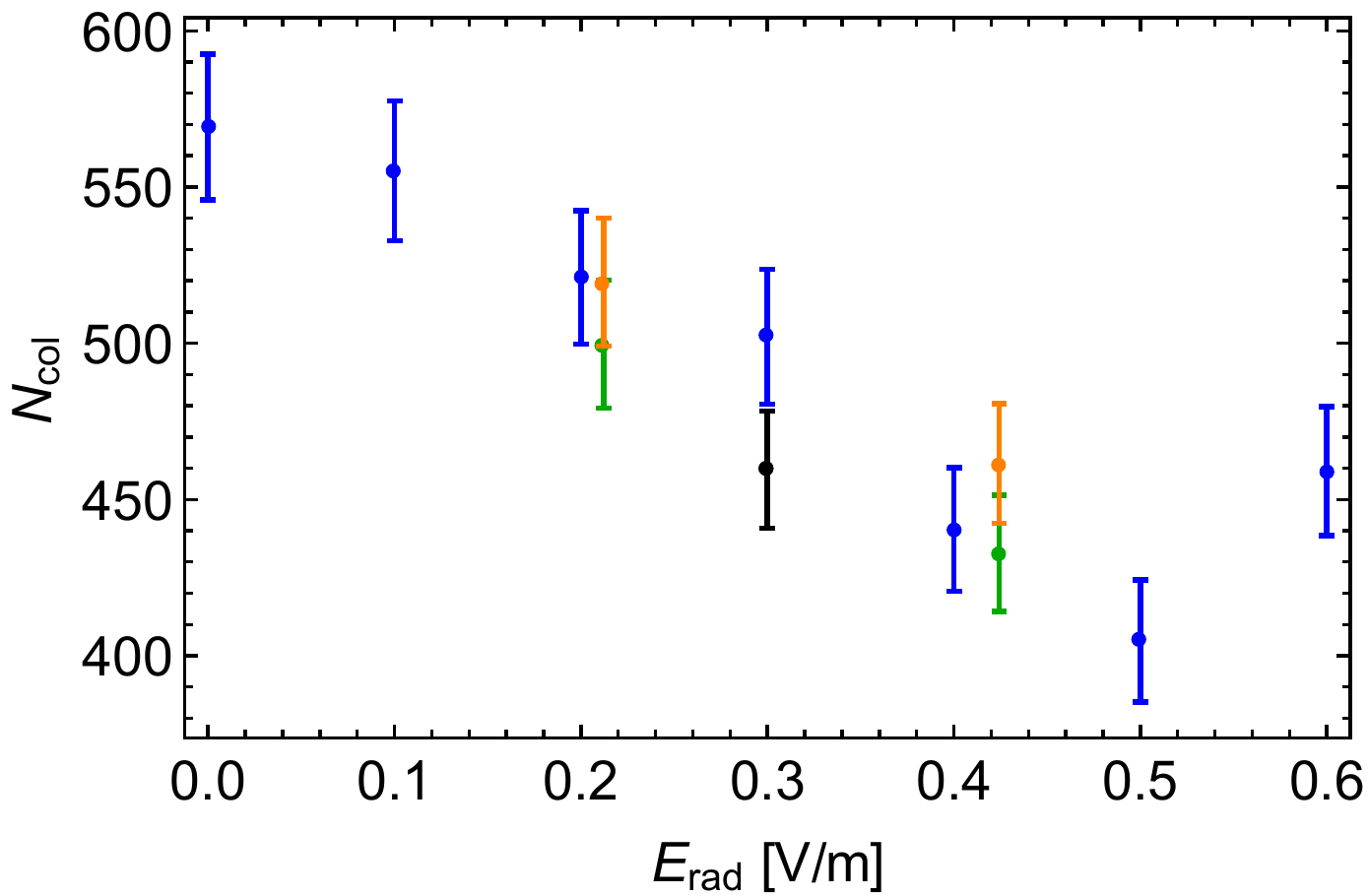}
\end{center} \caption{Equilibrium temperature of a single ion colliding with
  atoms at $2\,\mu$K (left) and number of collisions required to equilibrate
  (right) versus radial electric offset field. The results (blue points) were
  fit with a quadratic function (solid blue curve). The red points correspond to
  the average kinetic energy of an ion initialized at zero secular temperature
  without an atomic bath, along with a quadratic fit (red dashed curve) and the
  approximate theoretical amount of micromotion energy (blue dashed curve). The
  dashed gray lines indicate the $s$-wave temperature limit for
  $T_{\rm{a}}\rightarrow 0$.
  The inset shows the difference between the solid and dashed blue curve,
  resembling the micromotion induced heating. Other colors are explained in
  the text.}
\label{fig:oneionradial}
\end{figure}

In this paragraph, we investigate the influence of radial excess micromotion
caused by a stray electric field $\vec{E}_{\rm{rad}}$ on the average kinetic
energy of a single ion when immersed in a cold atomic bath of $2\,\mu$K. We
scanned $E_{\rm{rad}}$ over a range of 0.0 to 0.6\,V/m and determined the ion's
average kinetic equilibrium energy in units of $T_{\rm{kin}}$ and the typical
number of collisions required to equilibrate $N_{\rm{col}}$ according to
Eq.~$\ref{eq:expfit}$ by averaging over at least 300 individual runs for each
point. We additionally checked the influence of the radial direction of
$E_{\rm{rad}}$. The results are shown in Fig.~\ref{fig:oneionradial}.
The temperatures (blue) were calculated using a radial electric field in
$x$-direction only.  The results were fit with a quadratic function (solid blue
line),
\begin{equation}
  T_{\rm{kin}} = T_1 + \theta_{E_{\rm{rad}}} E_{\rm{rad}}^2\,,
\end{equation}
leading to a quadratic rise factor of $\theta_{E_{\rm{rad}}}
= 2680(15)\,\mu{\rm{K}}\cdot\left(\rm{V/m}\right)^{-2}$.
The dashed blue curve represents the approximate theoretical amount of kinetic
energy due to the presence of excess micromotion, according to
Eq.~\ref{eqn:enemm}, with a quadratic rise factor of
$2360\,\mu{\rm{K}}\cdot\left(\rm{V/m}\right)^{-2}$.  Also shown is the average
kinetic energy for an ion without an atomic bath present, initialized at zero
temperature (red points) along with a quadratic fit (red dashed line). The difference
between the solid blue curve and the dashed red curve corresponds approximately
to the amount of energy stored in the intrinsic micromotion and the secular
motion.  The point at $E_{\rm{rad}} = 0.3\,$V/m was simulated once with a factor
10 smaller tolerance parameter $p_{\rm{tol}}$ in the propagator to check for
numerical errors. The values in orange were taken using a dc field with equal
components $E_{\rm{rad},x} = E_{\rm{rad},y}$ in both radial directions, the
values in green (behind the orange points) with a dc field with opposite
components, $E_{\rm{rad},x} = -E_{\rm{rad},y}$, to check the influence of the
direction of $E_{\rm{rad}}$, showing no deviation from the fitted curve. The
number of collisions required to equilibrate (right) seems to slightly decrease
with increasing field amplitude.

\subsection{Influence of axial micromotion}\label{subs:axex}
\begin{figure}[t]
  \begin{center}
  \includegraphics[width=0.45\textwidth]{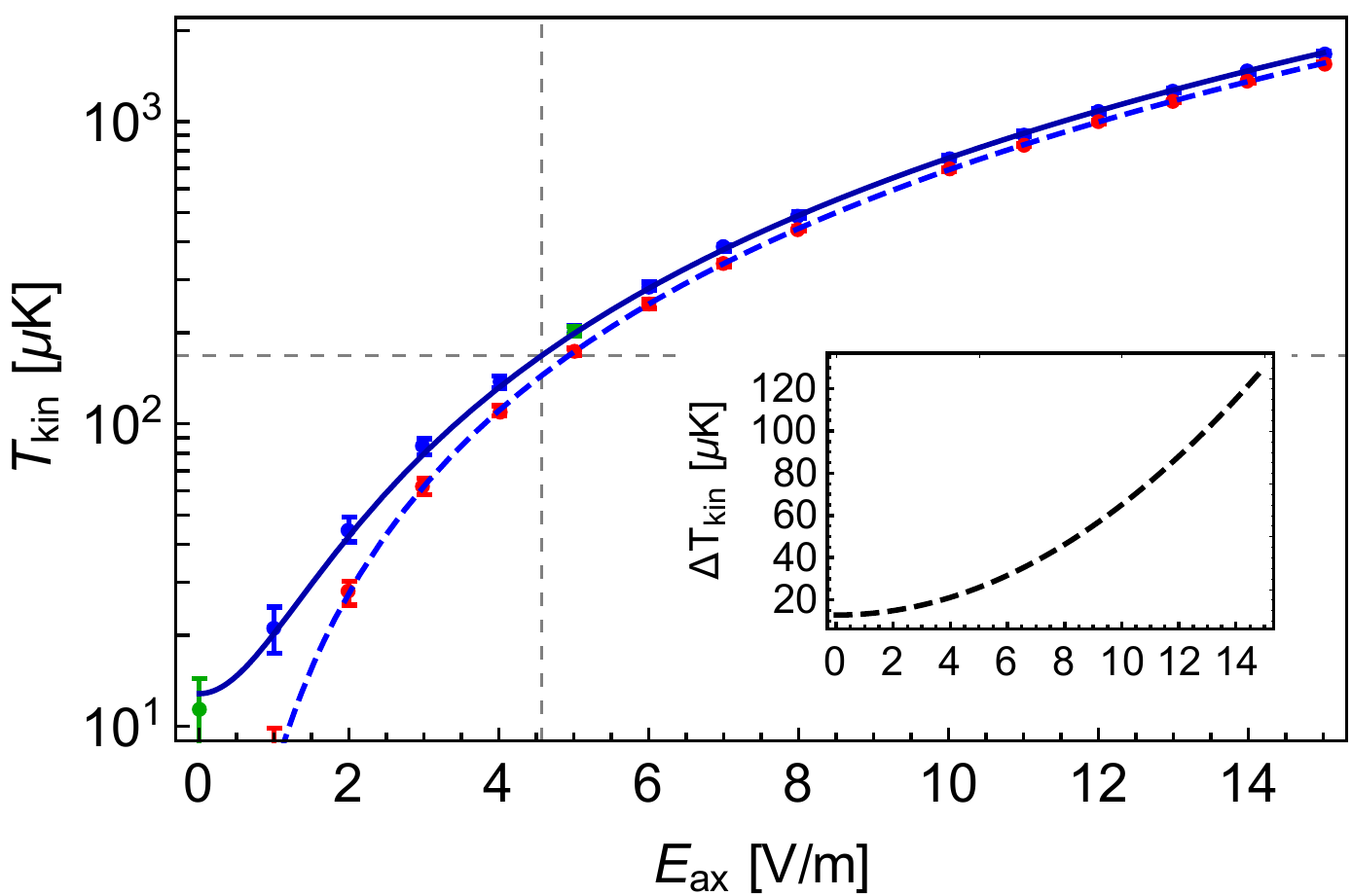}
  \hspace{0.05\textwidth}
  \includegraphics[width=0.45\textwidth]{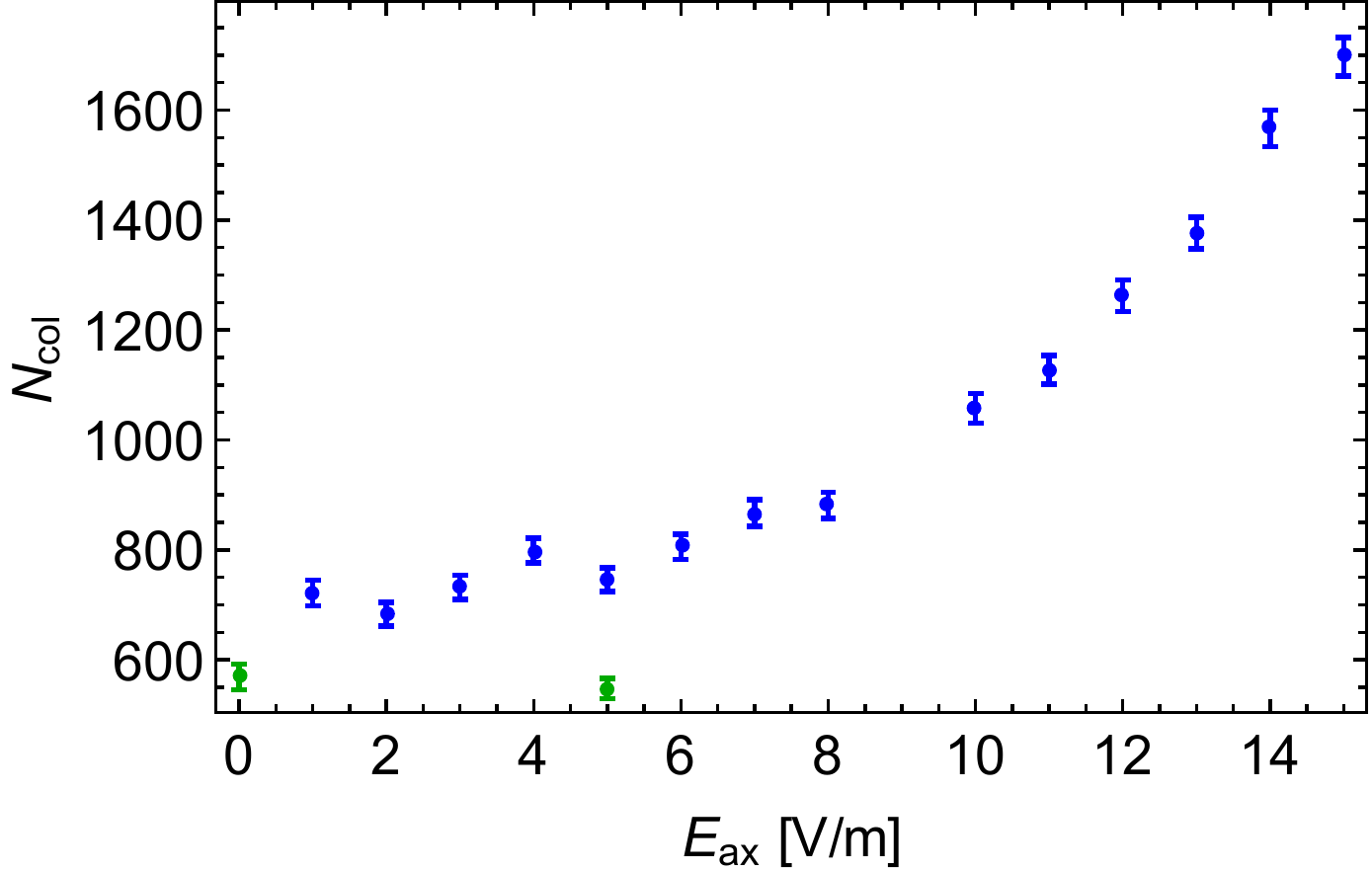}
\end{center} \caption{Equilibrium temperature of a single ion colliding with
  atoms at $2\,\mu$K (left) and number of collisions required to equilibrate
  (right) versus axial rf-field amplitude. The temperatures (blue) were fit with
  a quadratic function (solid blue curve). The red points correspond to an ion
  initialized at zero secular temperature without the presence of an atomic
  bath, in agreement with the approximate theoretical amount of micromotion
  energy (dashed blue curve). The dashed gray lines indicate the $s$-wave
  temperature limit.
  The inset shows the difference between the solid and dashed blue curve,
  resembling the micromotion induced heating.
  Colors are explained in the text.}
\label{fig:oneionaxial}
\end{figure}

In this paragraph, we investigate the influence of a homogeneous oscillating
electric field along the axial direction of the trap on the average kinetic
energy of a single ion when immersed in a cold atomic cloud at $2\,\mu$K. We
scanned the field amplitude $E_{\rm{ax}}$ from $0-15\,$V/m and determined the
ion's average kinetic equilibrium energy in units of $T_{\rm{kin}}$ and the
typical number of collisions required to equilibrate $N_{\rm{col}}$
according to Eq.~\ref{eq:expfit} by taking the average over at least 300
idividual runs for each point. The results are shown in
Fig.~\ref{fig:oneionaxial}.
The temperatures (blue) were fit with a quadratic function (solid blue curve),
\begin{equation}
  T_{\rm{kin}} = T_1 + \theta_{E_{\rm{ax}}} E_{\rm{ax}}^2\,,
\end{equation}
leading to a quadratic rise factor of $\theta_{E_{\rm{ax}}}
= 7.44(3)\,\mu{\rm{K}}\cdot\left(\rm{V/m}\right)^{-2}$. The dashed blue curve
represents the approximate theoretical amount of kinetic energy due to the axial
oscillating electric field, according to Eq.~\ref{eqn:enamm}, with a quadratic
dependence of $6.92\,\mu{\rm{K}}\cdot\left({\rm{V/m}}\right)^{-2}$, in agreement
with the points in red, showing the average kinetic energy of a crystal at zero
secular energy without atoms present.

Due to the large axial oscillation amplitudes at high values of $E_{\rm{ax}}$,
a fixed starting sphere causes the atoms to occasionally launch very close to
the ion, thus introducing unrealistic jumps in the potential energy that can
lead to unstable behavior. Therefore, the blue points were not obtained using
a starting sphere with fixed origin at the ion's equilibrium position, but
a comoving sphere around the ion's immediate position.  As a consequence, there
are events where the ion is moving away from the introduced atom such that the
atom is immediately registered as having escaped, leading to an increased number
of required collisions, (Fig.~\ref{fig:oneionaxial} right, blue points) as
compared to the non-comoving case (green points).  This effect seems to
increase with field amplitude.  A comoving starting sphere means that
especially very slow atoms that would usually cause a Langevin collision are
overseen. Therefore, the average contribution of the atom to the collision energy
increases. Since the ion temperature in this regime is dominated by the
micromotion energy anyways, this effect can be ignored.

\subsection{Influence of quadrature micromotion}\label{subs:phasemm}
\begin{figure}[t]
  \begin{center}
  \includegraphics[width=0.45\textwidth]{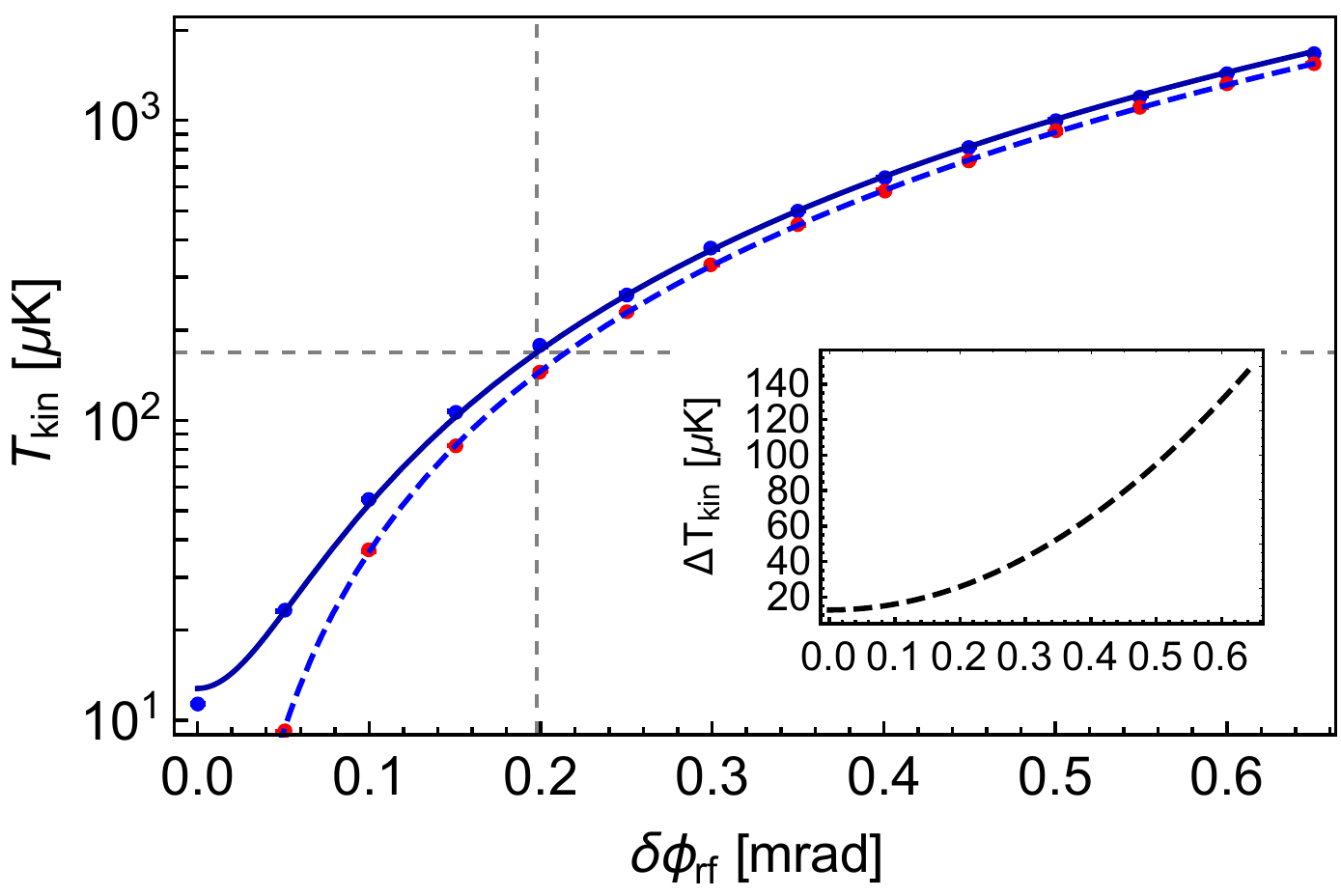}
  \hspace{0.05\textwidth}
  \includegraphics[width=0.45\textwidth]{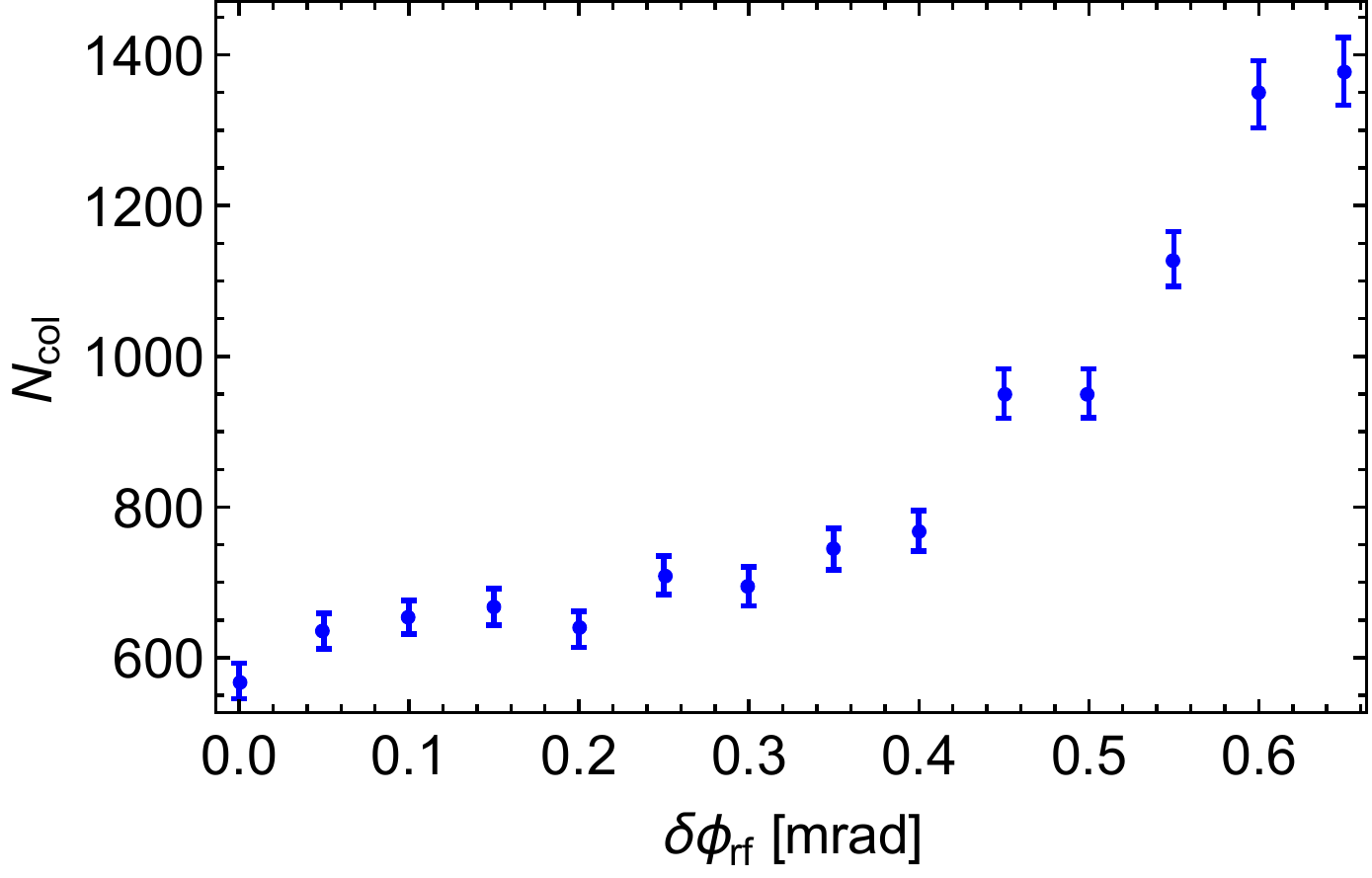}
\end{center} \caption{Equilibrium temperature of a single ion colliding with
  atoms at $2\,\mu$K (left) and number of collisions required to equilibrate
  (right) versus phase mismatch of the two radial rf-components of the potential.
  The temperatures (blue) were fit with
  a quadratic function (solid blue curve). The red points correspond to an ion
  initialized at zero secular temperature without the presence of an atomic
  bath, in agreement with the approximate theoretical amount of micromotion
  energy (dashed blue curve). The dashed gray
  lines indicate the $s$-wave temperature limit.
  The inset shows the difference between the solid and dashed blue curve,
  resembling the micromotion induced heating.
}
\label{fig:oneionphase}
\end{figure}

The effect of phase micromotion on the equilibrium average kinetic energy of
a single ion in an atomic gas of $2\,\mu$K is investigated. We scanned the
phase difference $\delta\phi_{\rm{rf}}$ from 0-0.65\,mrad, corresponding to the
expected experimental upper limit from the linewidth broadening measurement
as discussed in~\ref{sec:Exp}.
We determined the
resulting equilibrium average kinetic energy in units of $T_{\rm{kin}}$ as well
as $N_{\rm{col}}$ according to Eq.~\ref{eq:expfit} by averaging over at least
300 individual runs per point. The results are shown in
Fig.~\ref{fig:oneionphase}.
The temperatures (blue) were fit with a quadratic function (solid blue line),
\begin{equation}
  T_{\rm{kin}} = T_1 + \theta_{\delta\phi_{\rm{rf}}} \delta\phi_{\rm{rf}}^2\,,
\end{equation}
leading to a quadratic rise factor of $\theta_{\delta\phi_{\rm{rf}}}
= 3980(15)\,\mu{\rm{K}}\cdot{\rm{mrad}}^{-2}$.  Also shown is the approximate
theoretical amount of kinetic energy stored in the phase micromotion (dashed
blue), according to Eq.~\ref{eqn:enpmm}, with a quadratic increase of $3651.2
\,\mu{\rm{K}}\cdot{\rm{mrad}}^{-2}$ for the parameters used in the simulation.  As
in the case for axial micromotion, the red points show the average kinetic
energy of an ion without an atomic bath present, in agreement with the
dashed blue line. All points of the plot were simulated using a comoving start
and escape sphere for the atoms to prevent numerical instabilities, thus leading
to an increasing number of collisions required to equilibrate (right).

\section{Ion crystals}\label{sect:crystals1D}
In this section, we briefly introduce the theoretical and numerical
framework to describe the normal modes of oscillations in an ion crystal. We
present and test a numerical method to extract the energy stored in the secular
motion of each individual mode.

By treating the mutual Coulomb interaction of the ions as well as the trapping
itself in harmonic approximation, the ion crystal can be described as a system
of coupled harmonic oscillators. This system can be decomposed into normal mode
coordinates and frequencies. This procedure is described in detail in
e.g.~\cite{James:1998} for a linear ion crystal. For a given set of secular trap
frequencies and number of ions, the equation of motion reads
\begin{equation}
   m_{\rm{ion}} \ddot{\vec{r}}_{n} = \vec{F}_n = - m_{\rm{ion}} \hat{\omega}^2 \vec{r}_{n}
  + \frac{e^2}{4 \pi
  \epsilon_0}\sum_{m\neq n}
  \frac{\vec{r}_{n}-\vec{r}_{m}}
  {\left\|\vec{r}_{n}-\vec{r}_{m}\right\|^3}\,,
\end{equation}
where the first term describes the three-dimensional trapping of each ion with
the trap frequency matrix $\hat{\omega}
= {\rm{diag}}\left(\omega_x,\omega_y,\omega_z\right)$ and the second term is the
mutual Coulomb interaction of the $N$-ion system.  To obtain the transformation
matrix to transform the system into normal mode coordinates, one first has to
find the equilibrium positions $\vec{r}_{n}^{\,(0)}$ of the ions within the
trap, defined as $\vec{F}_n(\vec{r}_{n}^{\,(0)}) = 0\,$.  We do this in
a two-step process: First, we numerically simulate the cooling of an $N$-ion
system in our trap until it crystallizes by introducing an additional
velocity-dependent force in the equation of motion,
\begin{equation}\label{eqn:eomdamp}
  m_{\rm{ion}} \ddot{\vec{r}}_{n} = \vec{F}_n  - \kappa \dot{\vec{r}}_{n}
  \,,\quad\kappa>0\,.
\end{equation}
As a second step, we use the numerically obtained equilibrium positions as
a guess for numerically finding the positions where the force on the ions
disappears. This procedure was found to be more stable than immediate
minimization of force on the ions, especially for higher dimensional crystals.

Treating the coordinates of the ions as small deviations from their equilibrium
positions, $\vec{r}_{n} (t) \approx \vec{r}_{n}^{\,(0)} + \vec{\rho}_{n}(t)$,
the potential energy of the system can be expanded to second order in
$\vec{\rho}_{n}$ to
\begin{eqnarray}
  U &= \sum_{n=0}^{N_{\rm{ions}}} \frac{1}{2} m_{\rm{ion}}
  \bar\omega^2 \vec{r}_{n}^{\,2} +
  \frac{1}{2}\frac{e^2}{4\pi \epsilon_0}\sum_{n\neq m}\frac{1}{\left\|
  \vec{r}_{n}-\vec{r}_{m}
  \right\|}\\
  &\approx \frac{1}{2} m_{\rm{ion}} \omega_z^2 \sum_{i,j=1}^3\sum_{m,n=1}^{N_{\rm{ions}}}
  A_{3(m-1)+i,3(n-1)+j} \rho_{m,i} \rho_{n,j}\,.
\end{eqnarray}
With the $3 N_{\rm{ions}}\times 3 N_{\rm{ions}}$ Hessian matrix
$A_{3(m-1)+i,3(n-1)+j} = \left.\frac{\partial^2 U}{\partial r_{m,i} \partial
r_{n,j}}\right|_0$, where $r_{m,i}$ is the coordinate of ion $m$ in the $i$-th
direction and the 0 denotes its evaluation at equilibrium positions. For clarity
we rename the indices of A to $u= 3(m-1)+i$, $ v= 3(n-1)+j$, $u,v\in\{1,\dots,3
N_{\rm{ions}}\}$.  Diagonalization of the symmetric Hessian matrix leads to the
diagonal form $D_{u,v}$ that can be obtained from the transformation $D = S^{T}
A S$, where S is the matrix of eigenvectors of $A$. The $3 N_{\rm{ions}}$
eigenmode frequencies $f_{q_u}$ are then given by $2 \pi f_{q_u} = \omega_{q_u}
= \omega_z \sqrt{D_{u,u}}$ and the potential energy in secular approximation
reads
\begin{equation}\label{eqn:eigenmodeenergy}
  U_{\rm{sec}} = \frac{1}{2} m_{\rm{ion}} \sum_{u=1}^{3N_{\rm{ions}}}
  \omega_{q_u}^2 q_u^2\,,
\end{equation}
with $q_u = \sum_{n=1}^{N_{\rm{ions}}}\sum_{j=1}^3 S_{u,3(n-1)+j}\rho_{n,j}$ the
normal mode coordinates.

Once transformed to these coordinates, the trajectories stored in each kinetic
energy determination are Fourier transformed numerically using a standard
Cooley-Tukey fast Fourier transform (FFT) algorithm~\cite{cooley1965}.
The Fourier spectra of the normal coordinates then contain only a peak at the
respective mode frequency along with peaks at the micromotion sidebands. To
obtain the energy stored in each mode, we compute the average kinetic energy of
each normal coordinate $q_m$,
\begin{equation}
  \bar{E}_{m,{\rm{tot}}} = \frac{1}{2} m_{\rm{ion}} \frac{1}{N_{\rm{fft}}}
  \sum_{k=1}^{N_{\rm{fft}}} \dot{q}_m(t_k)^2\,,
\end{equation}
where $m$ is the mode index and $k$ the time index of the Fourier time grid of
spacing $\Delta t_{\rm{fft}}$.  Since this energy still contains micromotion, we
make use of the Fourier relation for time derivatives,
\begin{equation}
  (\mathcal F \dot{q}_m)(f) = - i 2 \pi f \tilde{q}_m (f)\,,
\end{equation}
where $\tilde{q}_m (f) = (\mathcal F q_m)(f)$ is the Fourier transform of the
normal coordinate $q_m(t)$, and Parseval's theorem for the discrete Fourier
transformation,
\begin{eqnarray}
  \sum_{k=1}^{N_{\rm{fft}}} ||\dot{q}_m(t_k)||^2\Delta t_{\rm{fft}}
  &= \sum_{k=1}^{N_{\rm{fft}}}
|| - i 2 \pi f_k \tilde{q}_m(f_k)||^2 \Delta f_{\rm{fft}}\\
&= (2\pi)^2 \Delta f_{\rm{fft}} \sum_{k=1}^{N_{\rm{fft}}} f_k^2 \tilde
{q}_m(f_k)\tilde{q}_m^*(f_k)\,,
\end{eqnarray}
with which we can replace the expectation value of the squared normal mode
velocity $\dot{q}_m(t)$  and obtain
\begin{equation}
  \bar{E}_{m,{\rm{tot}}}= \frac{1}{2}m_{\rm{ion}}(2\pi)^2 \Delta f_{\rm{fft}}^2
  \sum_{k=1}^{N_{\rm{fft}}}
  f_k^2\tilde{q}_m(f_k)\tilde{q}_m^*(f_k) = \frac{1}{2} k_B T_m\,,
  \label{eqn:tempmodefourier}
\end{equation}
using the identity for the Fourier frequency grid spacing $\Delta f_{\rm{fft}}
= (N_{\rm{fft}}\Delta t_{\rm{fft}})^{-1}$.  To test the validity of this method,
the total kinetic energy of all modes
\begin{equation}
  \bar{E}_{\rm{tot,fft}} = \sum_{m=1}^{3 N_{\rm{ions}}} \bar{E}_{m,{\rm{tot}}}
  = \frac{3 N_{\rm{ions}}}{2} k_{\rm{B}} T_{\rm{fft}}\,,
  \label{eqn:etotfft}
\end{equation}
can be compared with the average kinetic energy defined in
Eq.~\ref{eqn:ekinavg}, which is presented in section~\ref{subs:realityfft}.
Since typically all secular frequencies are separated far from the micromotion
frequency, the high frequency parts of the spectrum can be cut off easily by
reducing the limit of the sum in Eq.~\ref{eqn:tempmodefourier} to a value
$N_{\rm{c}} = f_{\rm{c}} / \Delta f_{\rm{fft}}$, where $f_{\rm{c}}$ is the
desired cut-off frequency.  To obtain only the secular energy part for each of
the modes $\bar{E}_{m,{\rm{sec}}}$, the cut-off frequency should be chosen
centered between the highest normal mode frequency and the lowest micromotion
sideband. We define the temperature of each mode by $T_{m,{\rm{sec}}}$ and the
total secular temperature as $T_{\rm{sec}}$ as
\begin{eqnarray}\label{eqn:tsec}
  \bar{E}_{\rm{sec}} &
  = \sum_{m=1}^{3 N_{\rm{ions}}} \bar{E}_{m,{\rm{sec}}}
  = \sum_{m=1}^{3 N_{\rm{ions}}} \frac{1}{2} k_{\rm{B}} T_{\rm{m,sec}}
 = \frac{3 N_{\rm{ions}}}{2} k_{\rm{B}} T_{\rm{sec}}\,.
\end{eqnarray}
The approximate eigenmode frequencies $f_{q_m}$ can be found by searching the
peak position of the Fourier spectrum for the respective mode within an accuracy
of the Fourier frequency grid size $\Delta f_{\rm{fft}} = 1/(N_{\rm{fft}} \Delta
t_{\rm{fft}})$ leading to a relative error typically on the order of
$\frac{1}{2}\Delta f_{\rm{fft}}/f_{q_m}$.

\begin{figure}[t]
  \begin{center}
  \includegraphics[width=0.45\textwidth]{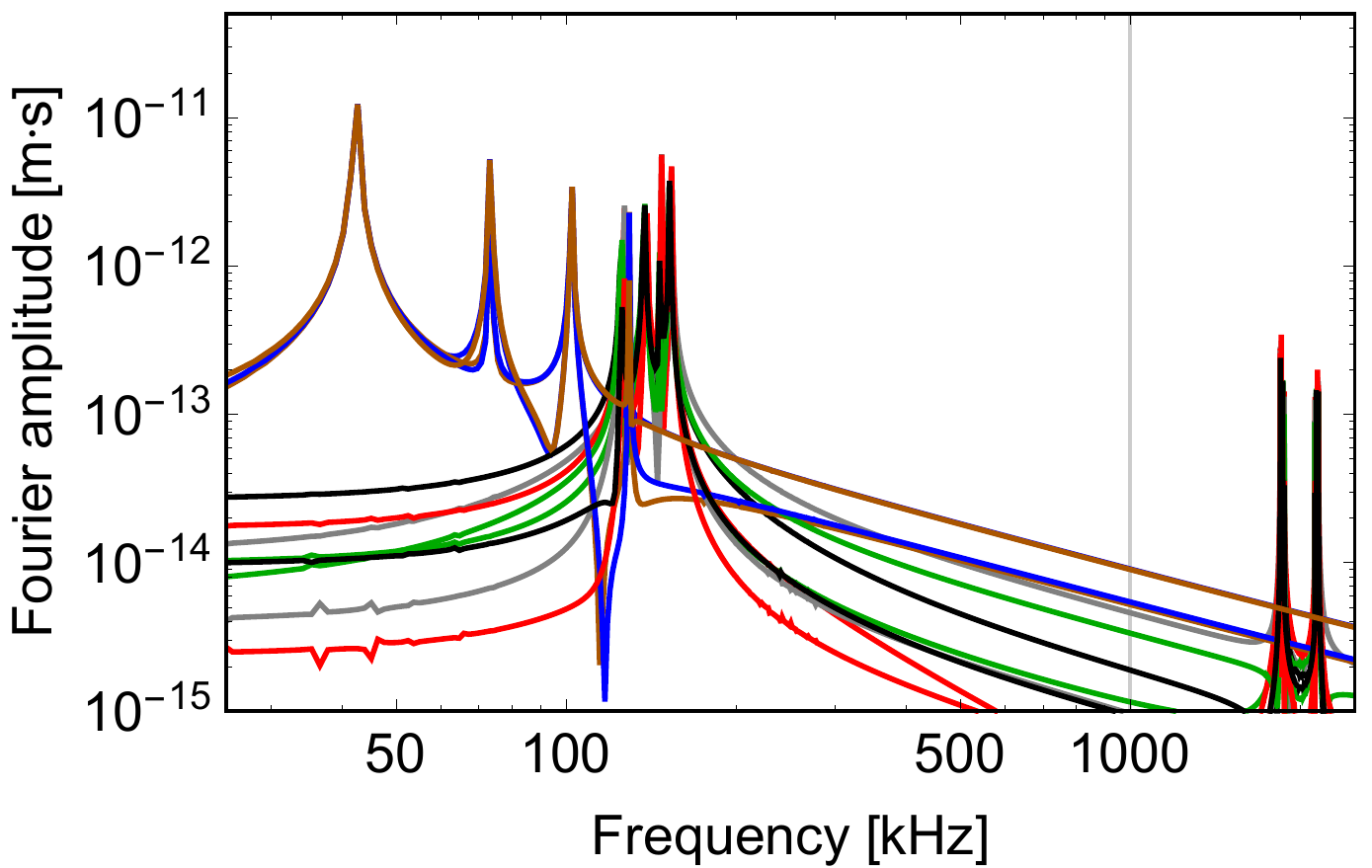}
  \hspace{0.05\textwidth}
  \includegraphics[width=0.45\textwidth]{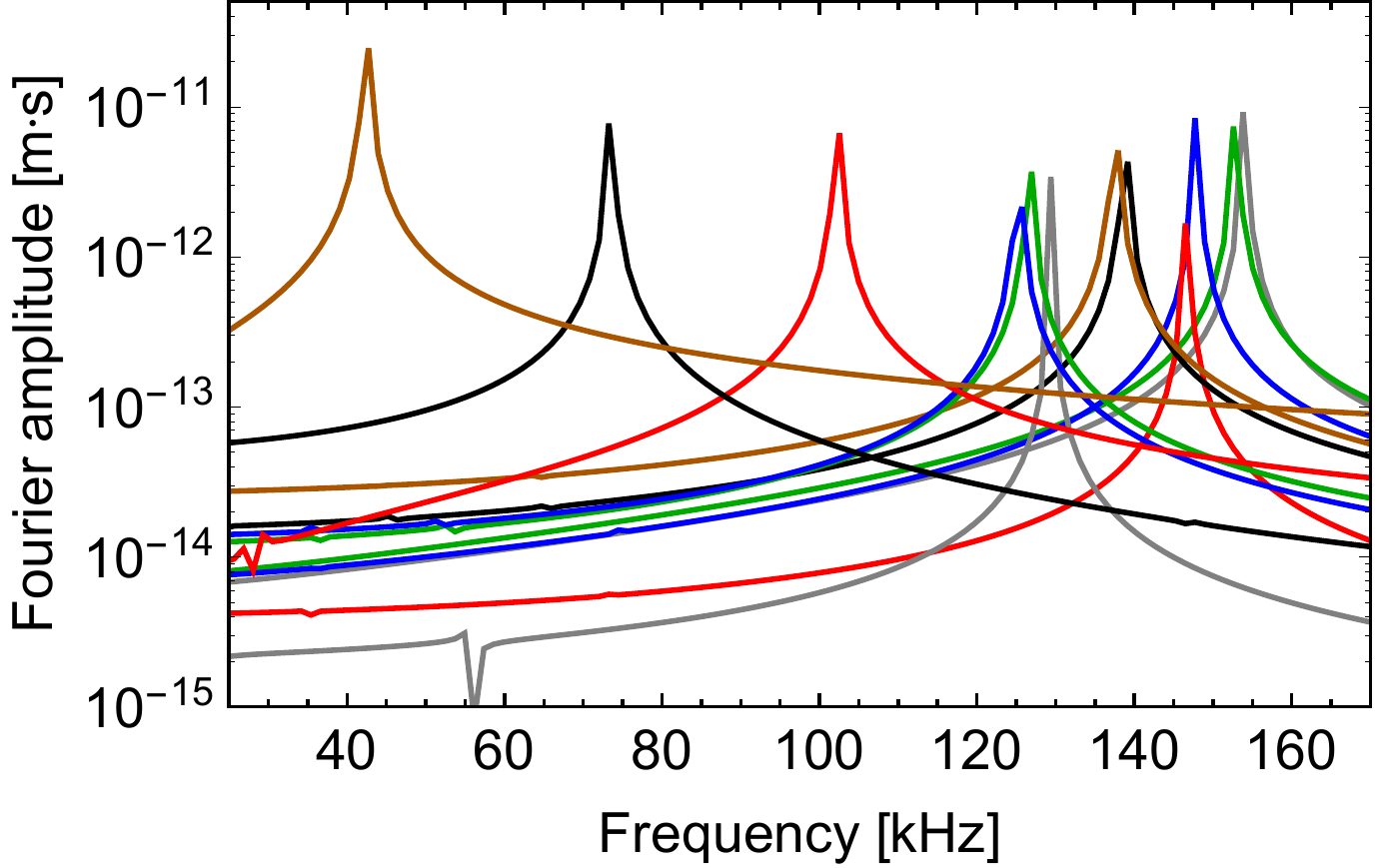}
\end{center} \caption{Fourier amplitudes for the twelve spatial coordinates
  (left) and for the twelve normal coordinates (right) for a simulated linear
  four-ion crystal at an average kinetic energy of around $5\,\mu$K. Each
  spatial coordinate shows contributions of multiple frequencies, whereas the
  normal modes are clearly decoupled and show only a single peak at the
  respective mode frequency. For the spatial coordinates, the micromotion
  sidebands around $f_{\rm{rf}} = 2$\,MHz and the cutoff frequency (gray bar)
  for the secular energy determination are shown as well. The Fourier spectra
  were obtained using $N_{\rm{fft}} = 16384$ steps of $\Delta t_{\rm{fft}}
  = 50\,$ns.}
\label{fig:fourierModesSpect}
\end{figure}

A typical spectrum of the Fourier amplitudes for a linear four-ion crystal is
shown in Fig.~\ref{fig:fourierModesSpect}. The Fourier spectra of all spatial
coordinates (left) show each multiple peaks at the twelve different mode
frequencies. The spectrum also contains the micromotion sidebands around the
trap drive frequency of $f_{\rm{rf}} = 2\,$MHz and a possible cutoff value (gray
bar) for the secular energy determination.  While some of the peaks at around
130\,kHz are too close to be distinguished, the Fourier spectra of the normal mode
coordinates (right) show only one peak each, allowing for the numerical
frequency and energy determination within each mode. Note that the plots are cut
off at the relevant eigenmode frequency scale, not showing the micromotion
sidebands around the trap drive frequency $f_{\rm{rf}} = 2\,$MHz.
The twelve normal modes of the four-ion crystal are visualized in
Fig.~\ref{fig:allModes} in Appendix~\ref{subs:realityfft}, along with their
respective frequencies obtained from the diagonalization of the secular case as
presented in this section and the frequency peak positions of
the fourier spectra. Typically, these modes are assigned with the names given in
the right column~\cite{Lemmer:2015}.

\section{Ion crystals in the cold buffer gas}\label{sec:Results_crystals1D}
In this section, we investigate the influence of the number of ions as well as
that of all types of micromotion in an ion crystal. We further analyze the case
where an additional oscillating electric quadrupole field in axial direction is present,
leading to a non-vanishing $q_z$-parameter, which is typically the case under
realistic experimental conditions. In this section, we assume that the entire
crystal is immersed in the atomic cloud, and each ion is equally likely to
collide with an atom. In particular, we dice the ion at which the atom is
introduced before calculating each collision event.

\subsection{Influence of the number of ions}\label{subs:nionsscan}
\begin{figure}[t]
  \begin{center}
  \includegraphics[width=0.45\textwidth]{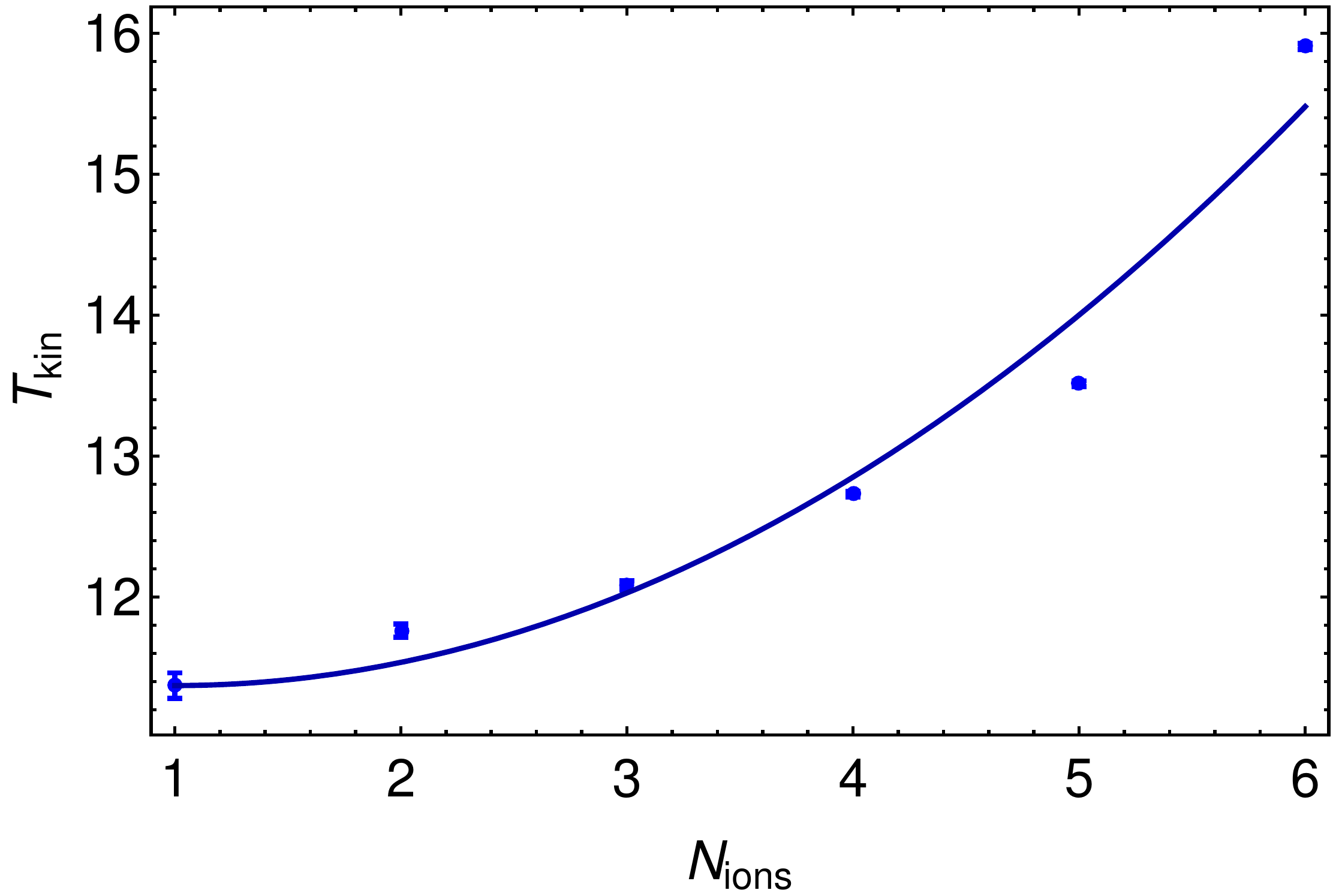}
  \hspace{0.05\textwidth}
  \includegraphics[width=0.45\textwidth]{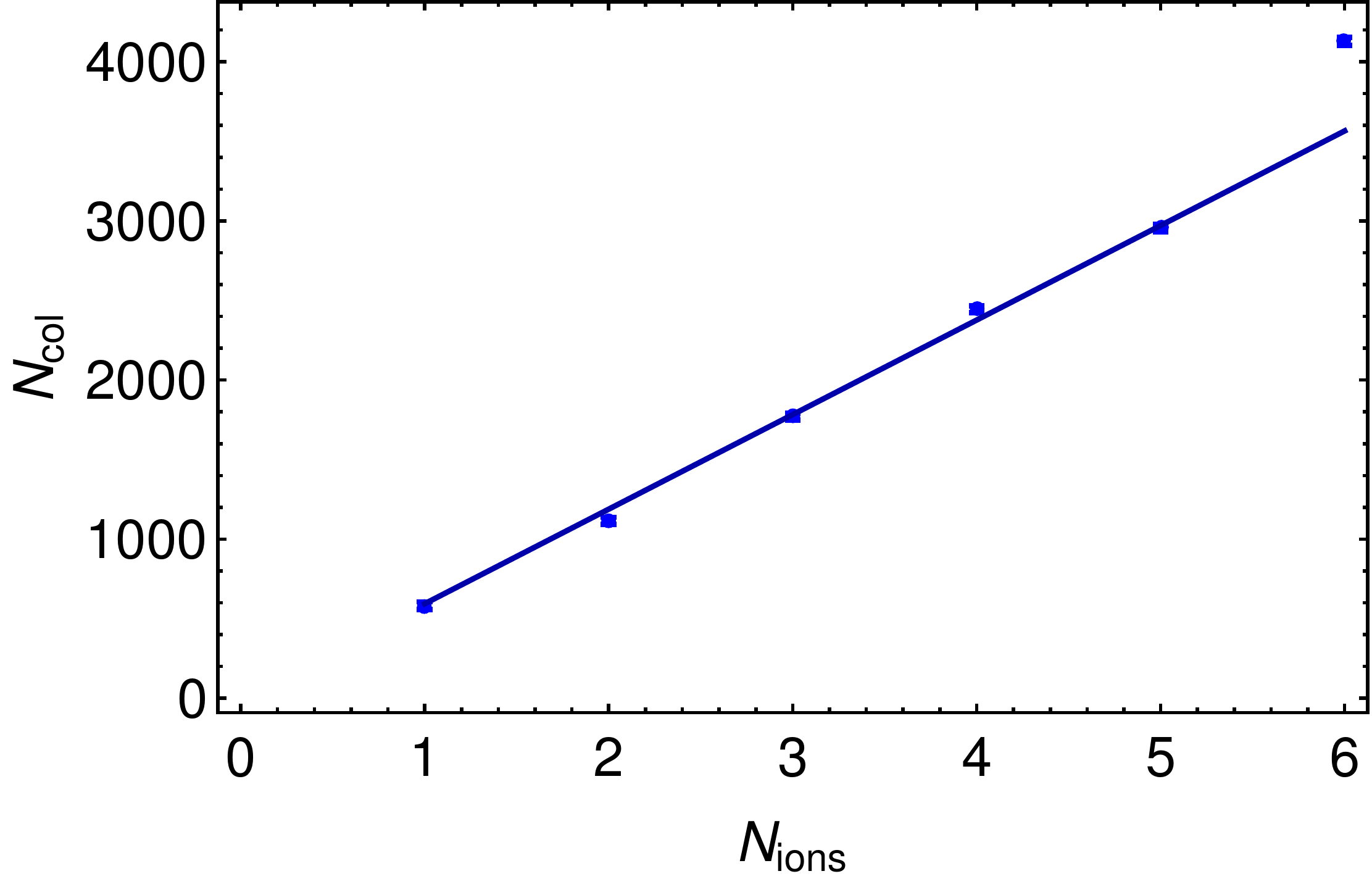}
\end{center} \caption{Final temperature $T_{\rm{kin}}$ (left) and the
  number of collisions required to equilibrate (right) versus number of ions in
  a linear crystal colliding with atoms at $2\,\mu$K. The kinetic energy shows
  a weak dependence on number of ions, whereas $N_{\rm{col}}$ is strictly linear.}
\label{fig:ionnum}
\end{figure}

First, the influence on the achievable temperature of the crystal $T_{\rm{kin}}$
(see Eq.~\ref{eqn:ekinavg}) and typical number of collisions required to
equilibrate $N_{\rm{col}}$ as defined in Eq.~\ref{eq:expfit} was investigated.
The results for one to six ions trapped using no axial or excess radial
micromotion is shown in Fig.~\ref{fig:ionnum}. For one and two ions at least 300
runs were averaged, whereas due to the computational effort for three to six
ions, only 40 runs each were simulated, thus leading to worse statistics and
thus larger errors.
For the final temperature of the crystal (left) a weak dependence on number of
ions can be observed. The results were fit with a heuristic fit function (blue
line),
\begin{equation}
  T_{\rm{kin}} = T_1 + \theta_{\rm{i}} \left(N_{\rm{ions}}-1\right)^2\,,
\end{equation}
leading to $T_1=11.4(2)\,\mu$K and a quadratic rise factor of
$\theta_{\rm{i}}=0.17(2)\,\mu{\rm{K}}$.
The number of collisions required for thermalization (right) is strictly linear
in number of ions. The linear fit (solid line) leads to an increase of 626(20)
collisions per additional ion. The behavior is to be expected since the number
of modes of the crystal that need to be cooled increases linearly as well. While
in the simulation only one atom is introduced at a time, in the experiment the
density of atoms ideally is the same all along the ion crystal, thus increasing
the actual collision rate by the factor $N_{\rm{ions}}$. Consequently, the
thermalization time for an $N_{\rm{ions}}$-crystal is expected to be the same as
for one ion.
\subsection{Influence of excess micromotion}\label{subs:4ionsemm}
Similar to the single ion case, the effect of radial excess micromotion as well
as axial micromotion and quadrature micromotion was investigated. Additionally, the
dependence of the secular energy was studied. The obtained results can be found
in~\ref{subs:emm4results}. The behavior of the final average kinetic
energy versus the scanned micromotion parameter is in perfect agreement with the
single ion case.
\subsection{Influence of a non-vanishing axial rf-gradient ($q_z\neq 0$)}\label{subs:qzscan}
\begin{figure}[t]
  \begin{center}
  \includegraphics[width=0.45\textwidth]{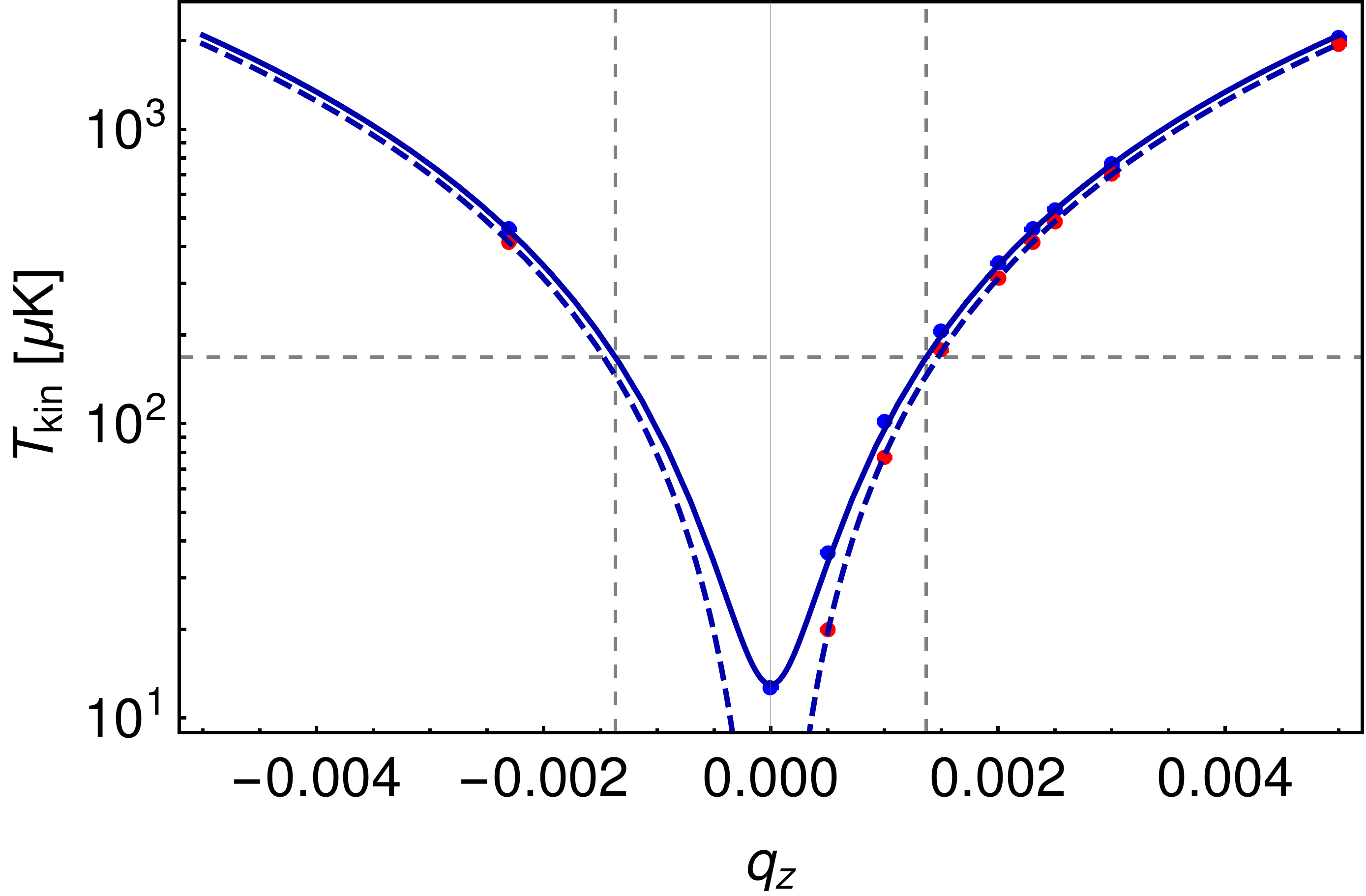}
  \hspace{0.05\textwidth}
  \includegraphics[width=0.45\textwidth]{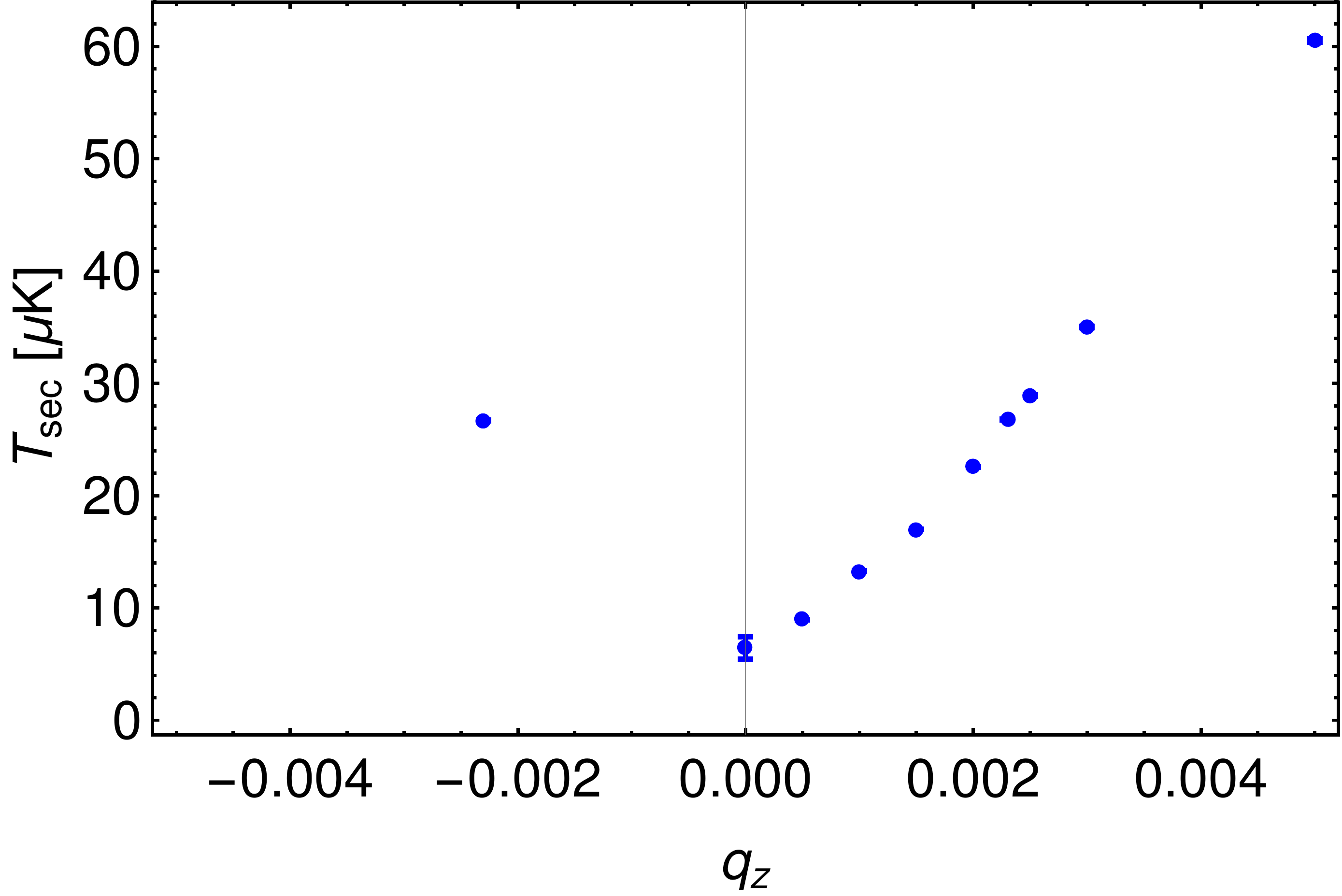}
\end{center} \caption{Equilibrium temperature of a linear four-ion crystal
  colliding with atoms at $2\,\mu$K (left) and secular temperature (right)
  versus $q_z$ parameter. The left points (blue) were fit with
  a quadratic function and an offset (solid curve). The dashed curve shows the
  approximate theoretical behavior as explained in the text. The dashed gray
  lines indicate the $s$-wave temperature limit for $T_{\rm{a}}\rightarrow 0$. The red dots were
  obtained from a simulation with zero secular energy and no atoms present.
  }
  \label{fig:qzscan}
\end{figure}

To study the effect of a non-vanishing $q_z$, the parameter was scanned from
0 to 0.005. The value in our ion trap is around $q_z^{\rm{exp}}=0.0023$ for
similar trapping parameters as used in the simulation.  The resulting
equilibrium Temperatures $T_{\rm{kin}}$ and $T_{\rm{sec}}$ are shown in
Fig.~\ref{fig:qzscan} (blue). The points were obtained by averaging over at
least 30 individual runs for each value of $q_z$ and fitting the averages
according to Eq.~\ref{eq:expfit}.
The results for the average kinetic energy (left) were fit using a quadratic function with offset
(solid line), $T_{\rm{kin}} (q_z) = T_1+\theta_{q_z} q_z^2$, leading to
a quadratic rise factor of $\theta_{q_z} = 8.29(6)\cdot10^7\,\mu{\rm{K}}$ with offset
$T_{\rm{1}} = 13.0(6)\,\mu{\rm{K}}$. The approximate theoretical dependence of
the average kinetic energy according to Eqs.~\ref{eqn:position}-\ref{eqn:enemm}
is shown as a dashed line. The quadratic rise of the theoretical curve is given
by $\theta_{q_z}^{\rm{theo}} = 7.80\cdot10^7\,\mu$K. The points in red show the
average
kinetic energies due to the influence of $q_z$ in the non-interacting case
where the ions were initialized without secular energy. A quadratic fit
of the red points lead to a rise factor of $\theta_{q_z}^{(0)}
= 7.81(1)\cdot10^7\,\mu{\rm{K}}$, in good
agreement with the prediction from the approximate solution, which is to be expected
as the approximation holds for $q_z^2 \ll 1$. The secular temperature (right)
shows an almost linear dependence on $q_z$ and resembles the actual influence of
the additional micromotion-induced heating due to a non-vanishing $q_z$.
\subsection{Micromotion-induced heating on the individual modes}\label{subs:simudecomp}
\begin{figure}[htpb]
  \begin{center}
  \includegraphics[width=0.45\textwidth]{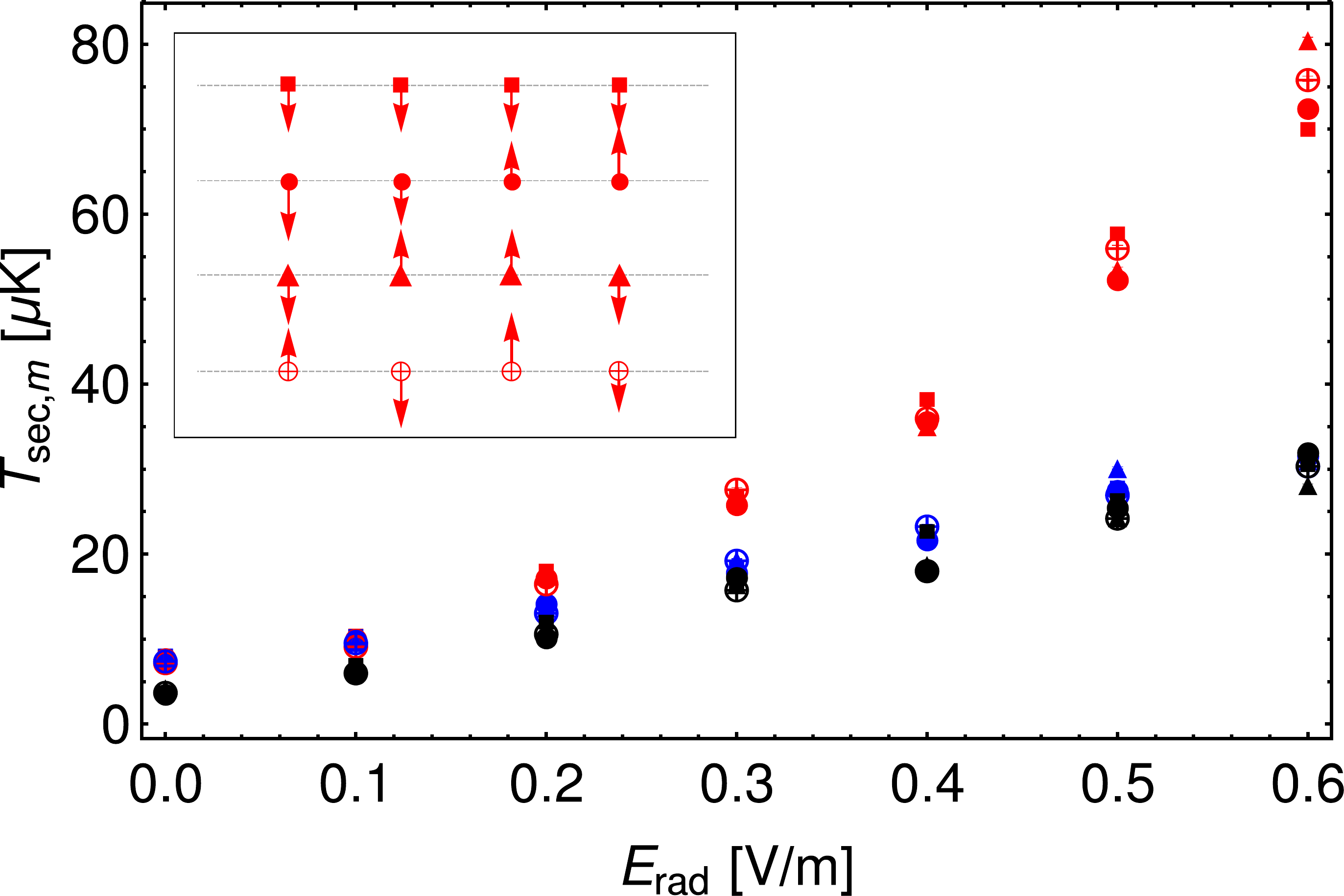}
  \hspace{0.05\textwidth}
  \includegraphics[width=0.45\textwidth]{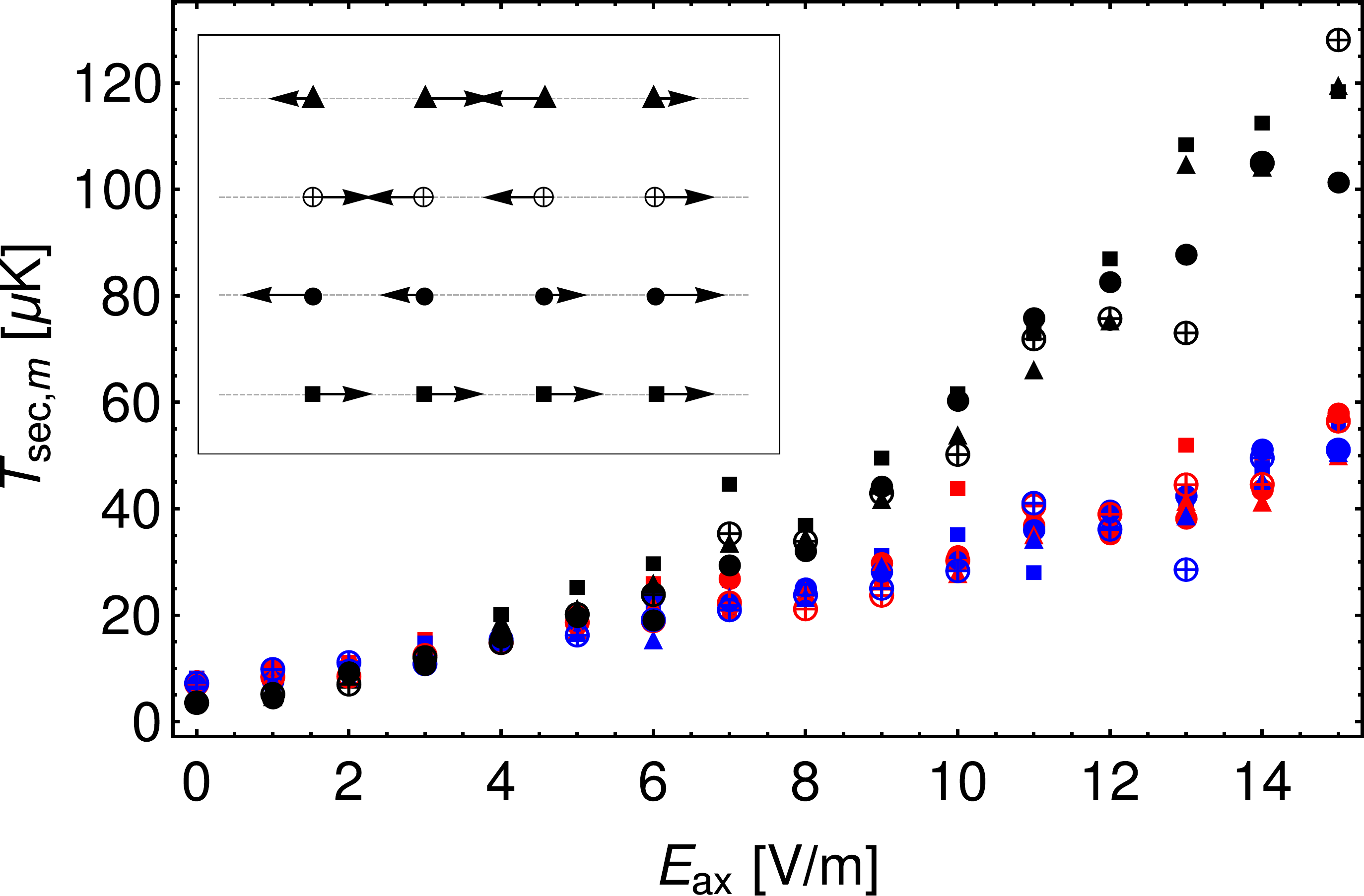}
    \vspace{0.05\textwidth}
  \includegraphics[width=0.45\textwidth]{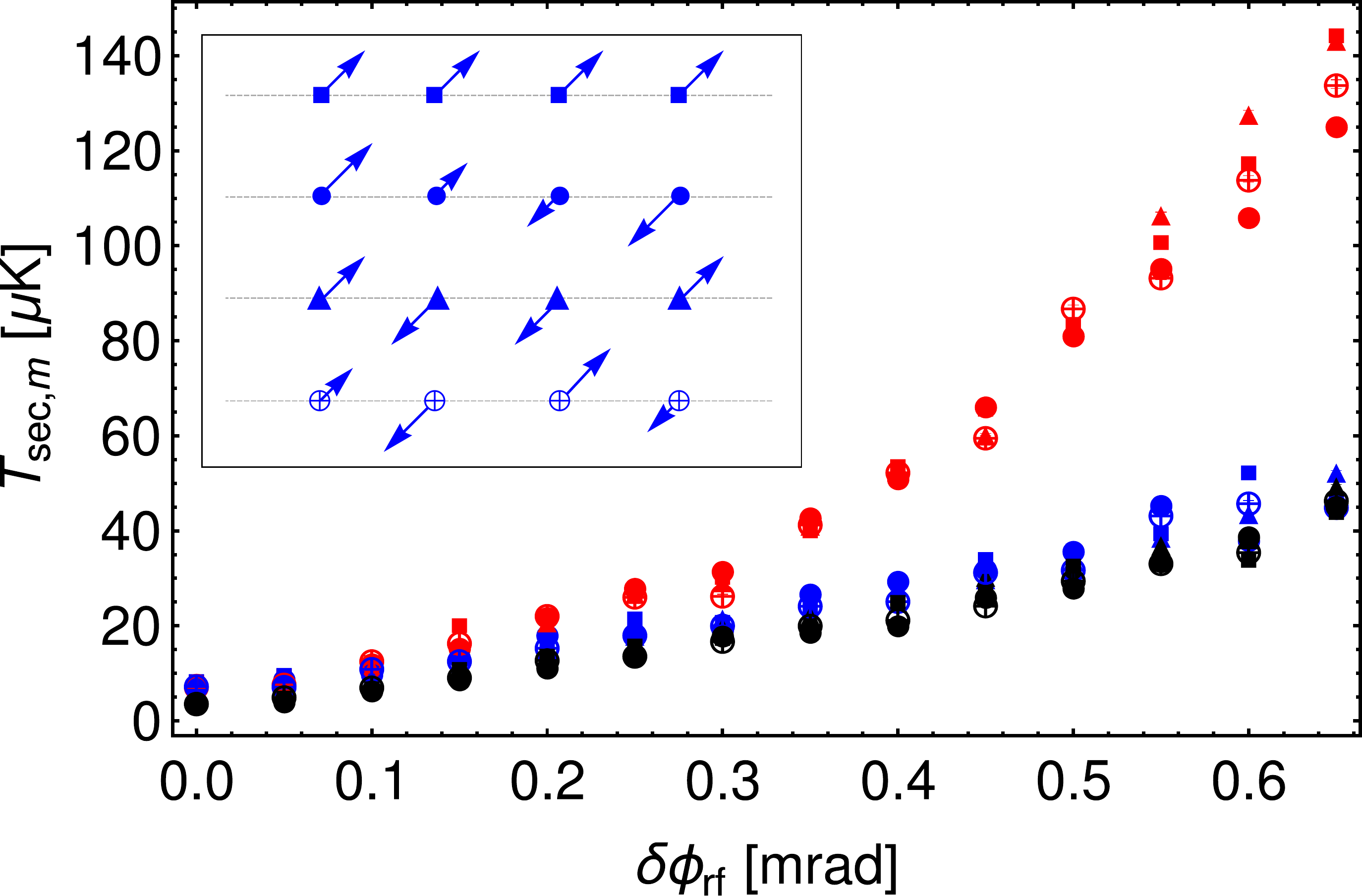}
  \hspace{0.05\textwidth}
  \includegraphics[width=0.45\textwidth]{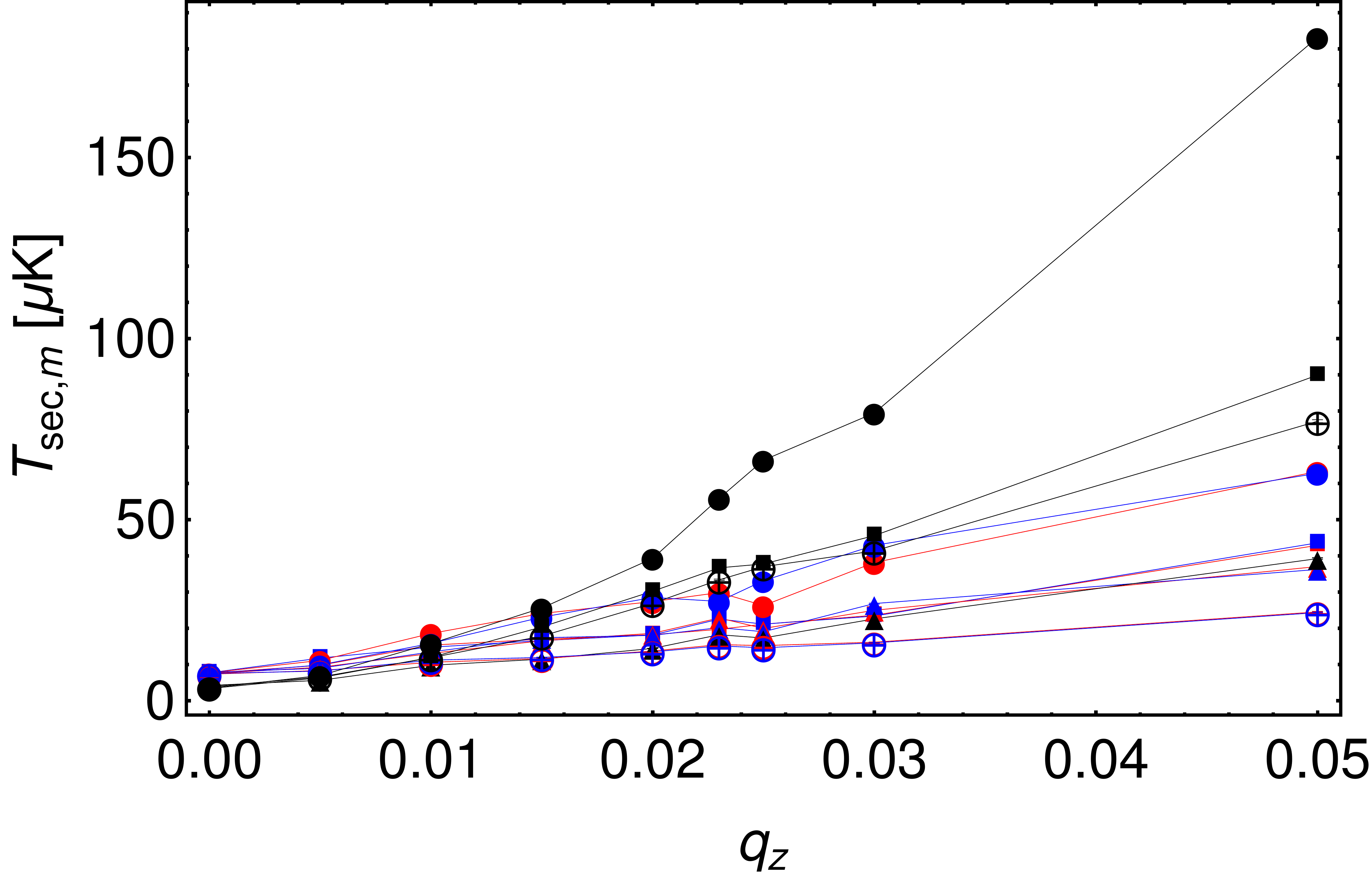}
  \end{center}
  \caption{Individual secular temperatures $T_{\rm{sec},m}$ of each normal mode
  of a linear four-ion crystal colliding with atoms at $T_{\rm{a}} = 2\,\mu$K in
  the case of a radial dc electric field $E_{\rm{rad}}$ in $x$-direction (left
  top), a homogeneous axial oscillating field (right) with amplitude
  $E_{\rm{ax}}$ (right top) or in the presence of a rf phase shift
  $\delta\phi_{\rm{rf}}$ between the rf electrodes (left bottom).  The results
  depicted in red and blue were obtained from the four modes oscillating in
  $x$- or $y$-direction respectively, whereas the results in black were obtained
  from the four axial modes. The plot on the down right shows the behavior of
  the twelve secular mode temperatures for a non-vanishing $q_z$ parameter. The
  insets illustrate the respective modes, in which the arrows indicate the
  direction and relative amplitude of motion.}
\label{fig:modesEmmAmm}
\end{figure}

In this section, we analyze the effect of each type of micromotion on
the individual modes of a four-ion crystal. The secular temperature of each
mode was obtained as described in section~\ref{sect:crystals1D} from the
simulations of the linear four-ion crystal in
section~\ref{subs:4ionsemm} and~\ref{subs:qzscan}. The resulting temperatures
for the twelve individual modes as presented in Fig.~\ref{fig:allModes}
are shown in Fig.~\ref{fig:modesEmmAmm} for radial excess micromotion (left top)
axial micromotion (right top) and quadrature micromotion (left bottom).
In each of the three cases the radial modes equilibrate to a slightly higher
temperature than the axial modes, when the scanned excess micromotion parameter
is low. For high values the temperature of the modes with excess micromotion
dominate, which is the $x$-direction (red) for both radial and quadrature
micromotion and the $z$-direction (black) in the case of axial micromotion.
A further sub-separation of the radial and axial modes is not resolved.

Interestingly, for a non-vanishing axial gradient, expressed by $q_z$, the situation is
quite different, as it is shown in Fig.~\ref{fig:modesEmmAmm}. In this
case, the modes separate for high $q_z$ into different groups, starting with the $x$ and $y$
zigzag modes (red and blue crossed circles) at the lowest temperature for $q_z=0.05$.
The next group is formed
by
the $x$ and $y$ center-of-mass modes
(red and blue squares) along with the drum modes (red and blue triangles) and the
$z$ anti-stretch mode (black triangles). Approximately located at mode average temperature the
two tilt modes (red and blue circles) are found. At higher temperature, the three
remaining axial modes Egyptian
(black crossed circles), center-of-mass (black squares) and stretch (black circles) are located.
This behavior is mainly reasoned by the participation of the outer ions to
these modes, since these can exchange the largest amount of energy during
a collision due to their large micromotion amplitudes. While the contribution of
the outer ion's motion to the zigzag modes is lowest and the mode is moving
perpendicular to the micromotion direction, the radial center-of-mass and drum
modes show larger and equal coupling as indicated by the arrow length in the mode visualization insets and in
table~\ref{fig:allModes}. The anti-stretch mode shows less coupling
strength for the outer ions but moves in the direction of micromotion, thus
enhancing the probability for a high energy exchange within a collision. The
strongest radial contribution of the outer ion's motion is to the two
tilt modes, leading to the highest radial mode temperatures. As in the case of
an homogeneous oscillating axial field, the highest temperatures are found
within axial modes, dominated by the one with the largest contribution of the
outer ion's motion, the stretch mode.

\section{Two-dimensional ion crystals}\label{subsec:planarsub}
By adjusting the axial and radial trapping fields,
it is possible to change the shape and dimensionality
to form two dimensional ion
crystals~\cite{Kaufmann:2012,Landa:2012,Shen:2014,Richerme2016}.
Even with
perfect micromotion compensation, there are always ions within any nonlinear crystal
that have their quasi-equilibrium position outside the radiofrequency node axis, thus
experiencing a non-vanishing oscillating electric field, leading to additional, unavoidable
micromotion. Therefore, immersing the complete ion crystal in a cloud of
ultracold atoms will always lead to micromotion-induced heating of the normal
modes. To avoid this effect, one can utilize the large spacing between the ions,
enabling the experimental possibility to overlap a dense and small
atomic cloud only with a single ion sitting at the axial radiofrequency node
within a larger ion crystal.

\begin{figure}[htbp]
  \begin{center}
  \includegraphics[width=0.8\textwidth]{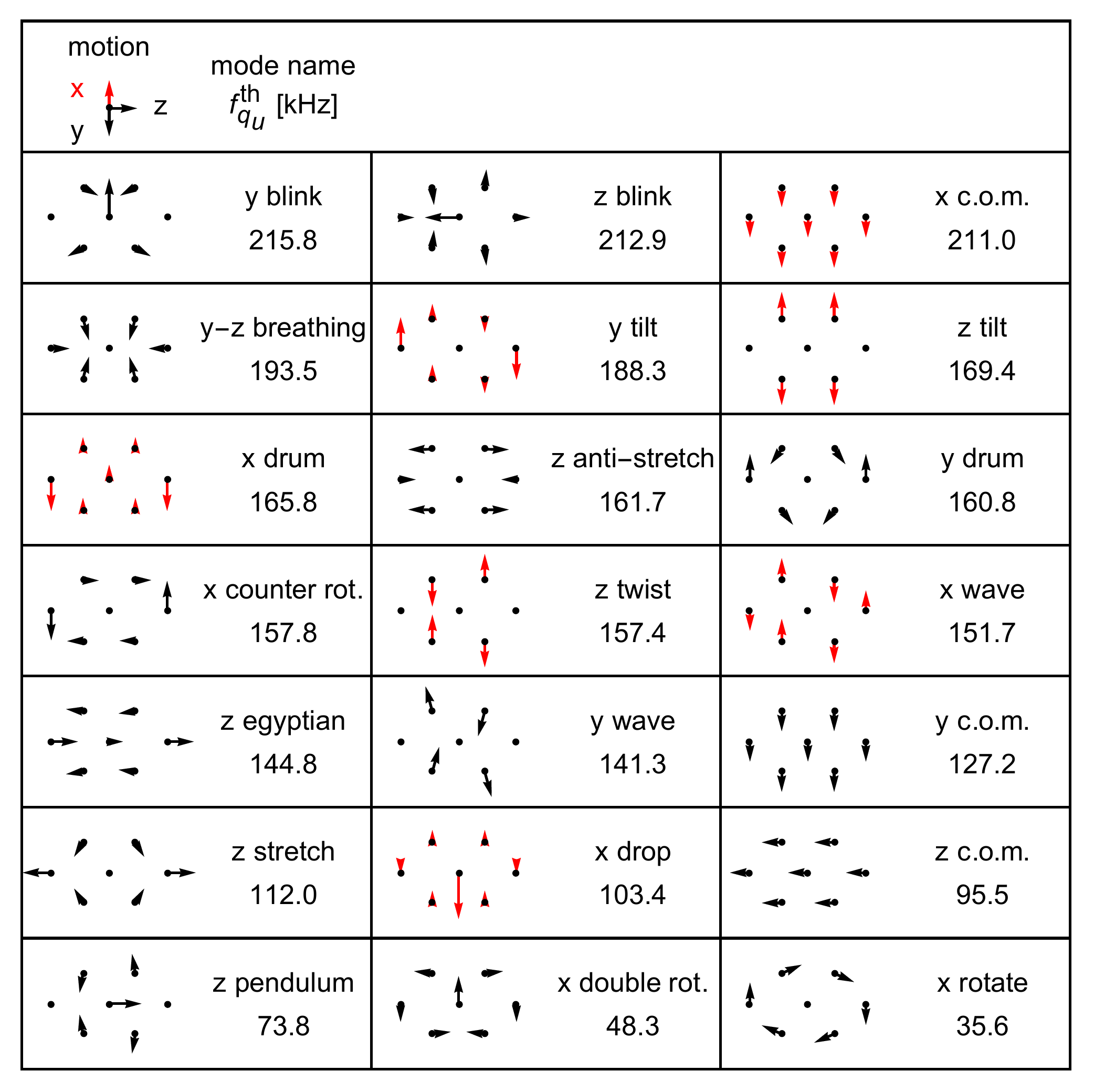}
  \end{center}
  \caption{Visualization of the normal mode movement for a trapped
  planar seven-ion crystal. The arrows indicate the direction
  and amplitude of the respective mode within the plane (black) and
  perpendicular to the plane (red).
  For each mode the respective eigenfrequency $f_{q_u}^{\rm{th}}$
  obtained from diagonalization of the secular approximation is shown.
}
\label{fig:allModesPlanar}
\end{figure}

To simulate a stable 7-ion hexagonal ion crystal, we change the trap parameters to
$f_z = 95.459\,$kHz,
$q_x = q_y = 0.261$,
and $\alpha_x = 1.0$, $\alpha_y = -2.0$ to achieve $f_x=211.002\,$kHz and
$f_y=127.229\,$kHz as radial secular trap frequencies, all within experimental
reach with the ion trap used in our experiment. Due to the stronger confinement
in the $x$-direction, the crystal forms in the $y-z$ plane.
Its geometry along with its approximate (secular) mode structure is depicted in Fig.~\ref{fig:allModesPlanar}.
Notably, in contrast to a linear crystal, the mode with the highest frequencies are not
center-of-mass modes, but the two planar $\textit{blink}$ modes, where the ion
density oscillates in $y$ and $z$ direction respectively. Also the mode with
the lowest frequency is not a center-of-mass mode but the \textit{x rotate}
mode, where all six ions defining the hexagon oscillate in phase
clockwise/counterclockwise
around the central ion within the crystal plane.

\begin{figure}[htpb]
  \begin{center}
  \includegraphics[width=0.45\textwidth]{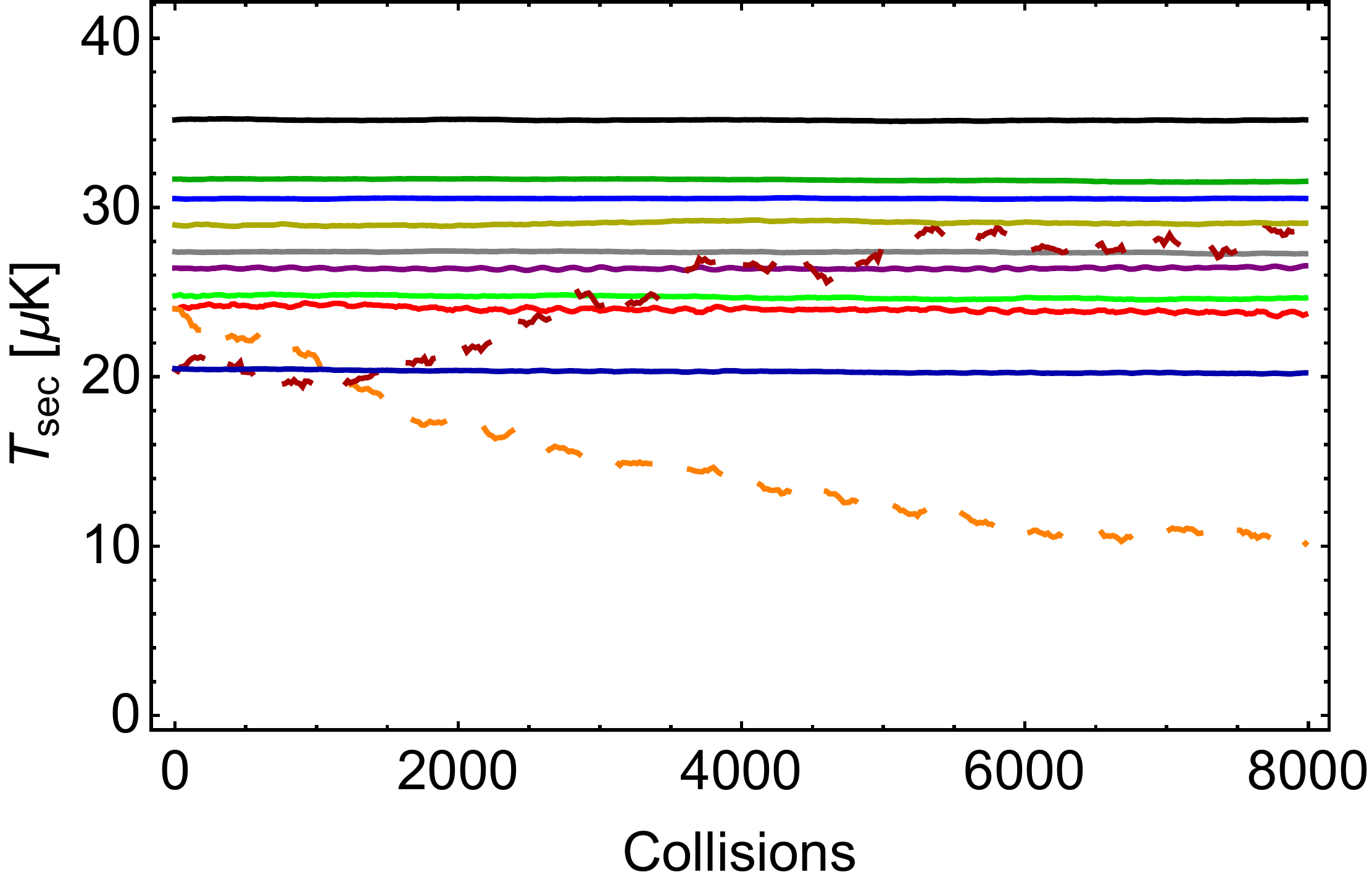}
  \hspace{0.05\textwidth}
  \includegraphics[width=0.45\textwidth]{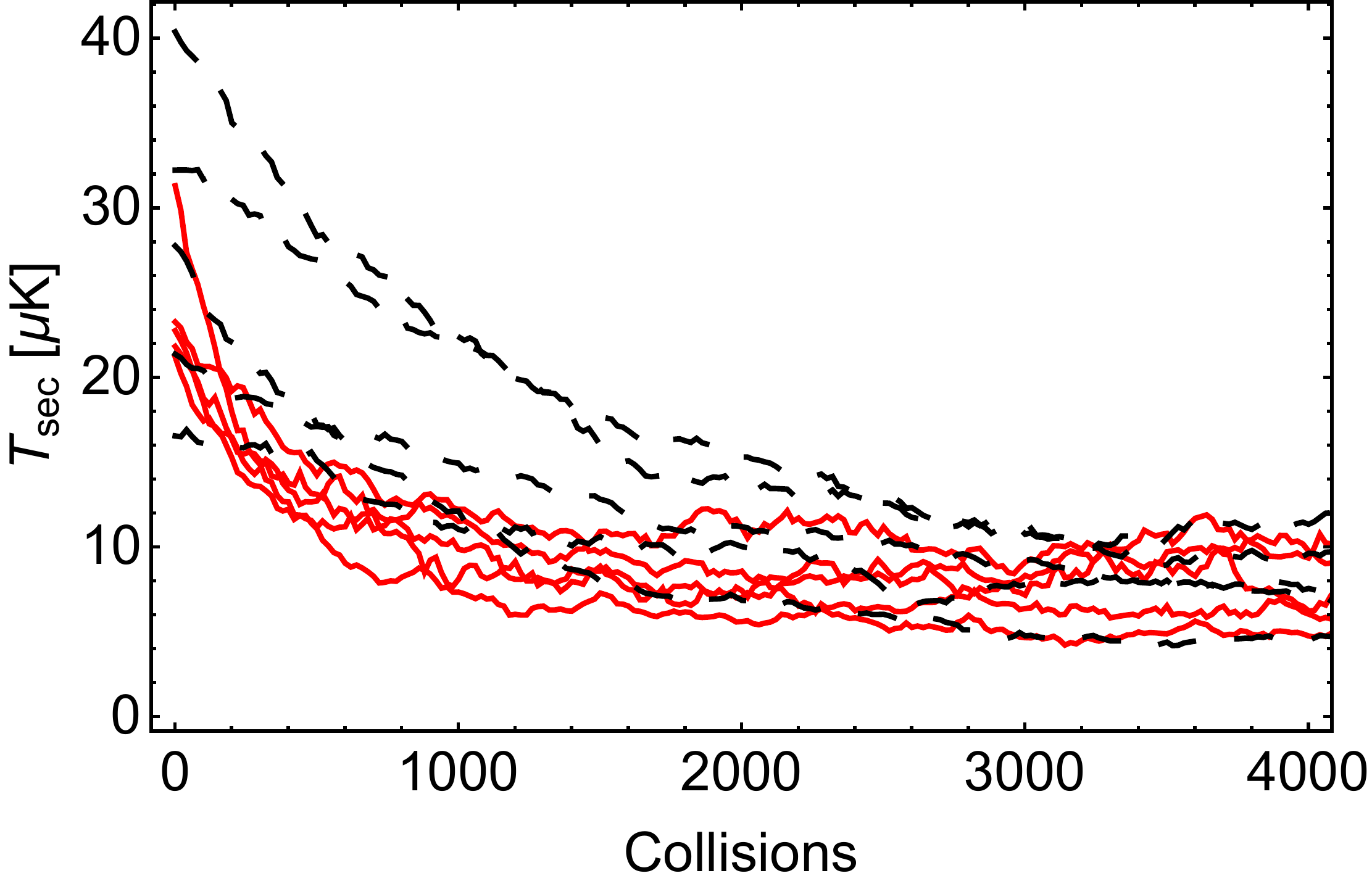}
\end{center}
\caption{Thermalization of the secular modes of a hexagonal planar seven-ion
  crystal initialized at a secular temperature of $25\,\mu$K when only the
  central ion is colliding with atoms at $2\,\mu$K. The modes without motional
  components of the central ion (left) show almost no dynamic, whereas all other
  modes (right) thermalize to temperatures below $20\,\mu$K.}
\label{fig:planartherm}
\end{figure}

To simulate the thermalization of the secular modes, we initialize the ion
crystal with negligible secular energy by first switching on a strong
velocity-dependent damping force as
defined in Eq.~\ref{eqn:eomdamp} that is adiabatically turned to zero. To give the
ion crystal an initial secular energy, we add to each ion's velocity components
a velocity sampled from a Maxwell-Boltzmann distribution at a given
temperature before the first collision occurs.
We only let the central ion collide with atoms at
$T_{\rm{a}}=2\,\mu$K. We obtain the secular temperatures for each mode
as in the case for the linear ion
crystal  by integrating over the Fourier spectra of the normal mode coordinates.
Due to the orders of magnitude larger micromotion sidebands around the trap
frequency of $2\,$MHz, it is necessary to increase the frequency resolution by
a factor of four and only take a narrow range around the respective peaks for
the integrals into account. Otherwise, the integrals suffer from
a non-neglegible micromotion floor of the Fourier spectra even around the
secular frequencies that can only be
suppressed by further increasing the Fourier resolution towards unfeasible
computational effort. To compensate for the
already large increase in computation time due to the large micromotion
amplitudes and increased number of particles compared to the four-ion linear
crystal, the atom start sphere size was chosen to be fixed and only $r_0 = 0.3\,\mu$m around the central ion,
thus increasing the likelihood of Langevin collisions but also cutting down the
propagation times during a collision.
The results for all 21 modes of a planar seven-ion crystal initialized at $25\,\mu$K are shown in
Fig.~\ref{fig:planartherm}. The values were averaged over 120 individual runs.
The thermalization of the modes can be classified into three different
groups.
\begin{itemize}
  \item The modes where the central ion's motion is not participating at all do
    not show significant cooling dynamics (left), besides the \textit{y drum} (orange dashed) and
    \textit{y wave} (dark red dashed) mode, showing a relatively slow cooling and
    heating, possibly due to enhanced nonlinear Coulomb interactions between the ions in
    these two modes.
  \item The modes where the central ion participates rather weakly (right, black dashed), as indicated by
  the length of the vectors in Fig.~\ref{fig:planartherm}, show a slow
  cooling dynamic over the observed number of collisions.
\item The modes where the central ion participates most (right, solid red), \textit{x/z blink},
  \textit{x drop}, \textit{z pendulum} and \textit{x double rotation},
    thermalize the fastest.
\end{itemize}
The different initial temperatures of each mode are caused by the different
coupling strength and
number of modes each ion is involved in and could in principle be corrected
for, but this is not necessary for the qualitative analysis of the behavior.
Remarkably, the achieved minimum temperatures of the modes that thermalize are all found to
be between $5\,\mu$K,
and $15\,\mu$K, comparable to the secular temperatures achieved using the linear
four-ion crystal at perfect micromotion compensation, although the average
kinetic energy of the planar crystal $T_{\rm{kin}}=700\,$mK is five orders of
magnitude larger due to the large micromotion amplitudes of the outer ions.

\section{Conclusions}\label{sec:Conclusions}
In this article we have presented numerical simulations of classical Yb$^+$-Li
collisions for ions trapped in a Paul trap. We presented and tested a numerical
framework to simulate and analyze the collisions using parameters that can be
achieved in our experiment, including all types of micromotion that are
observable in real ion traps. We analyzed the effect of the micromotion on the
achievable average kinetic energy of a single ion.  For an ion in an ideal Paul
trap and in the limit where $T_{\rm{a}}\rightarrow 0$, this energy is found to
be at $T_{\rm{kin}}=7.60(14)\,\mu$K. Owing to the large mass ratio, this leads
to a collision energy of $T_{\rm{col}}=$~0.4~$\mu$K which lies well below the
$s$-wave temperature limit. In this situation, the ion is cooled close to its
ground state of motion with $\bar{n}=1.2$ motional quanta remaining in the
secular motion on average.

For the limits for all types of excess micromotion
found in our experiment, the determined collision energies are a factor of 2-11
higher than the $s$-wave temperature limit, as it is shown in
Table~\ref{tab:mmrestab}. This indicates that better micromotion detection and
compensation is required there. In particular, using a narrow linewidth laser
would allow to put better limits on the axial and quadrature micromotion
amplitudes. Another option may be to use the atoms themselves for accurate
micromotion detection as described in Ref.~\cite{Harter:2013}.

The limits for each experimental parameter that lead to $s$-wave collisions
energies are also given in Table~\ref{tab:mmrestab}. Although all lie beyond the
limits of our current setup, they are not excessive, as e.g. H\"arter
\textit{et al.}~\cite{Harter:2013} report a field of $E_{\rm{rad}} \leq 0.02\,$V/m and
$E_{\rm{ax}} \leq 0.06\,$V/m in a similar system. For the quadrature
micromotion, we expect the given experimental limit of
$\delta\phi_{\rm{rf}}=0.65\,$mrad to be overestimated by at least an order of
magnitude due to the limitations of our detection techniques, as we show in
section~\ref{sec:Exp}. The rf phase shift mainly results from unequal length of
the connectors, which is approximately less than $\Delta x_
{\rm{rf}}\approx0.5\,$mm. Thus, we expect a phase mismatch on the order of
$\delta \phi_{\rm{rf}}\leq \frac{\delta
x_{\rm{rf}}}{v_{\rm{rf}}\Omega_{\rm{rf}}}\approx0.04\,$mrad for an assumed
signal propagation velocity of $v_{\rm{rf}}\approx c_{\rm{light}}/2$ half the
speed of light. Similarly, we expect that the true axial micromotion amplitude
lies significantly below the experimental limit stated. We conclude that
Yb$^+$/Li may reach the quantum regime with state-of-the-art micromotion
compensation.  We do note however that our present analysis is based on
classical theory. For excellent micromotion compensation, a quantum description
such as the one developed in~\cite{Krych:2013} should be generalized to include
excess mircomotion and used to predict thermalization in the ultracold regime.

We found that a buffer-gas cooled linear ion crystal
behaves similar as a single ion and the presence of
more than three modes of ion motion does not significantly influence the achievable
collision energies and thermalization rates. A non-vanishing axial gradient
expressed as a $q_z$-parameter leads to a collision energy of
$T_{\rm{col}}=26.3\,\mu$K for a four-ion crystal and the experimental value of
$q_z^{\rm{exp}}=0.0023$.
Also shown in the table are the mean secular energies of the single ion and four
ion case along with the mean thermal occupation numbers for the mode with the
lowest frequency (center-of-mass).

Within all simulations, we do not observe runaway heating, as expected, since
the mass of the ion is much larger than the mass of the atom. In the simulations
it takes around $N_{\rm{col}}\approx 550-600$ collisions for a single ion to
equilibrate within an atomic cloud with a density of $\rho_{\rm{a}}
= \frac{1}{4/3 \pi r_0^3} \approx 1.1\cdot10^{18}\,{\rm{m}}^{-3}$ (i.e.\ one atom
within the interaction sphere at a time). Within a simulation run using
a non-comoving sphere we observe an average flux $\Phi_{\rm{a}}$ of 10000
collisions within 120\,ms propagation time, which translates into
\begin{equation}
  \Gamma_{\rm L} t_{\rm{col}}
  = 2 \pi \rho_{\rm{a}}
  \sqrt{\frac{C_4}{\mu}} \frac{N_{\rm{col}}}{\Phi_{\rm{a}}}\approx 35-38
\end{equation}
Langevin collisions that are required for reaching
the equilibrium temperature. Luckily, the chance for an inelastic collision
happening during the interaction time,
leading to charge transfer or molecule formation is less than 0.76\,\%
as we recently measured~\cite{Joger:2017}.
The cooling rate for a linear ion crystal is comparable to the single ion case,
under the assumption of a homogeneous atomic density all along the ion crystal.
Interestingly, the secular modes of a linear ion crystal equilibrate to slightly
higher temperatures than average when moving in a micromotion direction.

We have shown that collisional cooling of
a planar seven-ion crystal by a
localized atomic cloud interacting with only the central ion should be possible. The technique enables cooling of all the ten
modes where the colliding ion participates in. The achieved
temperatures of these modes are all below $12\,\mu$K, corresponding to mode
occupation numbers of $\bar{n}_m = \frac{k_{\rm{B}} T_{m,{\rm{sec}}}}{\hbar
\omega_m} =2-11 $ phonons.
Shuttling the ion crystal to
overlap one of the outer ions with a small atomic cloud at the position of optimal
micromotion compensation should in principle increase the
number of cooled modes up to 18 out of the 21 total modes.
Such localized micro-clouds could be implemented by using a dimple
trap as it is described in Ref.~\cite{Serwane:2011}. There, the atomic cloud
is trapped by a strongly
focused laser beam with a waist of $\leq 1.8\,\mu$m, thus trapped in a volume
much smaller than the
interionic distance, e.g.\ 14.6\,$\mu$m for the ion crystal investigated in this
article.

Our results show that with modest improvements in micromotion compensation and
detection, reaching the quantum regime of atom-ion collisions can be achieved in
our experiment, enabling buffer-gas cooling of the trapped ion quantum platform
close to the motional ground state and the observation of atom-ion Feshbach
resonances.

\begin{table}[htpb]
	\centering
		\begin{tabular}{|c|c|c|c|c|c|c|c|}
			\hline
      &\multicolumn{1}{|c|}{Param.} &
      \multicolumn{1}{c|}{Value}&
      \multicolumn{1}{c|}{$T_{\rm{kin}} [\mu{\rm{K}}]$}&
      \multicolumn{1}{c|}{$T_{\rm{col}} [\mu{\rm{K}}]$}&
      \multicolumn{1}{c|}{$T_{\rm{sec}} [\mu{\rm{K}}]$}&
      \multicolumn{1}{c|}{$\bar{n}_{\rm{min}}$}&
      \multicolumn{1}{c|}{$\bar{n}_{\rm{max}}$}\\[0.3ex]
			\hline

    \multirow{4}{*}{\rotatebox[origin=c]{90}{\parbox[c]{0.5cm}{\centering
    single\\ion}}} &

      \multicolumn{1}{|c|}{$E_{\rm{rad}}$} &
       \multicolumn{1}{c|}{0.3\,V/m} &
       \multicolumn{1}{c|}{257(2)}&
       \multicolumn{1}{c|}{16(3)}&
       \multicolumn{1}{c|}{20.9(2)}&
       \multicolumn{1}{c|}{2.6(1)}&
       \multicolumn{1}{c|}{8.1(1)}\\

      &\multicolumn{1}{|c|}{$E_{\rm{ax}}$}&
      \multicolumn{1}{c|}{15\,V/m}&
      \multicolumn{1}{c|}{1686(8)}&
      \multicolumn{1}{c|}{89(4)}&
      \multicolumn{1}{c|}{86.5(5)}&
      \multicolumn{1}{c|}{7.0(1)}&
       \multicolumn{1}{c|}{76.5(1)}\\

       &\multicolumn{1}{|c|}{$\delta\phi_{\rm{rf}}$} &
       \multicolumn{1}{c|}{0.65\,mrad}&
       \multicolumn{1}{c|}{1694(7)}&
       \multicolumn{1}{c|}{89(4)}&
       \multicolumn{1}{c|}{75.7(7)}&
       \multicolumn{1}{c|}{6.4(1)}&
       \multicolumn{1}{c|}{21.7(1)}\\
       \hline		

    \multirow{5}{*}{\rotatebox[origin=c]{90}{\parbox[c]{0.5cm}{\centering
    four\\ions}}} &

      \multicolumn{1}{|c|}{$E_{\rm{rad}}$} &
       \multicolumn{1}{c|}{0.3\,V/m} &
       \multicolumn{1}{c|}{247(2)}&
       \multicolumn{1}{c|}{16(3)}&
       \multicolumn{1}{c|}{20.9(2)}&
       \multicolumn{1}{c|}{2.5(1)}&
       \multicolumn{1}{c|}{8.0(1)}\\

      &\multicolumn{1}{|c|}{$E_{\rm{ax}}$}&
      \multicolumn{1}{c|}{15\,V/m}&
      \multicolumn{1}{c|}{1685(7)}&
      \multicolumn{1}{c|}{89(4)}&
      \multicolumn{1}{c|}{86.5(5)}&
      \multicolumn{1}{c|}{7.3(1)}&
       \multicolumn{1}{c|}{57.7(1)}\\

       &\multicolumn{1}{|c|}{$\delta\phi_{\rm{rf}}$} &
       \multicolumn{1}{c|}{0.65\,mrad}&
       \multicolumn{1}{c|}{1706(6)}&
       \multicolumn{1}{c|}{90(4)}&
       \multicolumn{1}{c|}{75.7(7)}&
       \multicolumn{1}{c|}{6.3(1)}&
       \multicolumn{1}{c|}{22.0(2)}\\

       &\multicolumn{1}{|c|}{$q_z$} &
       \multicolumn{1}{c|}{0.0023}&
       \multicolumn{1}{c|}{452(4)}&
       \multicolumn{1}{c|}{26(4)}&
       \multicolumn{1}{c|}{26.9(1)}&
       \multicolumn{1}{c|}{2.5(1)}&
       \multicolumn{1}{c|}{18.0(1)}\\
\hline
    \multirow{5}{*}{\rotatebox[origin=c]{90}{\parbox[c]{0.5cm}{\centering
    desired}}} &
      \multicolumn{1}{|c|}{$E_{\rm{rad}}$} &
       \multicolumn{1}{c|}{$<$~0.24\,V/m} &
       \multicolumn{1}{c|}{168.4}&
       \multicolumn{1}{c|}{8.6}&
       \multicolumn{1}{c|}{-}&
       \multicolumn{1}{c|}{-}&
       \multicolumn{1}{c|}{-}\\

      &\multicolumn{1}{|c|}{$E_{\rm{ax}}$}&
      \multicolumn{1}{c|}{$<$~4.58\,V/m}&
      \multicolumn{1}{c|}{168.4}&
      \multicolumn{1}{c|}{8.6}&
      \multicolumn{1}{c|}{-}&
      \multicolumn{1}{c|}{-}&
       \multicolumn{1}{c|}{-}\\

       &\multicolumn{1}{|c|}{$\delta\phi_{\rm{rf}}$} &
       \multicolumn{1}{c|}{$<$~0.20\,mrad}&
       \multicolumn{1}{c|}{168.4}&
       \multicolumn{1}{c|}{8.6}&
       \multicolumn{1}{c|}{-}&
       \multicolumn{1}{c|}{-}&
       \multicolumn{1}{c|}{-}\\

       &\multicolumn{1}{|c|}{$q_{z}$} &
       \multicolumn{1}{c|}{$<$~0.0014}&
       \multicolumn{1}{c|}{168.4}&
       \multicolumn{1}{c|}{8.6}&
       \multicolumn{1}{c|}{-}&
       \multicolumn{1}{c|}{-}&
       \multicolumn{1}{c|}{-}\\
       \hline		
		\end{tabular}
	\caption{Simulation results for the different types of micromotion for the
  single and four-ion case. The collision energies $k_{\rm{B}} T_{\rm{col}}$ are
  all well above the $s$-wave energy of $E_s = k_{\rm{B}} 8.6\,\mu$K for the
  given experimental limits. Also shown are the corresponding minimum and
  maximum normal mode occupation numbers $\bar{n}_{\rm{min/max}}$ and the
  desired values to reach collision energies below $E_s$ for each micromotion
  case.}
	\label{tab:mmrestab}
\end{table}

\ack This work was supported by the EU via the ERC (Starting Grant 337638) and
the Netherlands Organization for Scientific Research (NWO, Vidi Grant 680-47-538
and Start-up grant 740.018.008) (R.G.). We gratefully acknowledge fruitful
discussions with Antonio Negretti.

\appendix
\addcontentsline{toc}{section}{Appendices}
\addtocontents{toc}{\protect\setcounter{tocdepth}{-1}}

\section{Reality checks of the simulation}\label{subs:reality}

\begin{figure}[t]
  \begin{center}
  \includegraphics[width=0.45\textwidth]{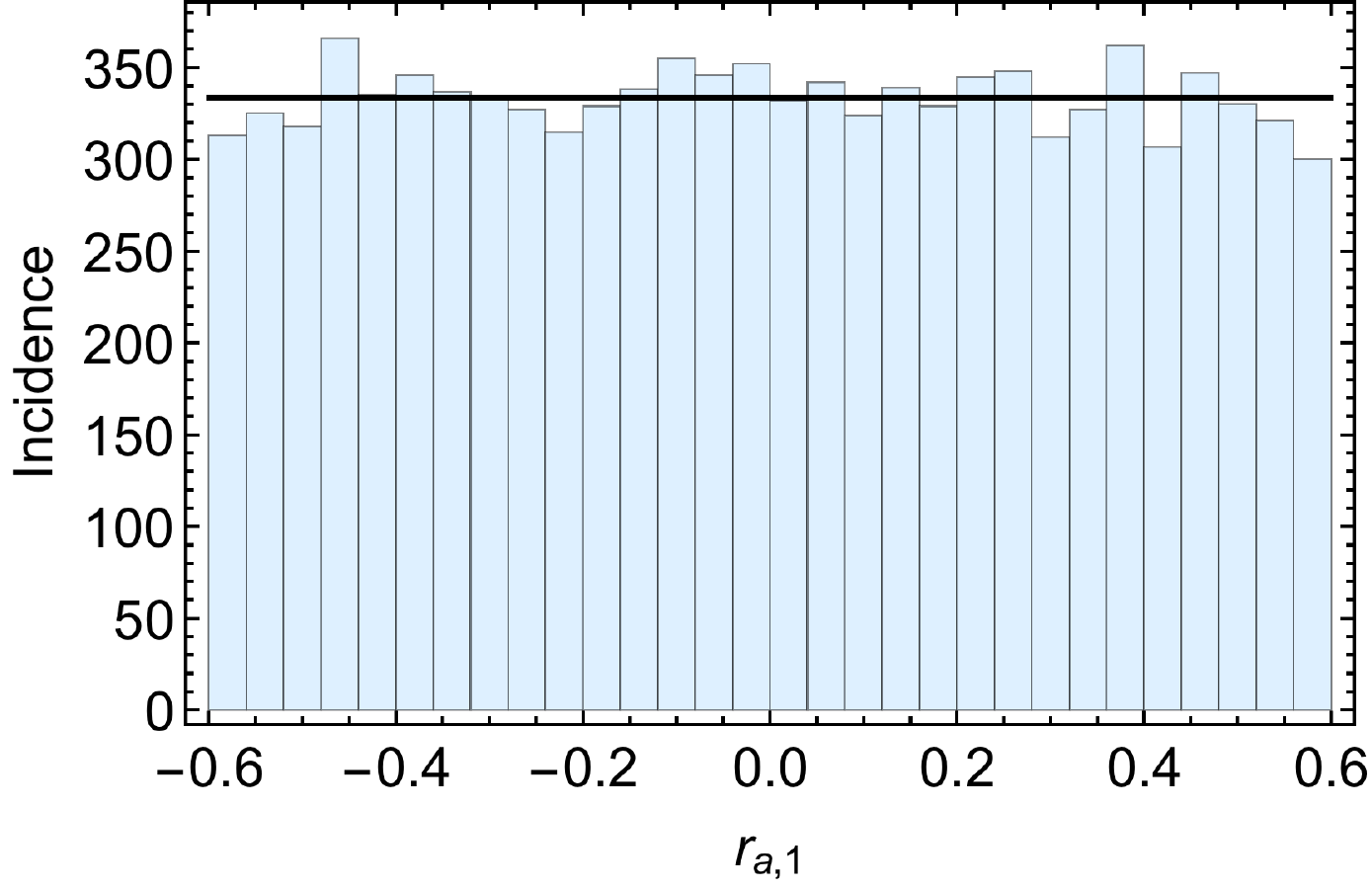}
  \hspace{0.05\textwidth}
  \includegraphics[width=0.45\textwidth]{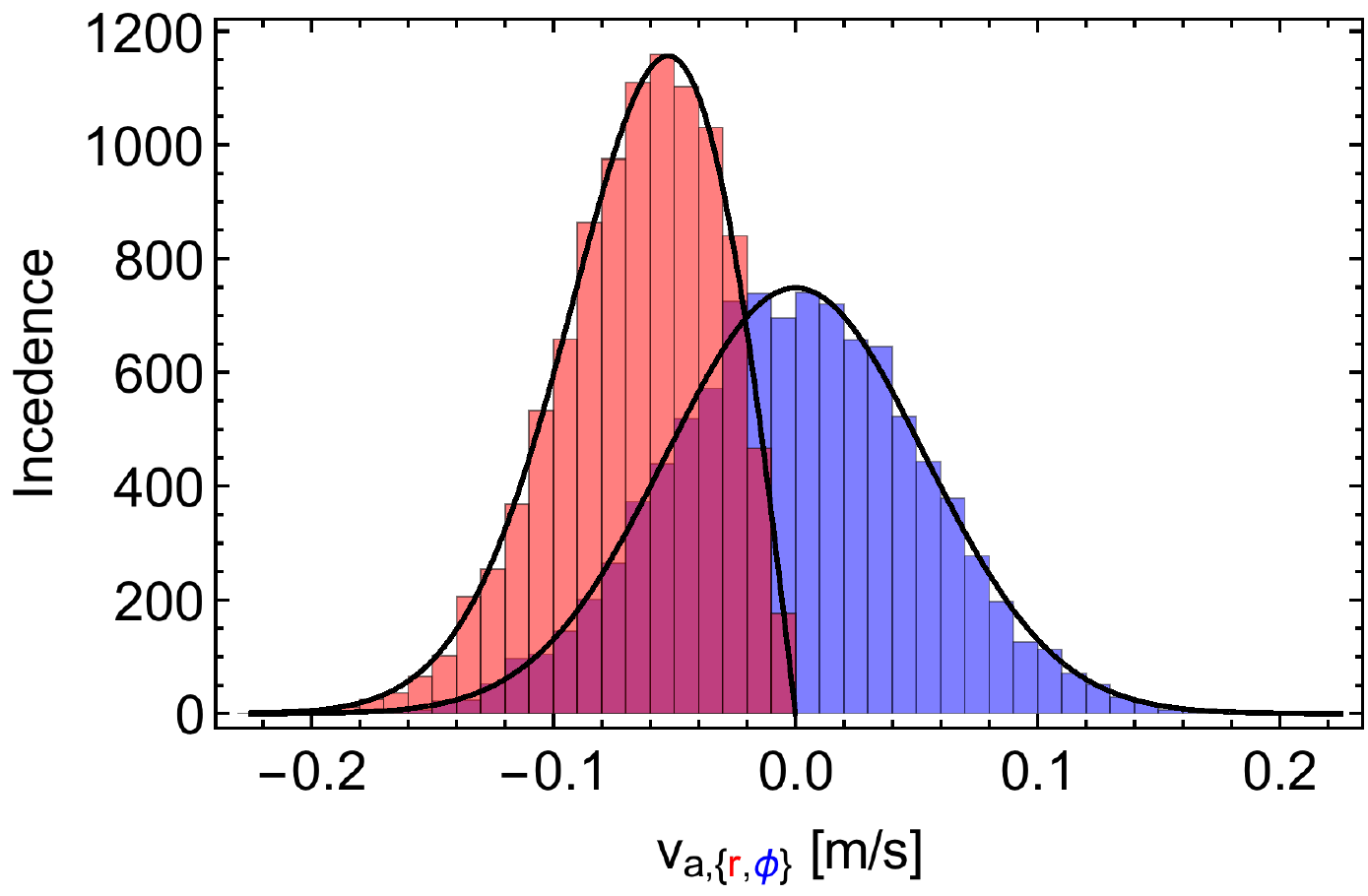}
\end{center} \caption{Spatial (left) and velocity (right) distributions of atoms
picked on a sphere with radius $0.6\,\mu$m and a temperature of $2\,\mu$K along
with the expected probability densities (black).}
\label{fig:randomnumbers}
\end{figure}

In this section, we check the accuracy of the simulation algorithm in detail
using realistic trapping fields that can be achieved in our
experiment. A summary of the parameters used in the simulations unless noted otherwise
can be found in table~\ref{tab:simparams}.

The functionality of the random number generation was checked by analysis of the
distributions of initial atom coordinates for 10000 events sampled at $T_{\rm{a}}
= 2\,\mu$K on a sphere of $r_{\rm{0}} = 0.6\,\mu$m. By definition, the spatial
coordinates automatically lie on the sphere. It is therefore sufficient to check
that each coordinate is uniformly distributed in the interval $[-r_0,r_0]$.
For the velocities, the distributions of Eq.~\ref{eqn:vatomsflux} must be obtained.
As an example, distributions for $r_{a,1}$, $v_{{\rm{a}},r}$ and $v_{{\rm{a}},\phi}$
are shown in
Fig.~\ref{fig:randomnumbers}.

\begin{figure}[t]
  \begin{center}
  \includegraphics[width=0.45\textwidth]{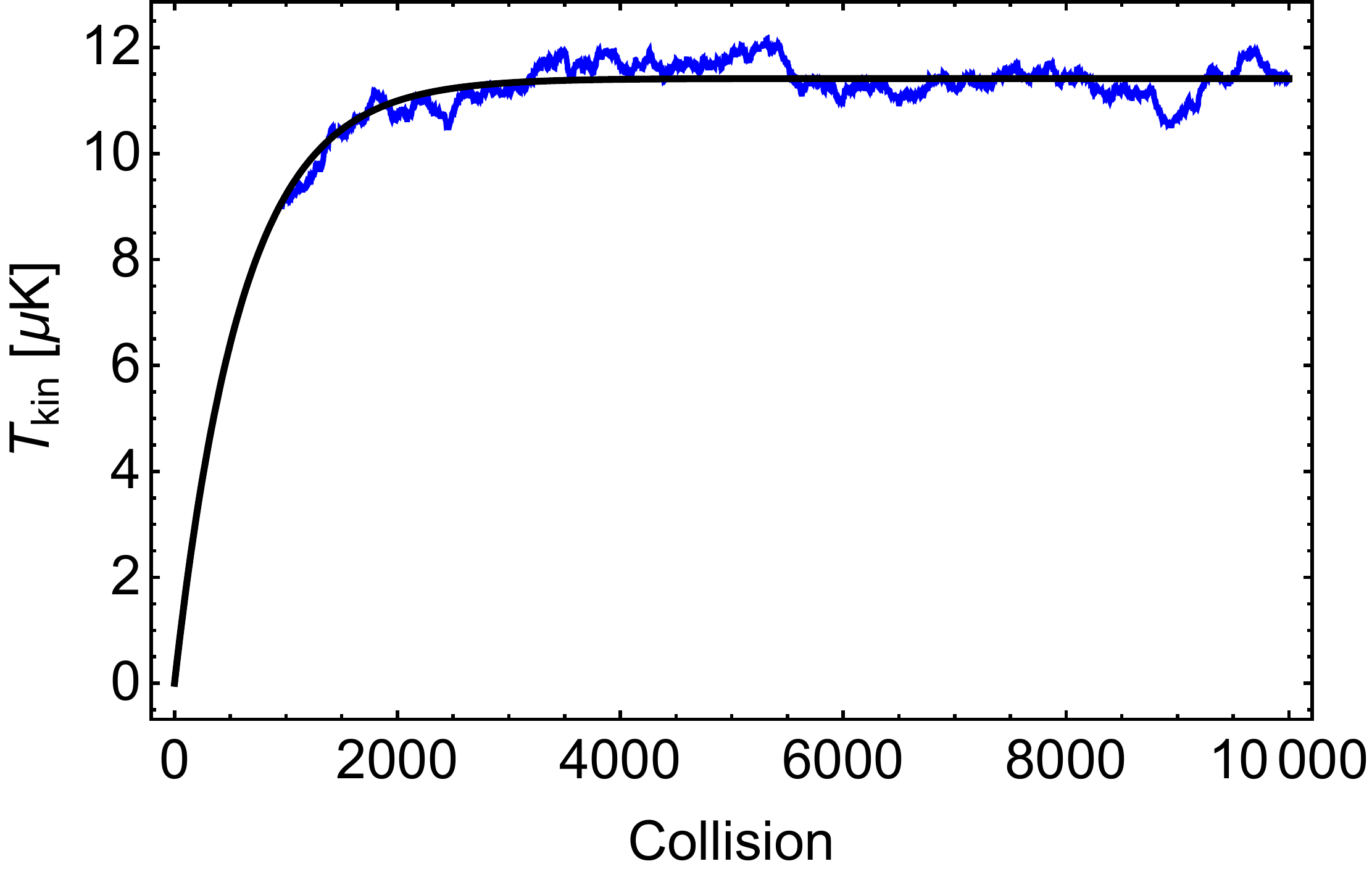}
  \hspace{0.05\textwidth}
  \includegraphics[width=0.45\textwidth]{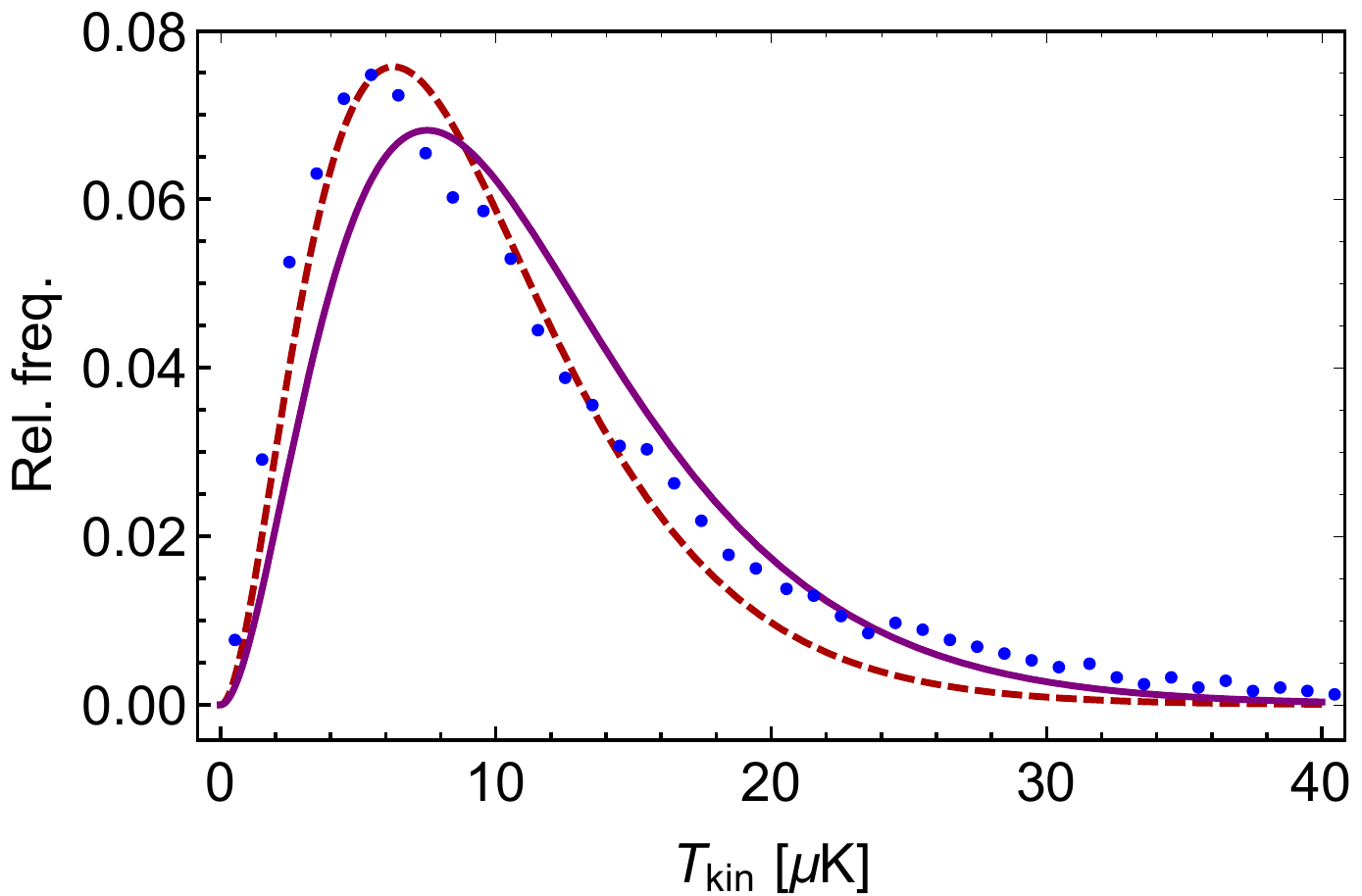}
\end{center} \caption{Average kinetic energy of an ion colliding with atoms at
$2\,\mu$K averaged over 656 runs to obtain the ion's temperature (left) from an
exponential fit (black) and
  distribution of the ion's energies in units of $T_{\rm{kin}}$ as defined in Eq.~\ref{eqn:ekinavg}
  after thermalization (right) along with
  a fitted thermal distribution (red, dashed) and a distribution where the
  temperature was fixed to the value obtained from the exponential fit (purple).}
\label{fig:thermalRad}
\end{figure}

Having a method for the energy determination at hand, it is of
importance to check the negligible influence of the start- and escape sphere
sizes for the atoms. If the sphere radii are picked at the same order as
the range of the atom-ion interaction, the immediate change in potential energy
after the insertion and extraction of an atom leads to unrealistic kicks in
the force on the ion.
To check the influence of the sphere radii on the ion temperature, the inner
sphere radius $r_0$ was scanned between 0.2 and 1.8 $\mu$m. The thermalization of
a single trapped ion initially at rest with a thermal cloud of atoms at $2\,\mu$K
was simulated. The outer
sphere radius $r_1$ was chosen to be 0.5\,\% bigger than $r_0$.
An example for a thermalization curve (blue points) is shown in Fig.~\ref{fig:thermalRad}
(left). The curve was obtained by averaging over 656 individual runs and fitted
with an exponential (see Eq.~\ref{eq:expfit})
(black line)
leading to an equilibrium temperature of $T_{\rm{kin}} = 11.4(1)\,\mu$K on
the characteristic time scale of $N_{\rm{col}}=607(2)$ collisions, using $T_0
= 0$ as the initial ion's temperature.
The ion's energy distribution after thermalization is shown in
Fig.~\ref{fig:thermalRad} (right). The blue points were obtained from all ion
energies of the 656 runs between collision 5000 and 10000 and fitted with a
thermal distribution (red, dashed) leading to a temperature
of $9.4(2)\,\mu$K and a thermal distribution with fixed temperature (purple)
obtained from the exponential fit (left). The ion's energies deviates quite a bit
from the thermal distributions, showing a longer tail towards high energies,
which is a well known
behavior~\cite{Chen:2014,Weckesser:2015,Meir:2016,Rouse:2017},
caused by the additional kinetic energy due to the micromotion of the
ion.

\begin{figure}[htpb]
  \begin{center}
  \includegraphics[width=0.45\textwidth]{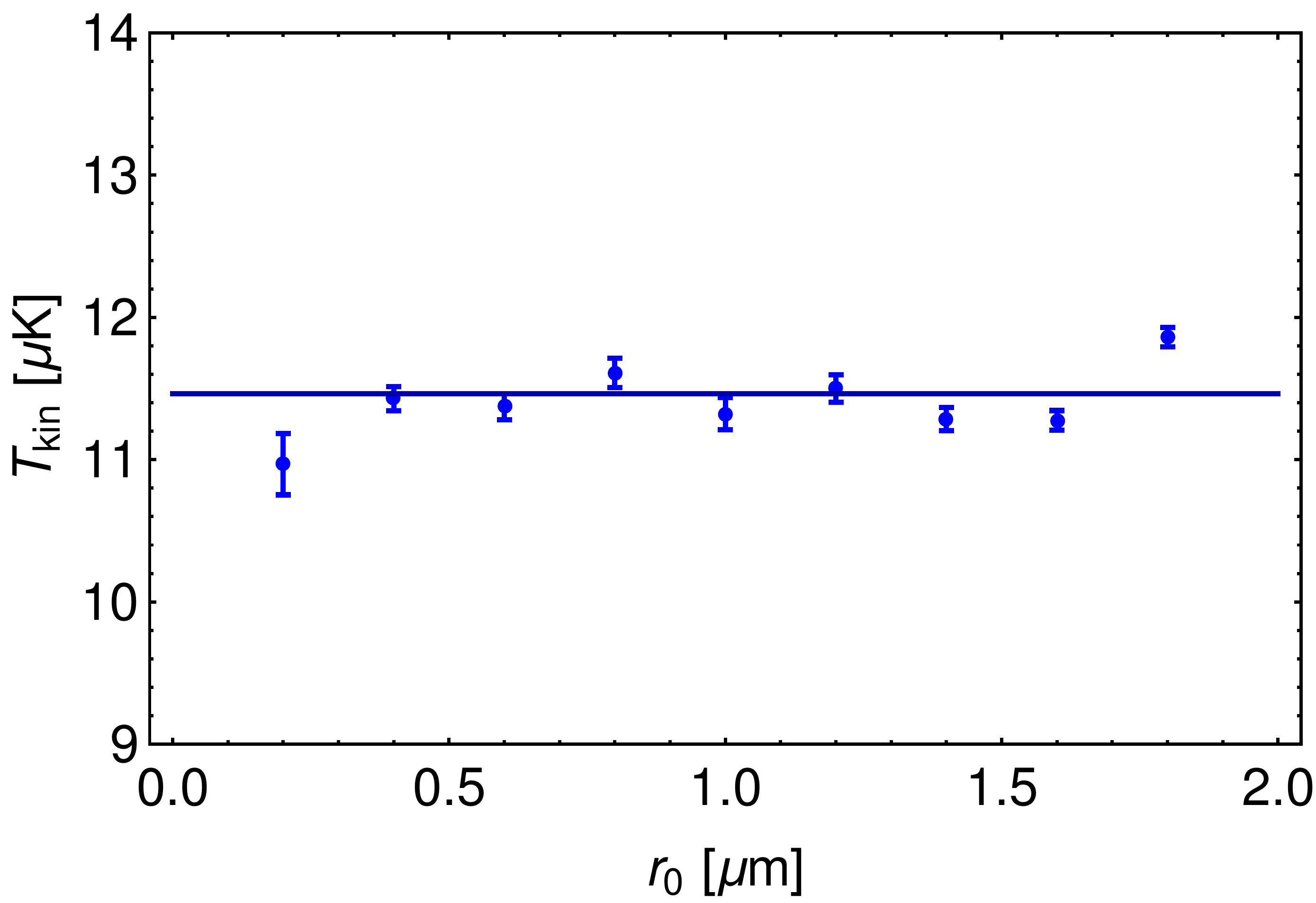}
  \hspace{0.05\textwidth}
  \includegraphics[width=0.45\textwidth]{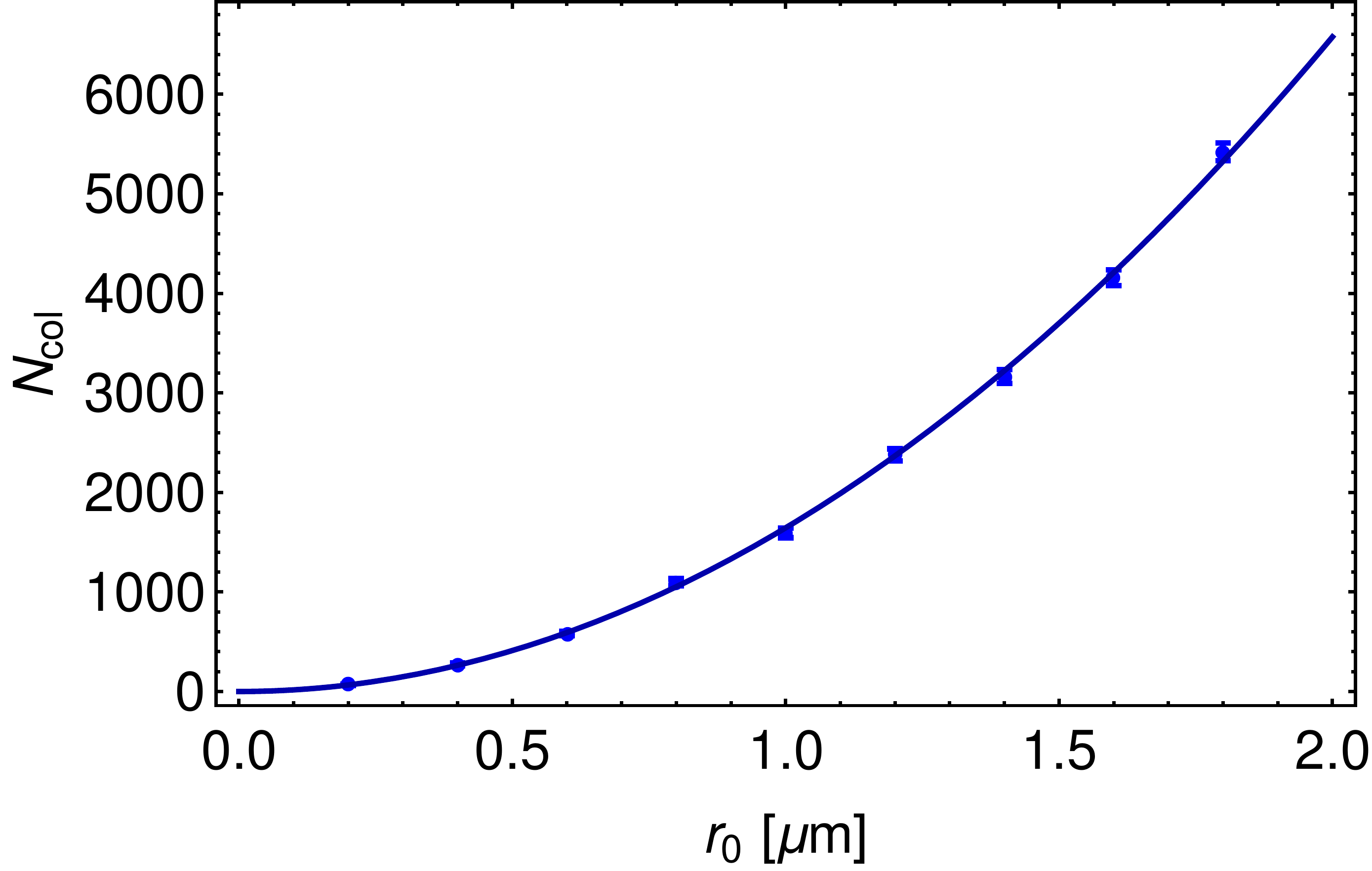}
  \end{center} \caption{Equilibrium temperature $T_{\rm{kin}}$ as defined in
  Eq.~\ref{eqn:ekinavg} (left) and characteristic number of collisions $N_{\rm{col}}$
  (right) for an ion colliding with atoms at
$2\,\mu$K versus the starting distance $r_0$ between atom and ion. While
  the equilibrium temperature does not depend on $r_0$ in the scanned regime, the
  characteristic number of collisions increases quadratically. The lines show
  a constant (left) and quadratic fit (right).}
\label{fig:thermalRadScan}
\end{figure}

The final temperatures and characteristic number of collisions $N_{\rm{col}}$
required for equilibration
for the different starting radii
are shown in Fig.~\ref{fig:thermalRadScan} and were obtained using the
exponential fit model given by Eq.~\ref{eq:expfit}. For each point, at
least 300 runs were averaged.
The equilibrium temperature $T_{\rm{kin}}$ of the ion shows no dependence on the
starting sphere size $r_0$, whereas $N_{\rm{col}}$ shows a quadratic behavior
over the scanned range. This behavior can be qualitatively explained by the
nature of Langevin collisions. For a given collision energy $E_{\rm{col}}$, every
atom with an impact parameter smaller than $b_c=(2 C_4/E_{\rm{col}})^{1/4}$ undergoes
a Langevin collision and can therefore cause a large energy and momentum transfer that
contributes to the thermalization process. The fraction of atoms that undergo
a Langevin collision $P_{\rm{L}}$ and therefore fly into
the solid angle element defined by $b_c$ is then given by $P_{\rm{L}}
= (1-\cos\left(b_c/r_0\right))\approx 1/2 (b_c/r_0)^2$. For an increasing $r_0$ this
automatically demands for a quadratic increase in the required number of total
collisions to equilibrate.
Unless noted otherwise, $r_0=0.6\,\mu$m is used in all further simulations as
a trade-off between simulation time and realistic atomic densities (see e.g.~\cite{Gross:2016}).
Demanding
only one atom at a time inside the sphere around one of the ions results in a
density of
$\rho_{\rm{a}} = \frac{1}{4/3 \pi r_0^3 N_{\rm{ions}}} < 1.1\cdot10^{18}\,{\rm{m}}^{-3}$.

\begin{figure}[htpb]
  \begin{center}
  \includegraphics[width=0.45\textwidth]{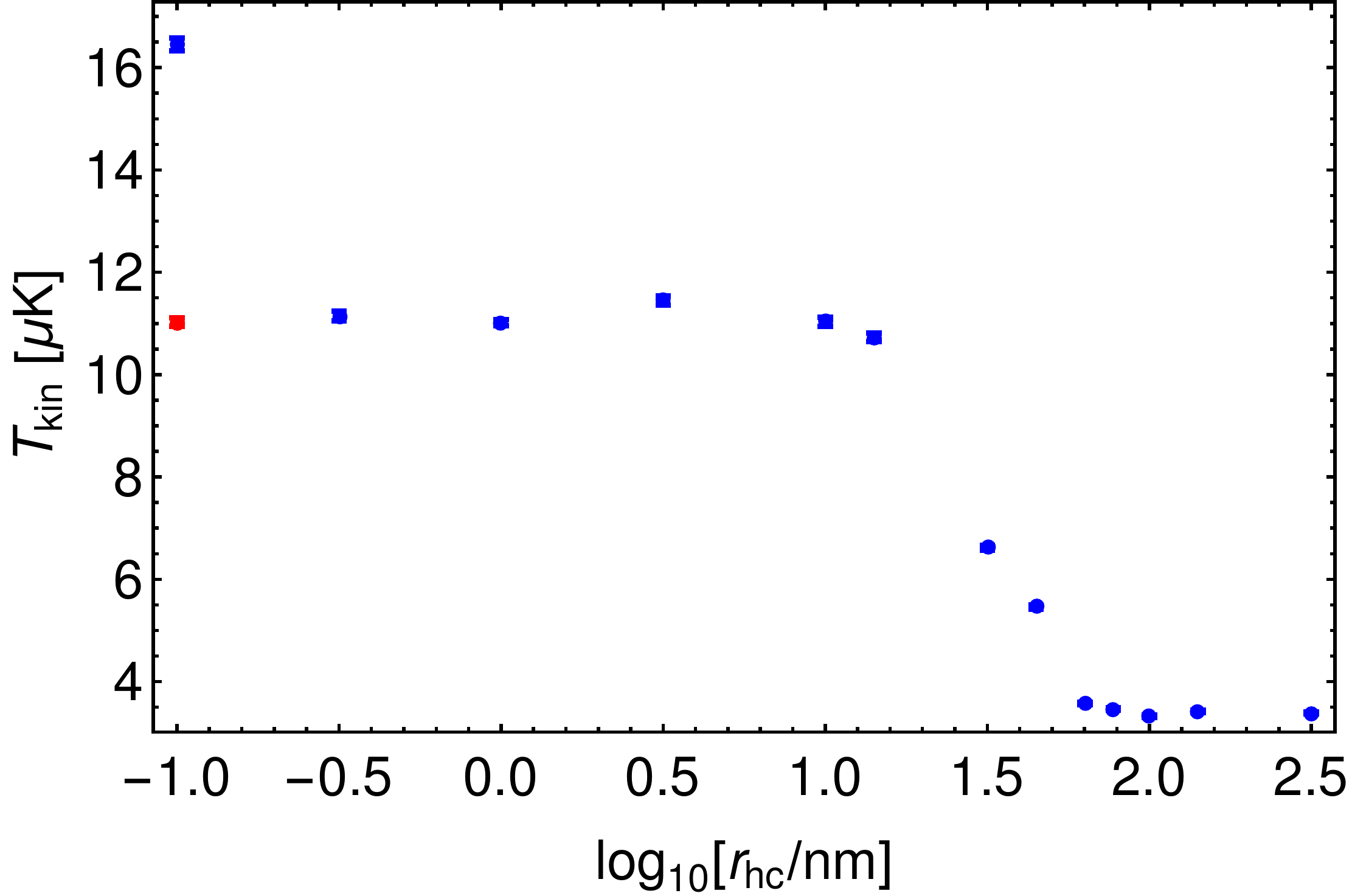}
  \hspace{0.05\textwidth}
  \includegraphics[width=0.45\textwidth]{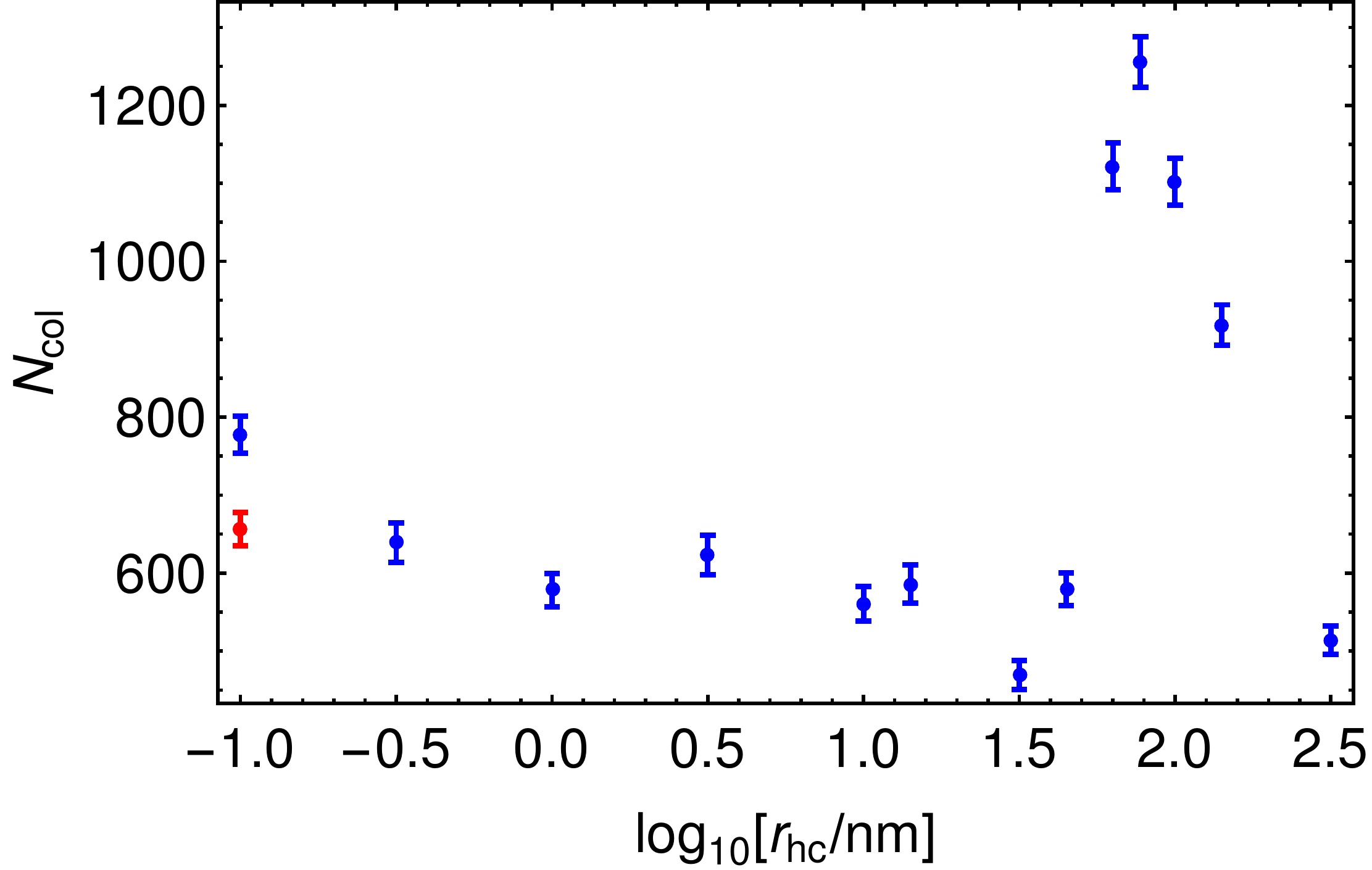}
  \end{center} \caption{Equilibrium temperature $T_{\rm{kin}}$ as defined in
  Eq.~\ref{eqn:ekinavg} (left) and characteristic number of collisions $N_{\rm{col}}$
  (right) for an ion colliding with atoms at
  $2\,\mu$K versus the repulsive barrier radius $r_{\rm{hc}}$. The
  red points were obtained using a higher numerical precision as explained in
  the text.}
\label{fig:c6scans}
\end{figure}

To realistically model the atom-ion interaction, one needs to check as well that
the temperature of the ion does not strongly depend on the choice of the hard-core
radius
parameter $C_6$ as introduced in Eq.~\ref{eqn:aipot}. In reality, a repulsive
barrier is expected to be at a distance, where the electronic wavefunctions of
the atom and ion begin to significantly overlap, typically in the range of
hundreds of picometers to a few nanometers. The parameter $C_6$ was therefore
scanned in a range between $5\cdot 10^{-14}\,{\rm{m}}^2$ to $5\cdot
10^{-21}\,{\rm{m}}^2$, effectively varying the position of the classical turning
point $r_{\rm{hc}} = \sqrt{2 C_6}$ between $0.1\,$nm and $316\,$nm.  The results
for both final ion temperature $T_{\rm{kin}}$ and collisions required for
equilibration $N_{\rm{col}}$ are shown in Fig.~\ref{fig:c6scans}.  The points
were obtained by averaging the ion's average kinetic energy over at least 300
runs and fitting it according to Eq.~\ref{eq:expfit}.
For a broad range of barrier radii the final temperature of the ion remains at
the same level. For values bigger than $r_{\rm{hc}} = 10\,$nm the potential is more and more
dominated by the repulsive term proportional to $C_6$, preventing Langevin
collisions and therefore the ion from micromotion-induced heating as we
describe it in our work Ref.~\cite{Secker:2017}, where a repulsive barrier
is utilized to prevent exactly this heating mechanism.
For the smallest value of $r_{\rm{hc}} = 0.1\,$nm, the ion
temperature seems to be a factor of 1.5 higher than in the regime between 0.3 to
10\,nm, which can be explained by
numerical errors due to the increasing steepness of the hard core barrier for
low values of $r_{\rm{hc}}$ leading to large changes in acceleration in a hard core collision.
Therefore, this point was simulated again
with a five times smaller tolerance in the adaptive step-size Runge-Kutta propagator,
leading to the red points, in
agreement with the values for larger $r_{\rm{hc}}$. The number of collisions
required for thermalization $N_{\rm{col}}$ seems to first slightly decrease for higher
values of $r_{\rm{hc}}$ but shows a dramatic increase by around a factor of two
at $r_{\rm{hc}}=77\,$nm. Note that at this point the potential energy minimum
caused by the attractive $C_4$-term of the potential becomes comparable
to the collision energy, dominated by the atom temperature of $2\,\mu$K.
Therefore, the intermediately
released kinetic energy during a Langevin collision becomes negligible. For even
higher values of $r_{\rm{hc}}$ the thermalization process speeds up again due to
the quadratically increasing geometric cross section for repulsive collisions.
For all further simulations, $r_{\rm{hc}}=1\,$nm is used, which is around three times
larger than the classical turning point of the Li-Yb$^+$ system~\cite{Joger:2017,Furst:2018}
but still produces similar results with less numerical effort due to the weaker
forces involved.

\begin{figure}[htpb]
  \begin{center}
  \includegraphics[width=0.45\textwidth]{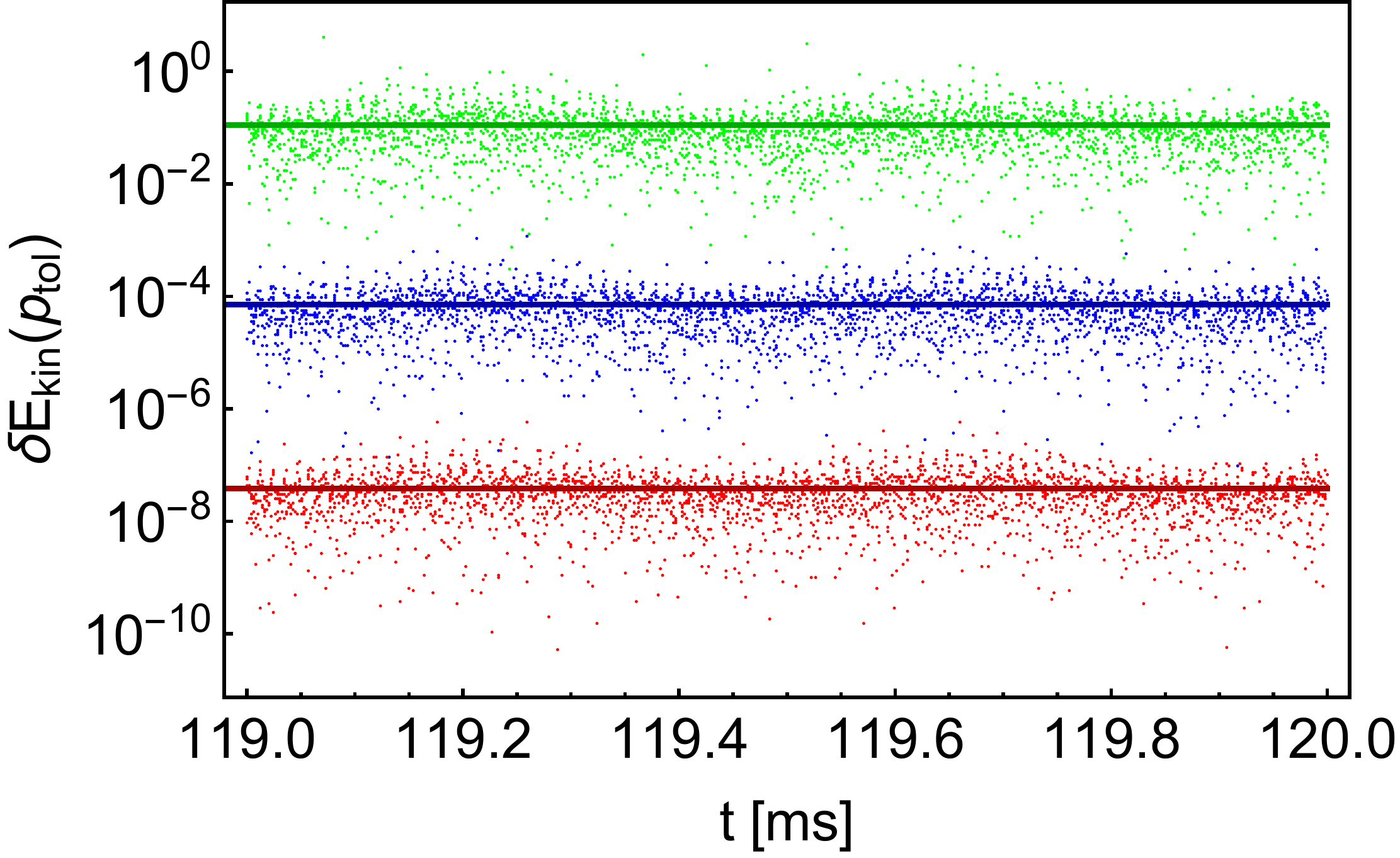}
  \hspace{0.05\textwidth}
  \includegraphics[width=0.45\textwidth]{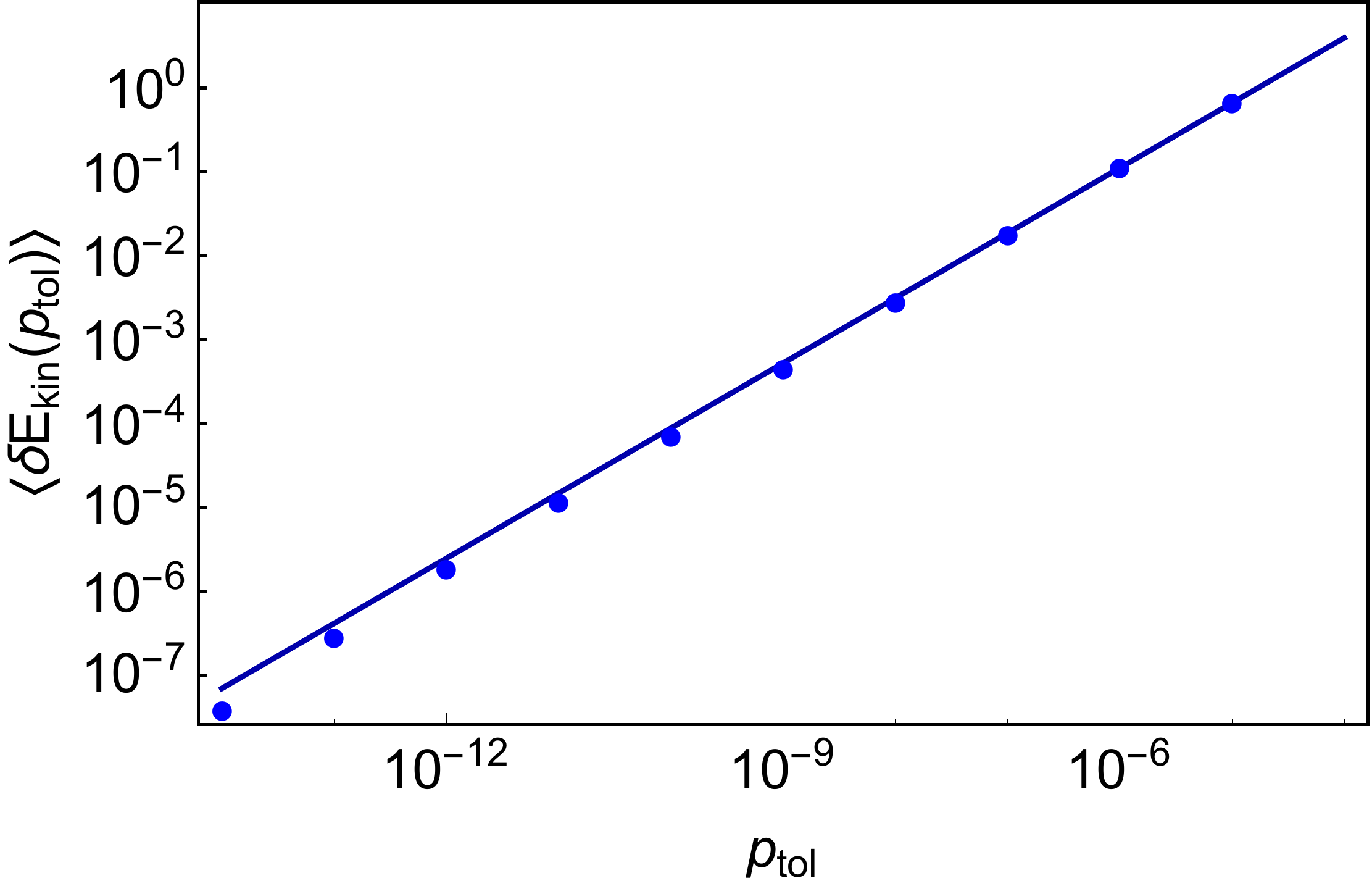}\end{center}
  \caption{Relative deviations in kinetic energy for a single trapped ion for
  different values of the adaptive step-size algorithm tolerance $p_{\rm{tol}}$
  versus propagation time (left). For clarity only the values for $p_{\rm{tol}}
  = 10^{-6}$ (green), $10^{-10}$ (blue) and $10^{-14}$ is shown in the left
  plot, along with the time averages (lines). The time averages for the other
  scanned tolerances are shown on the right, along with an exponential
  fit.}\label{fig:toltest1}
\end{figure}

During propagation, the Runge-Kutta propagator adjusts the size of the time
steps in order to stay below a given relative accuracy parameter $p_{\rm{tol}}$.
It therefore propagates the system once by a full time step and once by two half
time steps and compares the relative difference in propagated coordinates
between both methods. If the maximum relative difference between one of the
coordinates (including velocity) is bigger than the desired tolerance, the
propagation step is repeated using an adjusted time step.
To ensure a sufficiently small tolerance $p_{\rm{tol}}$, further tests were
performed. Firstly, the allowed tolerance was scanned
from $p_{\rm{tol}} = 10^{-5}$ to $10^{-15}$ as a parameter for the propagation of
a single ion starting at a randomly chosen kinetic energy sampled from a thermal Distribution at
$T_{\rm{kin}}=13\,\mu$K, leading to $E_{\rm{kin}}/k_{\rm{B}} = 20.5\,\mu$K in the
presented case. The trajectories including the velocities for the individual runs were
stored to compute the relative deviation in kinetic energy for each tolerance with
the one from the smallest value\footnote{Note that
values of $p_{\rm{tol}} < 10^{-15}$ can cause numerical instabilities due to
the close by machine
precision limit $\epsilon$ for which the numerical addition/subtraction $1.0\pm\epsilon=1.0$.
On a 64-bit computer, $\epsilon\approx 2.22\cdot 10^{-16}$ for double precision
floating point numbers, according to the IEEE-754 standard.},
$p_{\rm{tol}} = 10^{-15}$
,
\begin{equation}
  \delta E_{\rm{kin}}(p_{\rm{tol}}) = \left|\frac{E_{\rm{kin}}
  (p_{\rm{tol}})-E_{\rm{kin}}(10^{-15})}{E_{\rm{kin}}(10^{-15})}\right|\,.
\end{equation}
Because collisions with atoms can cause a dramatically
different change in
trajectory for each tolerance, no atoms were introduced in this test. The ions
were propagated for $120\,$ms, a timescale that typically corresponds to 10000
collisions in the simulation. Due to the large amount of data, the trajectories
were stored only during the last millisecond of propagation.
The resulting relative deviations $\delta E_{\rm{kin}}(p_{\rm{tol}})$ are shown
in Fig.~\ref{fig:toltest1} (left).
Due to the adaptive step-size algorithm, it is not
possible to have the trajectories for each tolerance stored at the exact same
time steps each, therefore the kinetic energy $\delta E_{\rm{kin}}(10^{-15})$
was interpolated using cubic polynomials to match the time grid of the other
tolerances, possibly leading to a small amount of interpolation noise. For
clarity, only the values for $p_{\rm{tol}} = 10^{-6}$ (green), $10^{-10}$ (blue)
and $10^{-14}$ (red)
are shown along with their time averages (straight lines). In
Fig.~\ref{fig:toltest1} (right) the time averaged deviations for the other
values of $p_{\rm{tol}}$ are shown, approximately following an exponential
behavior (solid line) with exponent $ n\approx0.78$\,.
While for a tolerance of $p_{\rm{tol}} = 10^{-6}$ the time averaged relative
deviation is $< 11\,\%$, $p_{\rm{tol}}=10^{-8}$ delivers
acceptable values of $\langle \delta E_{\rm{kin}}(p_{\rm{tol}})\rangle \leq
0.2\,\%$ already.

\begin{figure}[htpb]
  \begin{center}
  \includegraphics[width=0.45\textwidth]{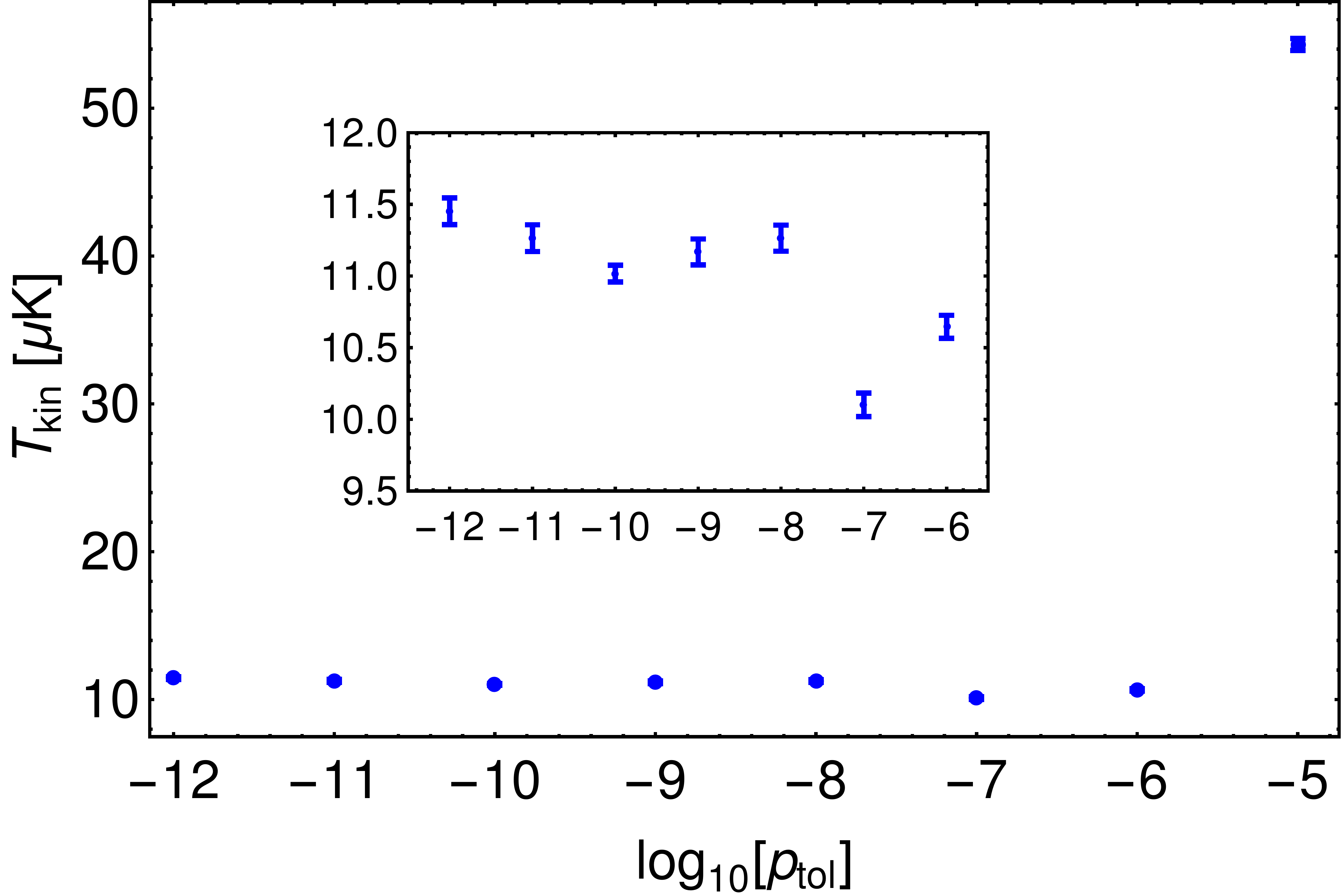}
  \hspace{0.05\textwidth}
  \includegraphics[width=0.45\textwidth]{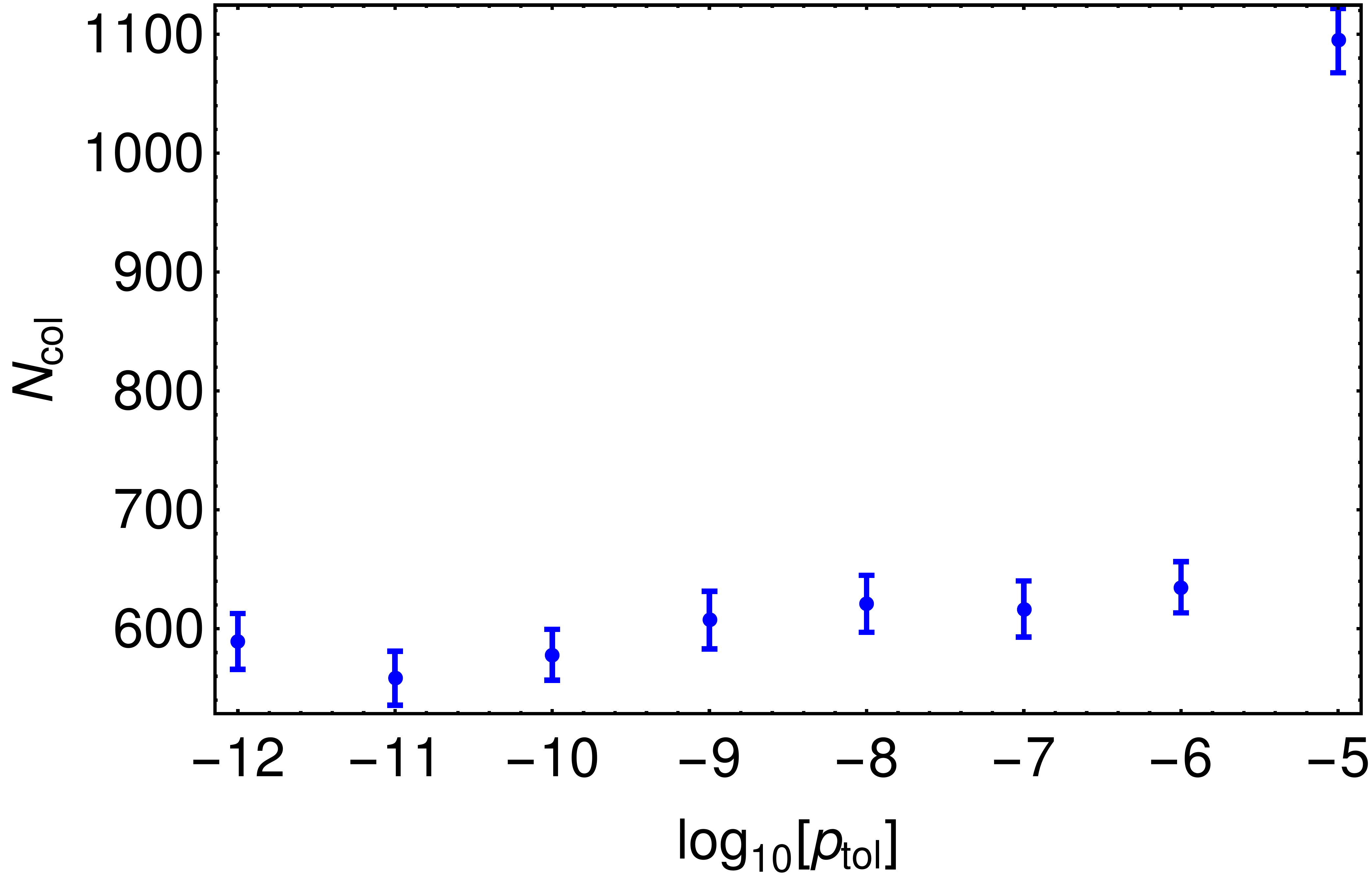}
  \end{center} \caption{Final ion temperature $T_{\rm{kin}}$ (left) and
  characteristic number of collisions required to equilibrate $N_{\rm{col}}$
  (right) for an ion colliding with atoms at $2\,\mu$K
  versus tolerance
  parameter $p_{\rm{tol}}$ used in the adaptive step-size propagator.
  The inset shows a magnified version of the plot from $p_{\rm{tol}} = 10^{-12}$ to $10^{-6}$.}\label{fig:toltest2}
\end{figure}

Similar to the tests for $C_6$ and $r_0$, also the influence of the tolerance
parameter $p_{\rm{tol}}$ on
the final ion temperature $T_{\rm{kin}}$ and required collisions $N_{\rm{col}}$
to equilibrate was investigated. The results are shown in Fig.~\ref{fig:toltest2}.
Each point was obtained from taking the average of $\bar{E}_{\rm{kin}}$ over at least 300
individual runs and fitting the curves according to Eq.~\ref{eq:expfit}.
Both observables do not change significantly from $p_{\rm{tol}} = 10^{-12}$ to
$10^{-6}$, only the point at $p_{\rm{tol}} =  10^{-5}$ shows a dramatic increase
in both $T_{\rm{kin}}$ and $N_{\rm{col}}$ due to increasing numerical errors.
For all further simulations $p_{\rm{tol}} = 10^{-10}$ is used (unless noted otherwise)
as a trade-off between precision and computational effort.

\begin{figure}[htpb]
  \begin{center}
  \includegraphics[width=0.45\textwidth]{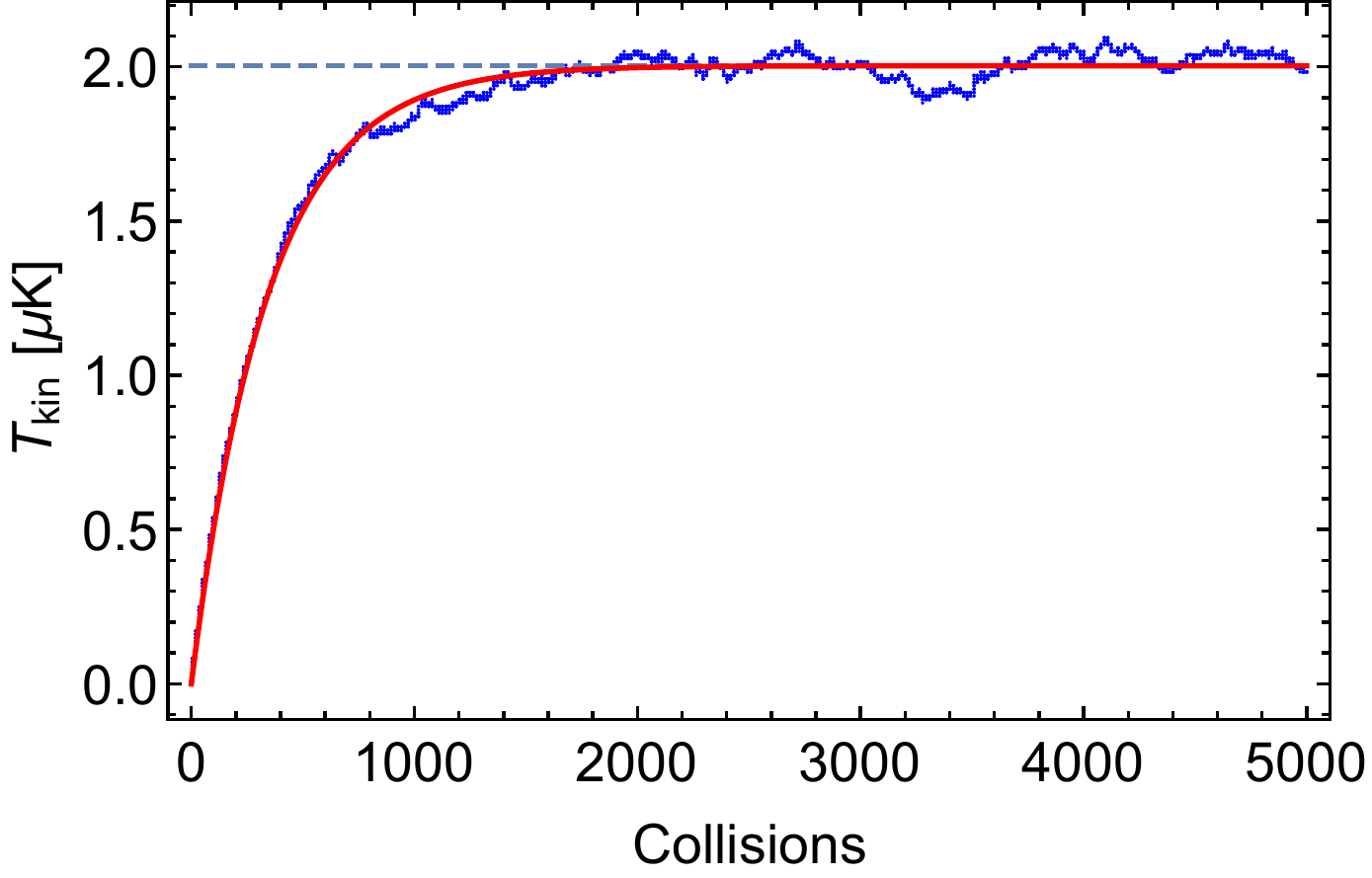}
  \hspace{0.05\textwidth}
  \includegraphics[width=0.45\textwidth]{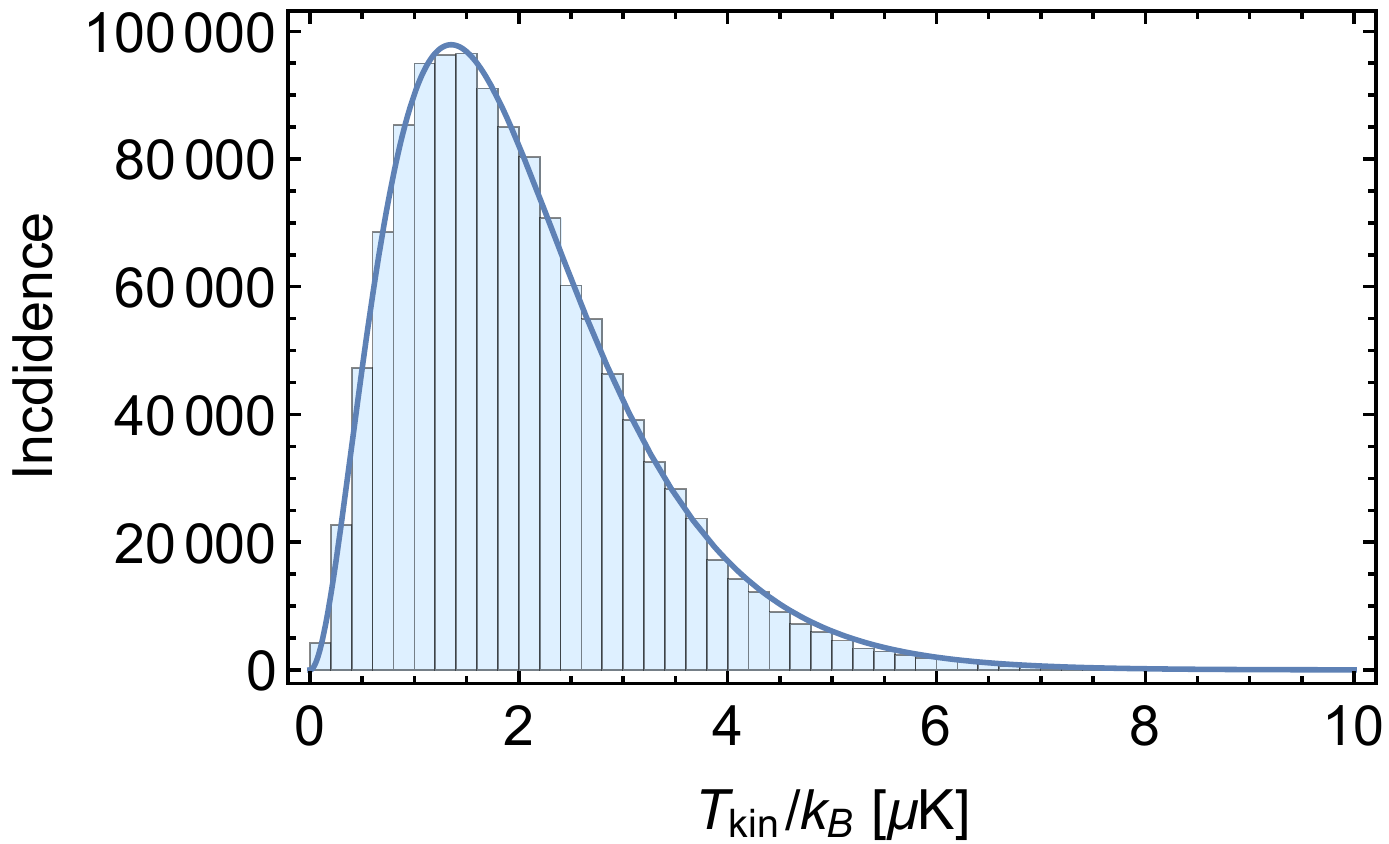}
\end{center} \caption{Thermalization curve of a single ion trapped in harmonic
  oscillator potential colliding with atoms at $2\,\mu$K,
  using the secular trap frequencies of the employed Paul trap potential (left).
  The distribution of average kinetic energies according to
  Eq.~\ref{eqn:ekinavg} in units of $T_{\rm{kin}}$ (right) shows a perfect
  thermal behavior. The exponential fit (left) as well as the fitted thermal
  distribution lead to a final temperature of $T_{\rm{ion}} = 2\,\mu$K.
  }
\label{fig:hamostherm}
\end{figure}

A final check for both energy conservation of the propagator
during collisions as well as physical
behavior of the system is to investigate the secular case, where the time-dependent
trapping potential of the Paul trap is replaced by a 3D harmonic oscillator
potential with the secular trap frequencies of the Paul trap. From
a thermodynamic point of view, the ion should then thermalize to the same
temperature as the atomic bath and the total energy during each collision should
be conserved since no micromotion energy can be transferred to the secular
oscillation. The resulting thermalization curve, averaged over 608 individual
runs along with a histogram of the energy distribution is shown in
Fig.~\ref{fig:hamostherm}.
The histogram was taken from all points between collision 3000 and
5000 and is in perfect agreement with a thermal distribution (solid line) at
$2\,\mu$K, the same temperature as the atomic bath.
Also the exponential fit of the thermalization curve (left) leads to
the same value, thus indicating a correct physical
behavior of the numerical model.

\begin{figure}[t]
  \begin{center}
  \includegraphics[width=0.45\textwidth]{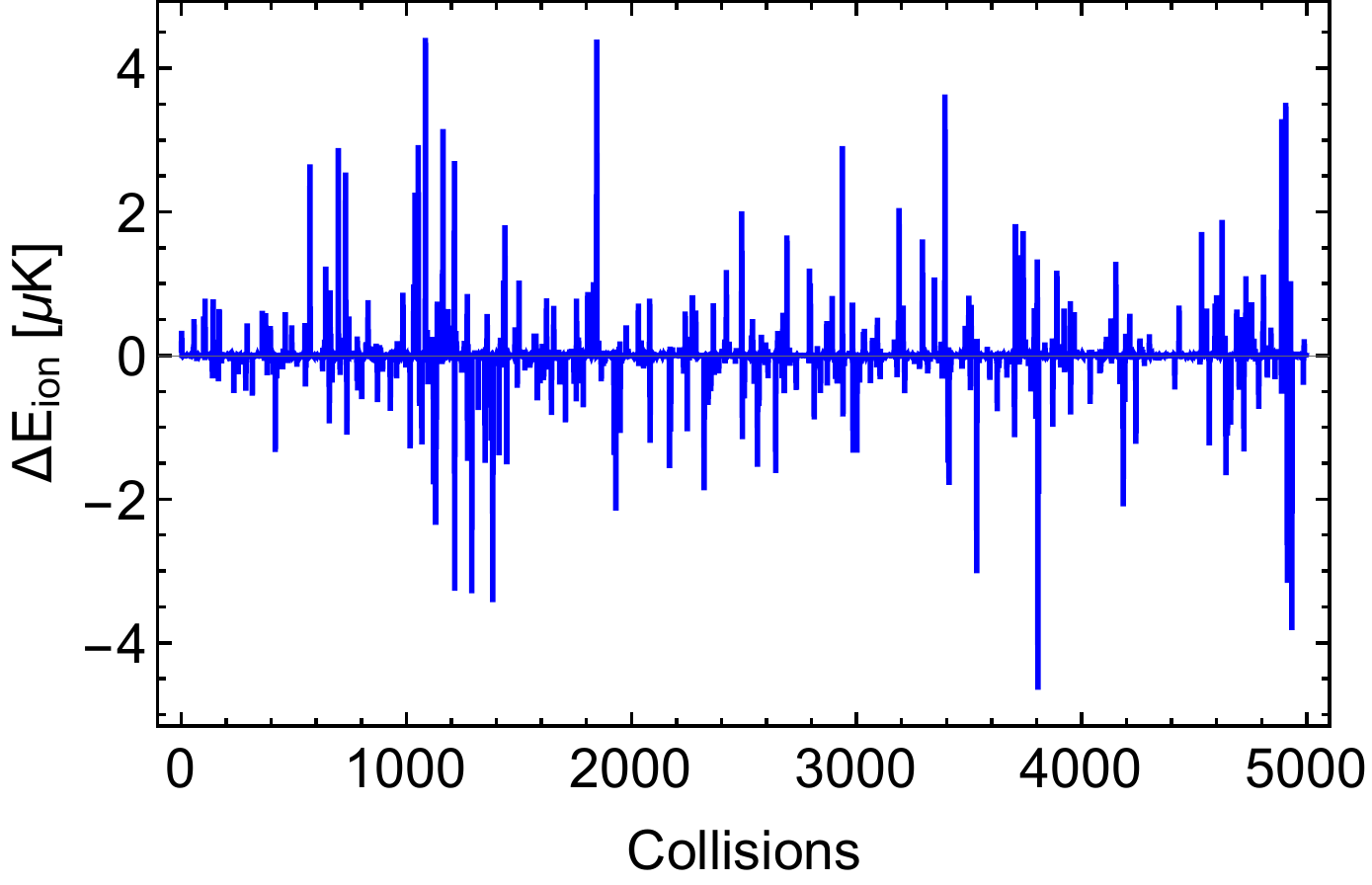}
  \hspace{0.05\textwidth}
  \includegraphics[width=0.45\textwidth]{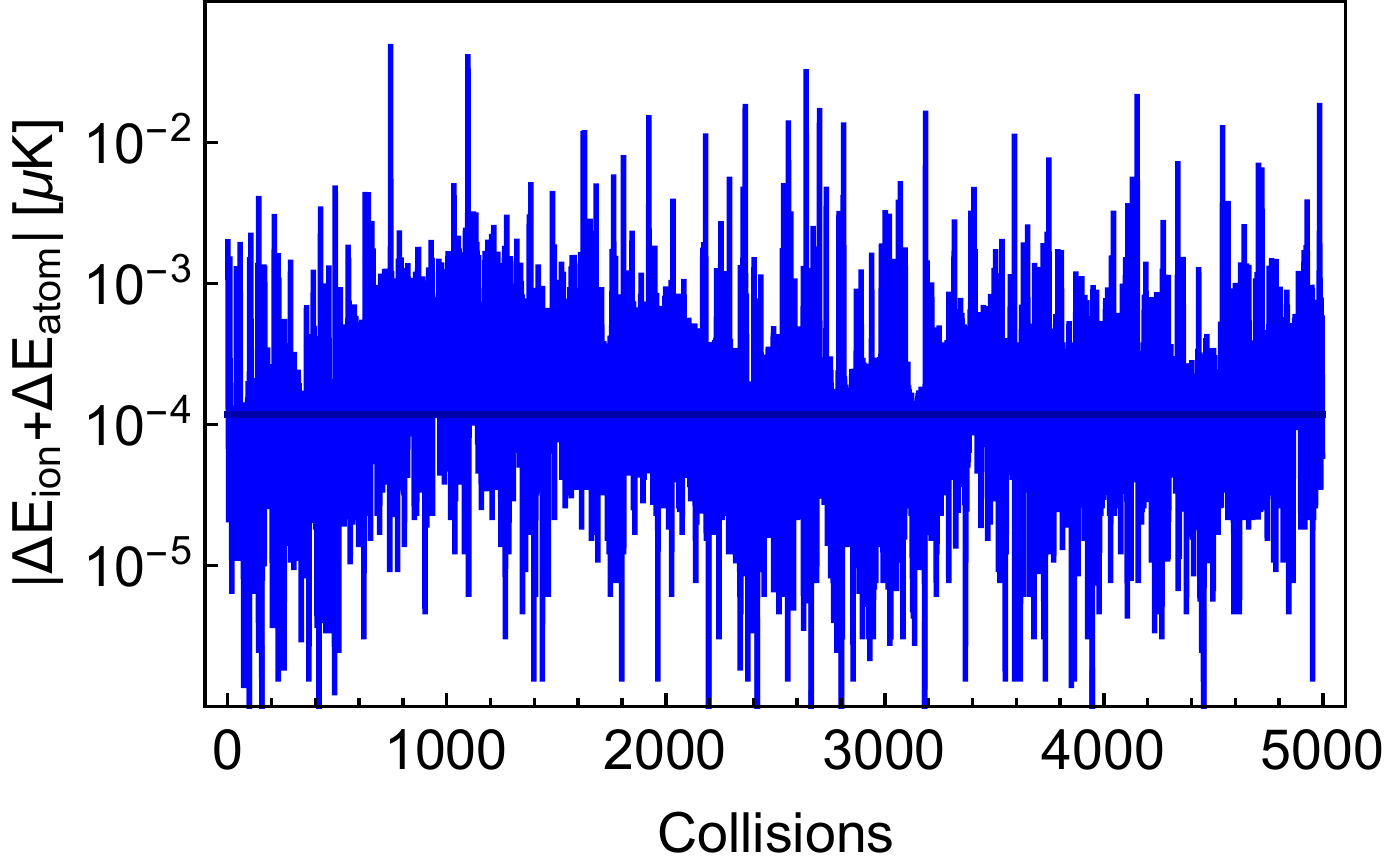}
\end{center} \caption{Change in energy of a harmonically trapped ion within each
  collision with atoms at $2\,\mu$K (left) and total change in energy of the
  atom-ion system for each collision (right). The solid line shows the averaged
  energy gain of the system per collision.
  }
\label{fig:hamosenergy}
\end{figure}

To finally investigate the energy conservation of the collisions, the energy
transfer between atom and ion in each collision was investigated by comparing
the atom and ion energies before and after a collision, at the points in time
$t_0$
when an atom is introduced on the sphere with radius $r_0$ with the point in time $t_{\rm{1}}$ when that atom escapes the sphere defined by $r_1$.
The
energy transfer on an ion trapped in the harmonic oscillator potential is
shown in Fig.~\ref{fig:hamosenergy}~(right), taken from one of the 608
individual runs from the simulation used for Fig.~\ref{fig:hamostherm}~(left).
The plot shows the ion's energy transfer $\Delta
{E}_{\rm{ion}} = {E}_{\rm{ion}}(t_1)-{E}_{\rm{ion}}(t_0)$ for each collision and
ranges on scales limited by the atom energies. For the atom, a corresponding
curve can be obtained. In Fig.~\ref{fig:hamosenergy}~(right) the level of energy
conservation $\left|\Delta E_{\rm{ion}} + \Delta E_{\rm{atom}}\right|$ is shown.
The averaged error in total energy in each step
is less than 0.12\,nK (dark blue line) and therefore negligible on the typical energy scales of
the simulations. Note that this error is mainly caused by the sudden but tiny jump
in potential energy when the atom is introduced and extracted. With reasonable
effort this could be corrected in the energy determination and atom
injection
scheme, but is only of interest for much higher densities and lower temperatures
that may anyways require a quantum mechanical treatment.
We therefore conclude that the employed propagator produces physical results
with reasonable precision.

\section{Reality checks of the Fourier method}\label{subs:realityfft}
In this section, we check the accuracy of the presented Fourier method for
determining the average kinetic energy of an ion crystal. Unless stated otherwise, we use a
linear chain of four ions at around $100\,\mu$K and let them thermalize by
collisions with a cloud of atoms at $2\,\mu$K.

\begin{figure}[htpb]
  \begin{center}
  \includegraphics[width=0.45\textwidth]{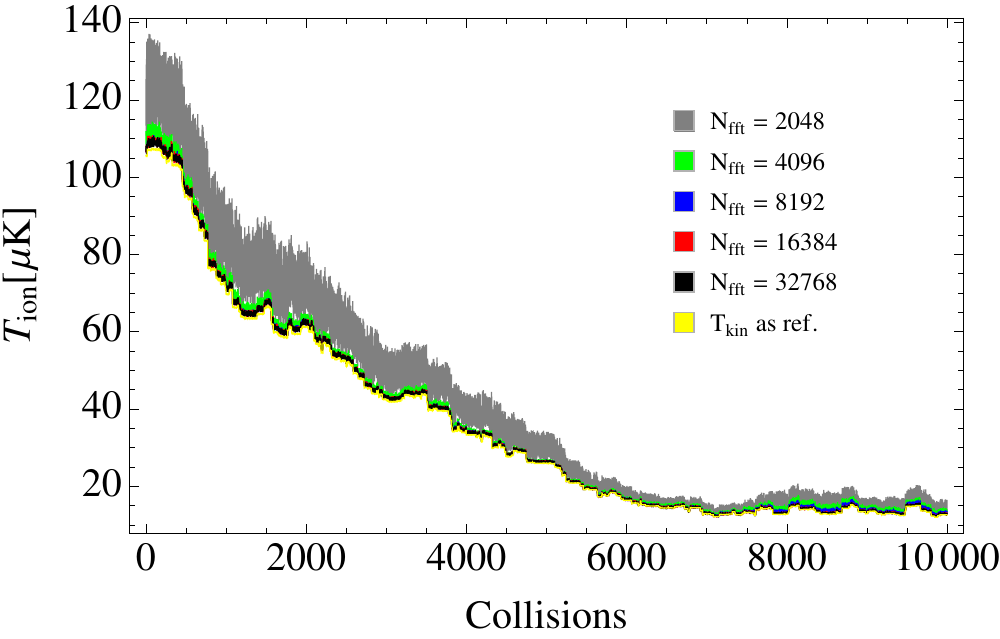}
  \hspace{0.05\textwidth}
    \includegraphics[width=0.45\textwidth]{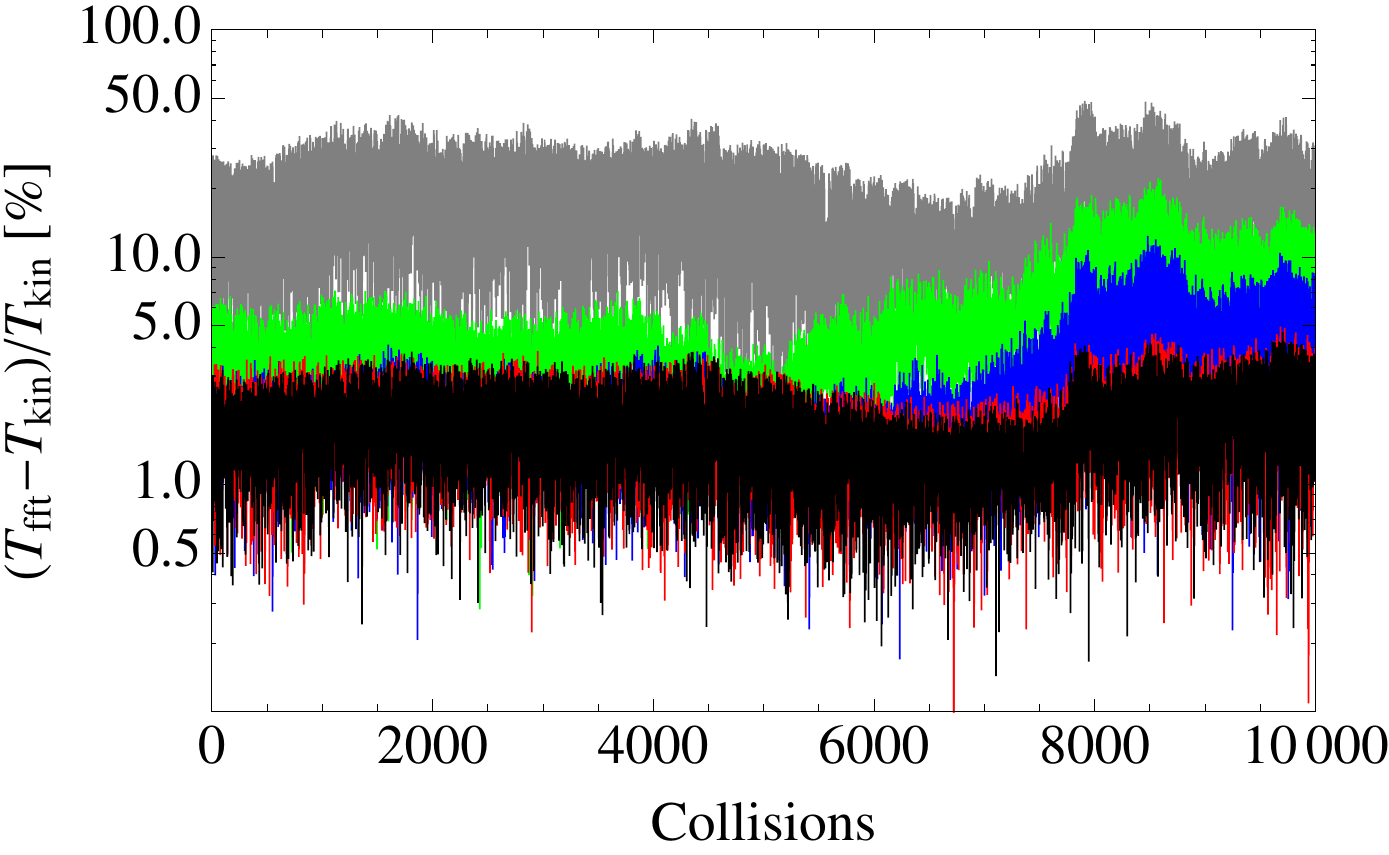}
\end{center}
\caption{Average kinetic energy of a four-ion crystal colliding with thermal atoms at $2\,\mu$K
  (left) obtained by the Fourier method
  (see Eq.~\ref{eqn:tempmodefourier},\ref{eqn:etotfft}) The ions start at an initial temperature of around
$100\,\mu$K. The Fourier spectra were obtained at different $N_{\rm{fft}}$ and
a constant $\Delta t_{\rm{fft}} = 50\,$ns, thus effectively varying the frequency spacing.
As a reference, the temperature obtained from $\bar{E}_{\rm{kin}}$
(see Eq.~\ref{eqn:ekinavg}) is shown (yellow), averaging over $8\,$ms in steps of $5\,$ns.
The curves for $N_{\rm{fft}}= 16384$ (red) and
32768 (black) steps are almost on top of each other, as it can be also seen in
the relative
deviation from $T_{\rm{kin}}$ (right).}
\label{fig:fftsteps}
\end{figure}

To test the Fourier analysis method for obtaining the temperature of an ion
crystal, we compare the temperature $T_{\rm{fft}}$
of Eq.~\ref{eqn:etotfft} with the temperature obtained from the average kinetic
energy $T_{\rm{kin}}$ (Eq.~\ref{eqn:ekinavg}) as shown in
Fig.~\ref{fig:fftsteps}. For a step-size of $\Delta t_{\rm{fft}} = 50\,$ns,
sufficient to resolve frequency components of up to $f_{\rm{max}} = 20\,$MHz,
there is no
significant improvement when increasing the number of steps from 16384 (red) to
32768 (black), the relative deviation from $T_{\rm{fft}}$ to $T_{\rm{kin}}$
is at around $2.5\,\%$ on average over a broad range of temperatures. This
leads to the conclusion that a frequency resolution of $\Delta f_{\rm{fft}}
= 1/(N_{\rm{fft}} \Delta t_{\rm{fft}}) \approx 1.2\,$kHz is a good choice.

\begin{figure}[htpb]
  \begin{center}
  \includegraphics[width=0.45\textwidth]{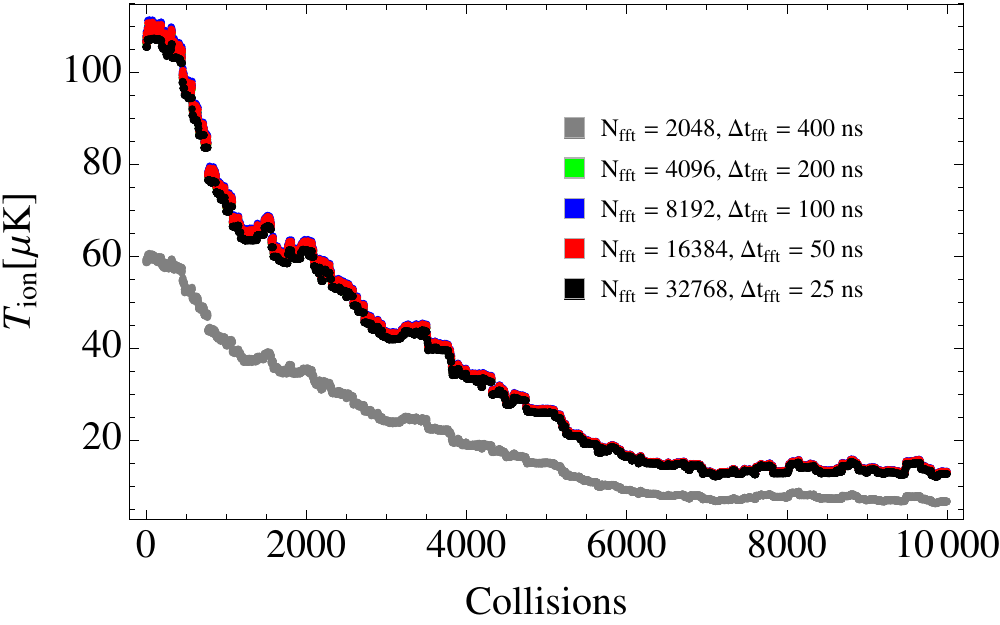}
  \hspace{0.05\textwidth}
  \includegraphics[width=0.45\textwidth]{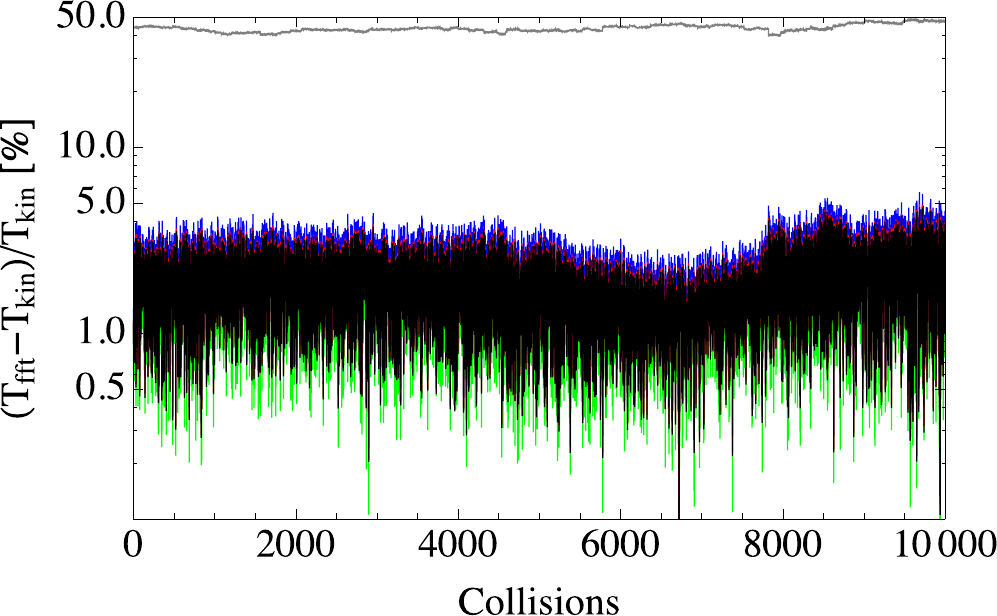}
\end{center}
\caption{Average kinetic energy of a four-ion crystal colliding with atoms at $2\,\mu$K
  (left) obtained by the fourier method for several combinations of
  fourier grid sizes $N_{\rm{fft}}$ and grid spacings $\Delta t_{\rm{fft}}$,
  resembling a variation of the maximally resolvable frequency $f_{\rm{max}}$.
  All curves besides for $N_{\rm{fft}}= 2048$ (gray) lie on top of each other,
  deviating from the average kinetic energy $T_{\rm{kin}}$ by less than 5\,\% (right).}
\label{fig:fftgridsize}
\end{figure}

To
find a sufficient number of grid points while
leaving the frequency resolution constant, we vary $\Delta t_{\rm{fft}}$
inversely with $N_{\rm{fft}}$, as shown in Fig.~\ref{fig:fftgridsize}. For all
combinations with $\Delta t_{\rm{fft}} \leq 200\,$ns, the relative deviation from
$T_{\rm{kin}}$ is approximately the same. At $\Delta t_{\rm{fft}} = 400\,$ns
(gray)
the maximum resolvable frequency is $f_{\rm{max}} = 2.5\,$MHz, being too close
to the micromotion sidebands at around $f_{\rm{rf}} = 2.0\,$MHz and therefore
leading to a much lower energy, dominated by only the low frequency parts.
To be on the safe side, we chose the combination $\Delta t_{\rm{fft}} = 50\,$ns and
$N_{\rm{fft}} = 16384$ for our system.

While during the
collision processes very fast dynamics demanding for an adaptive step
size algorithm may occur, the fastest timescale during the temperature
determination is set by the micromotion oscillation at
$f_{\rm{rf}} = 2\,$MHz in the case for $q_z^2 \ll 1$. Therefore, a fixed step-size propagator with
$\Delta t_{\rm{kin}} \ll 1/f_{\rm{rf}}$ is sufficient.
In order to save computation time, we
use the same fixed step-size propagator
for the Fourier transformation energy determination as for obtaining the
average kinetic energy.
We therefore choose the time grid to be integer subdivisions of the Fourier grid,
$\Delta t_{\rm{kin}} = \Delta t_{\rm{fft}}/n, n \in \mathbb{N}$.
To find a sufficiently small
$\Delta t_{\rm{kin}}$ to resolve the micromotion oscillations at
$f_{\rm{rf}}=2\,$MHz, we compare the kinetic temperature as defined in
Eq.~\ref{eqn:ekinavg} for different propagation time steps $\Delta t_{\rm{kin}}$
with the temperature obtained using the smallest time step $2.5\,$ns as a reference.
For time steps up to $40\,$ns we obtain relative deviations of less than $0.05\,\%$
from the kinetic energy derived using time steps of $2.5\,$ns when averaged over
$8\,$ms propagation time over the whole temperature range. To be on the safe
side, we chose $\Delta t_{\rm{kin}}= \Delta t_{\rm{fft}}/10 = 5\,$ns.

\begin{figure}[htbp]
  \begin{center}
  \includegraphics[width=0.8\textwidth]{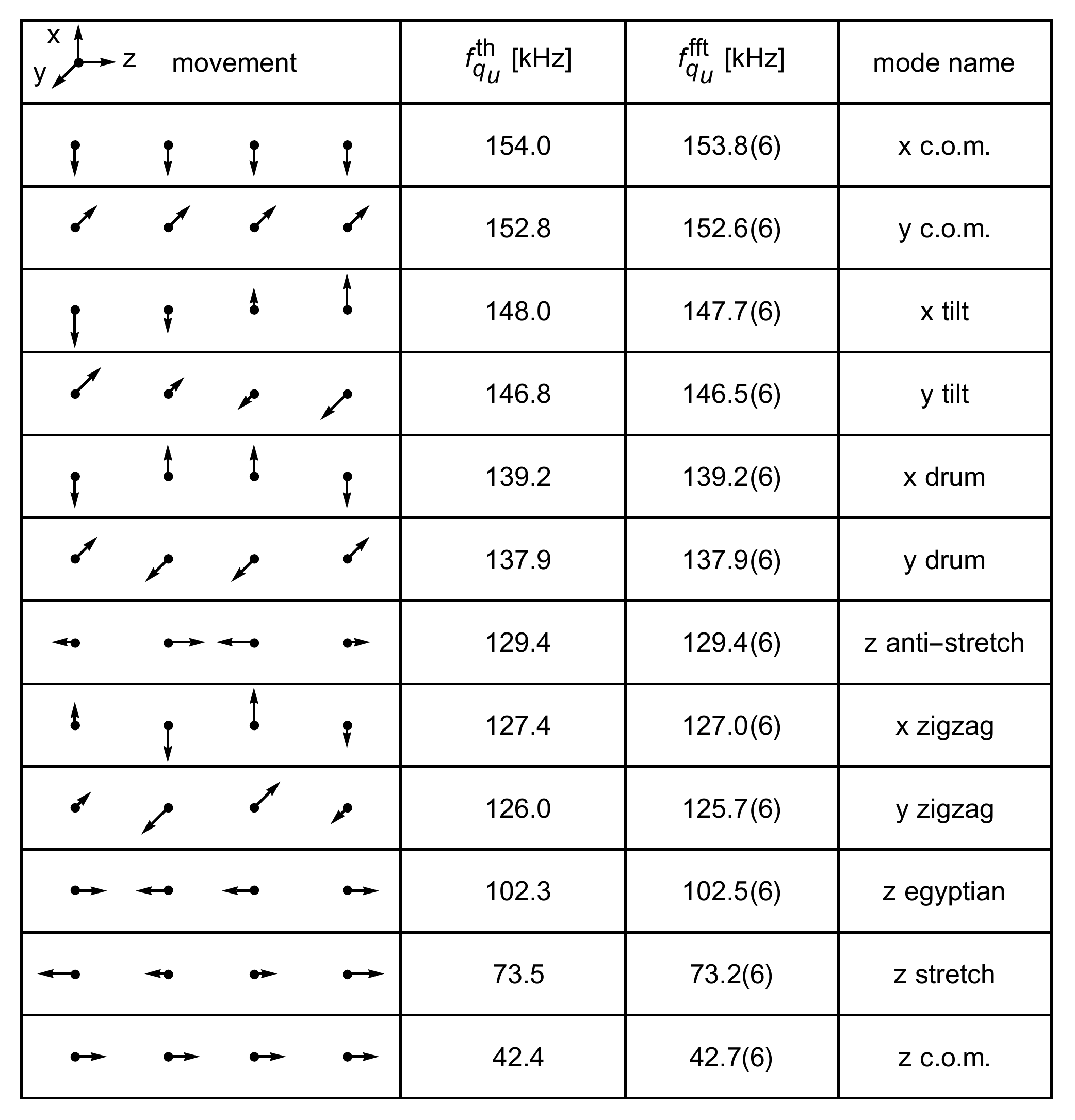}
  \end{center} \caption{Visualization of the normal mode movement for a trapped
  linear four-ion crystal in descending order with respect to the eigenmode
  frequency. The arrows indicate the direction
  and amplitude of the respective mode. The modes are shown along with
  their respective eigenfrequencies $f_{q_u}^{\rm{th}}$ obtained from the
  diagonalization of the secular approximation and the values
  $f_{q_u}^{\rm{fft}}$ obtained numerically from Fourier analysis of the mode
  spectra.  Typical names of the modes are shown on the right column. The
  center-of-mass (c.o.m.) modes represent the upper and lower limit of the
  frequencies.}
\label{fig:allModes}
\end{figure}

\section{Excess micromotion in a linear four-ion crystal}\label{subs:emm4results}
\begin{figure}[htpb]
  \begin{center}
  \includegraphics[width=0.45\textwidth]{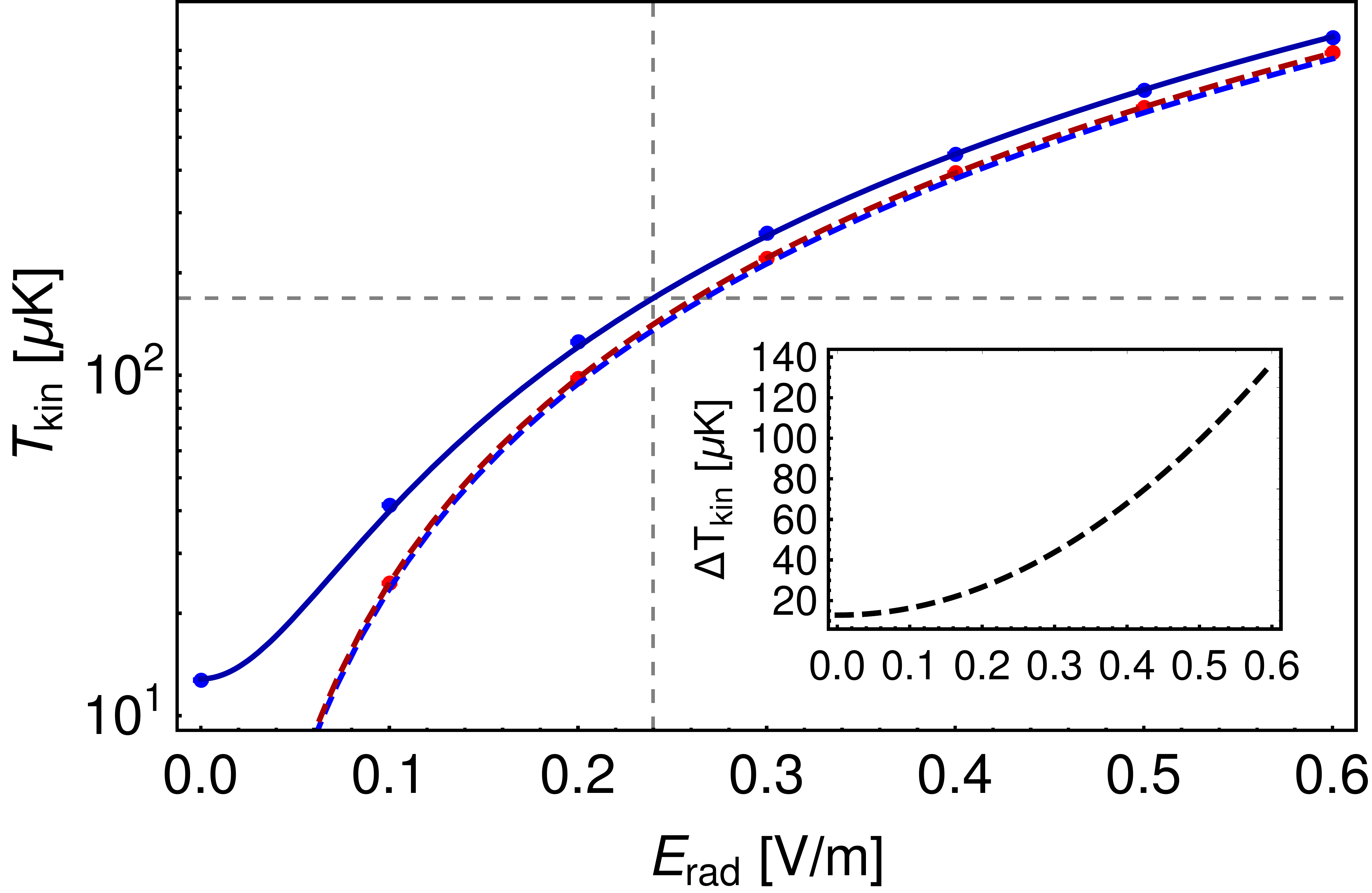}
  \hspace{0.05\textwidth}
  \includegraphics[width=0.45\textwidth]{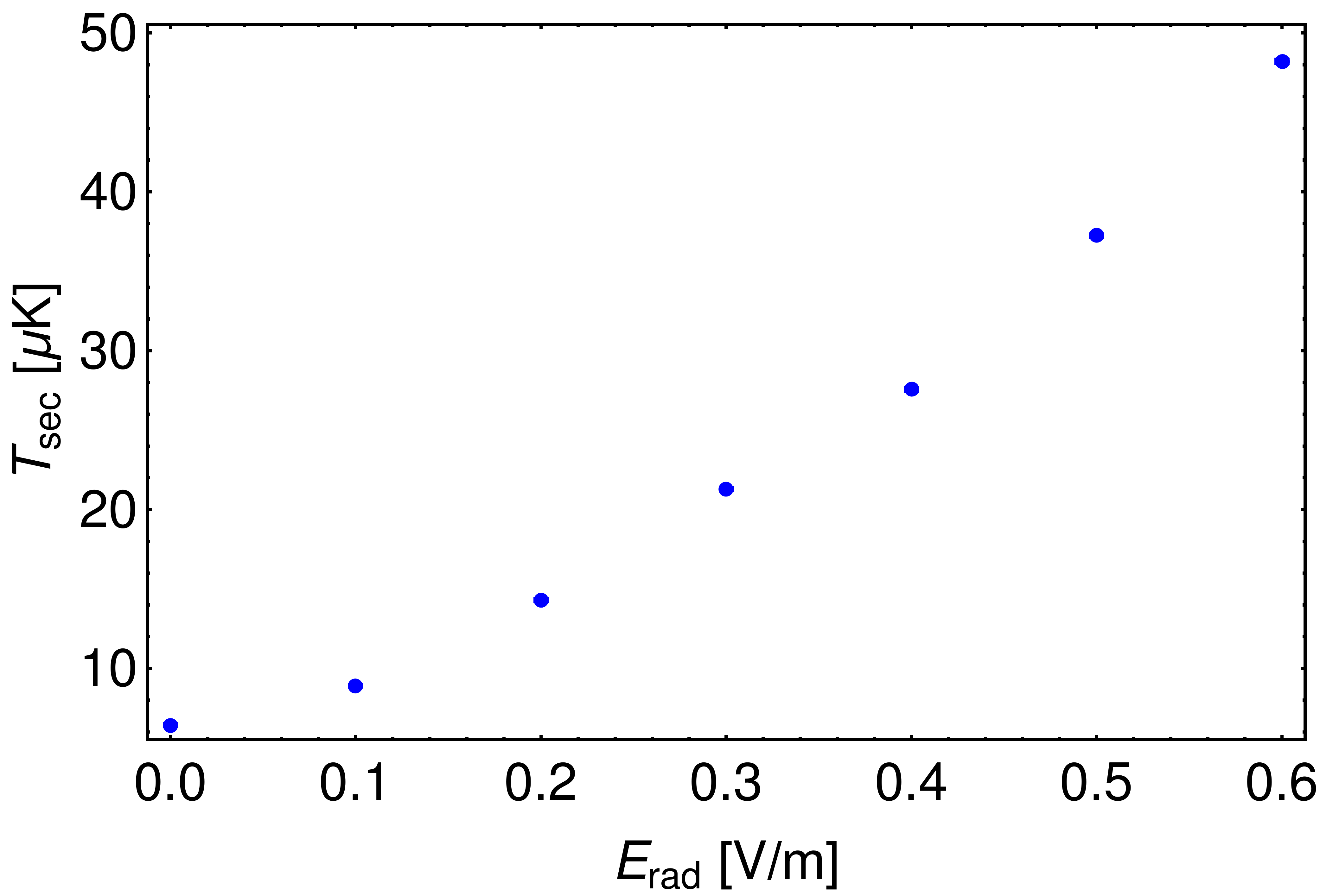}
    \vspace{0.05\textwidth}
  \includegraphics[width=0.45\textwidth]{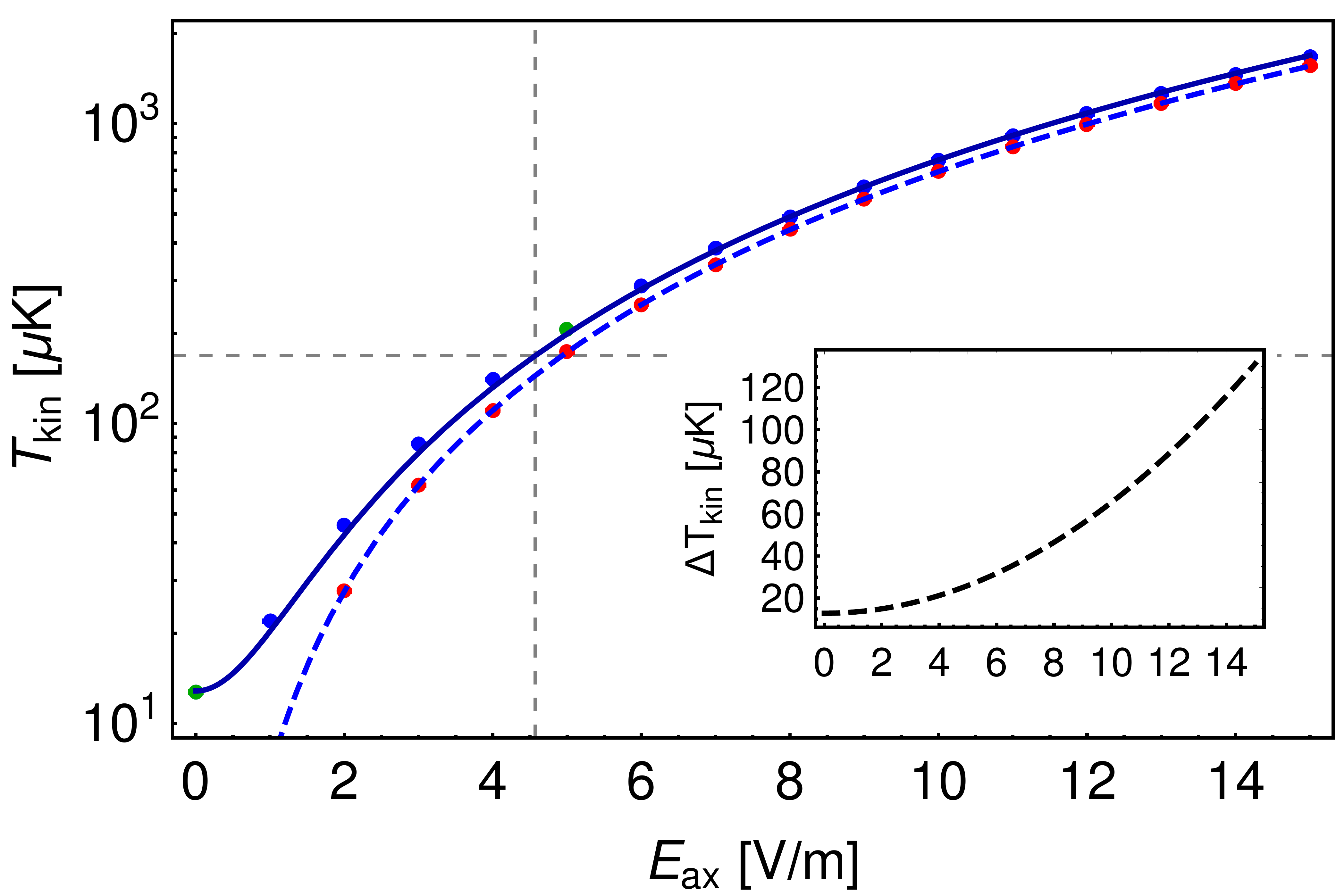}
  \hspace{0.05\textwidth}
  \includegraphics[width=0.45\textwidth]{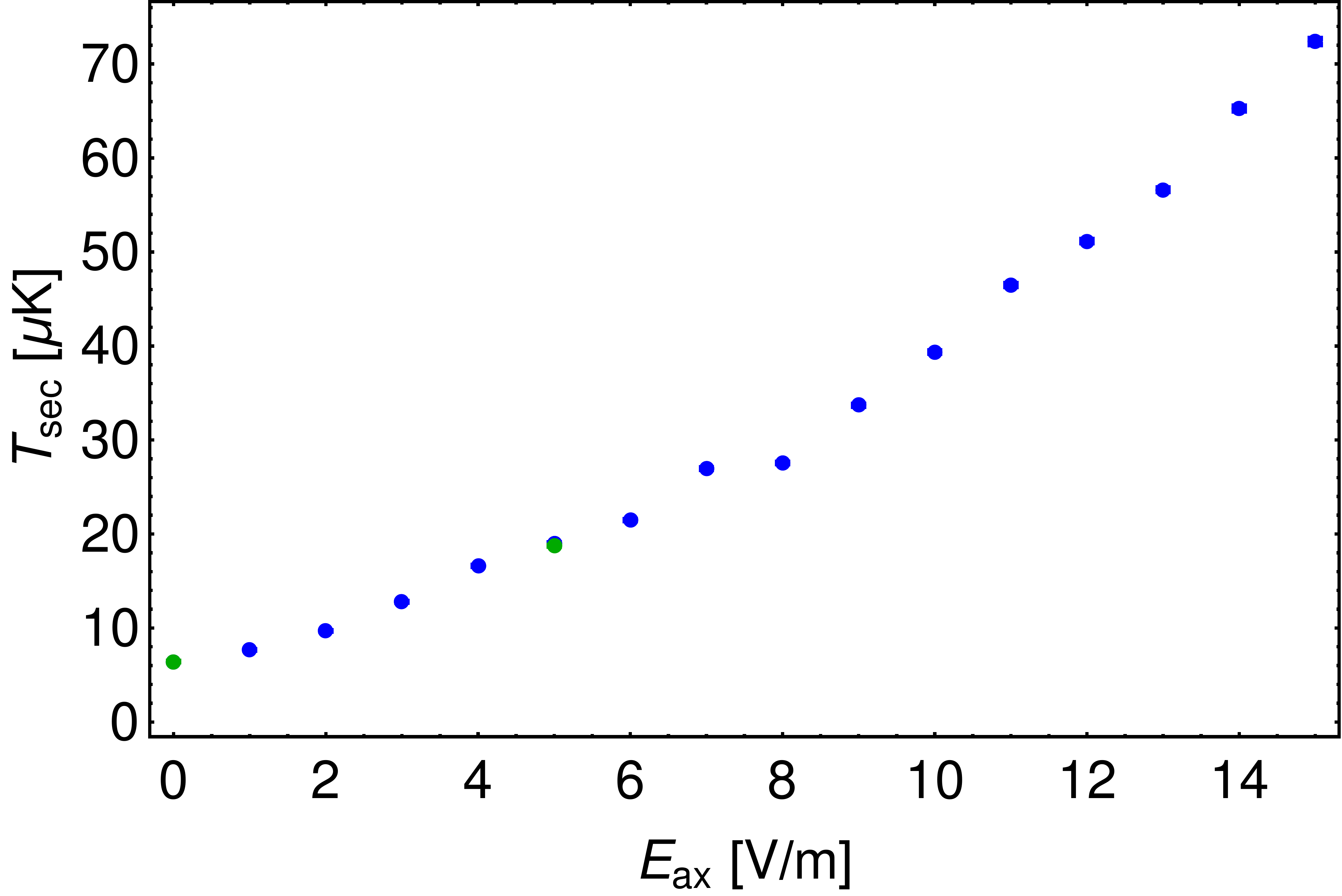}
    \vspace{0.05\textwidth}
    \includegraphics[width=0.45\textwidth]{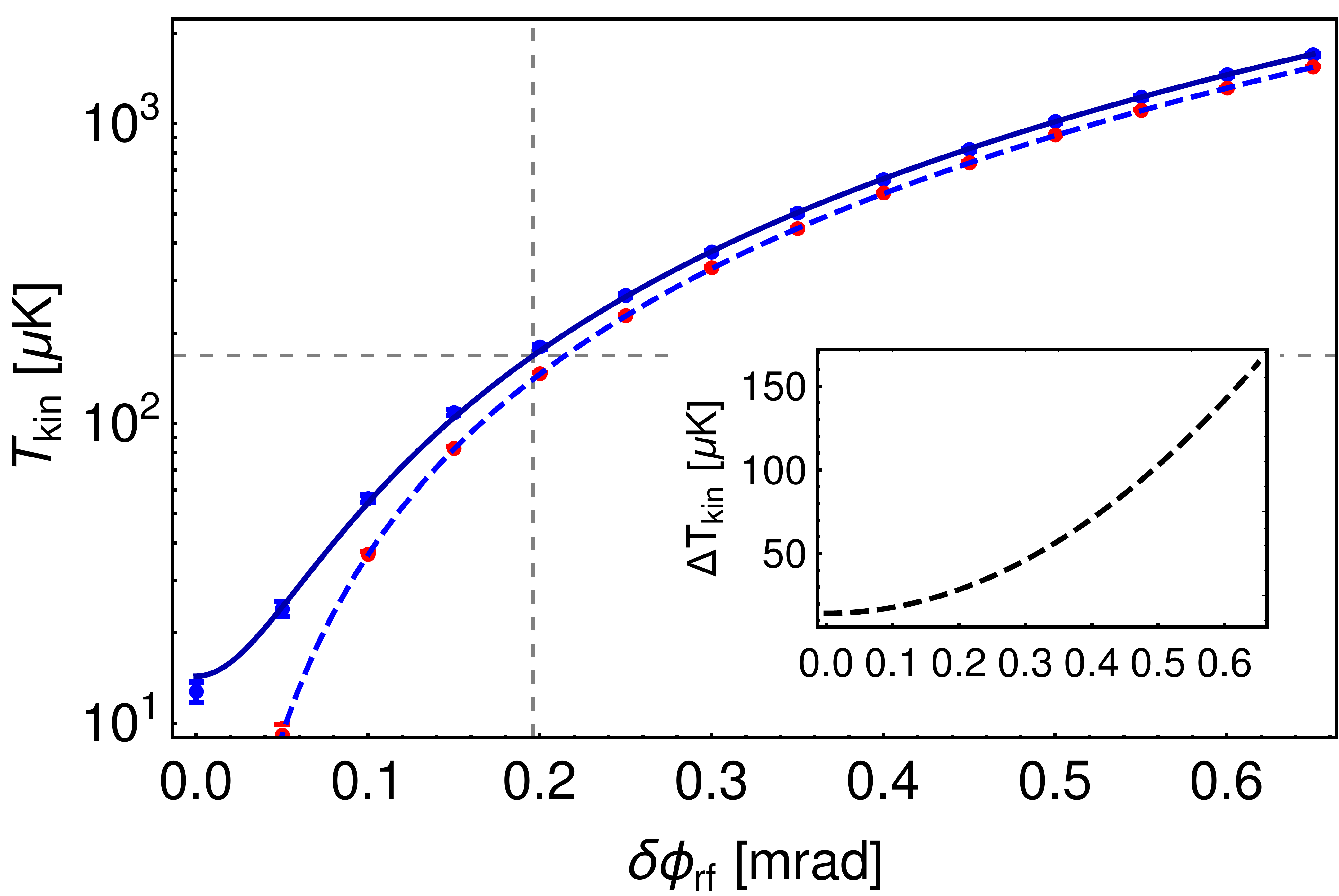}
  \hspace{0.05\textwidth}
  \includegraphics[width=0.45\textwidth]{fig22d.pdf}
\end{center}
  \caption{Average kinetic energy of a linear four-ion crystal colliding with
  atoms at $2\,\mu$K (left, blue) expressed as $T_{\rm{kin}}$ and temperature of
  the secular motion (right) versus micromotion causing parameters. The results
  for the kinetic energy were fit with quadratic function (solid blue curves).
  The dashed blue curves show the approximate theoretical amount of energy
  stored in the excess micromotion according to
  Eqs.~\ref{eqn:enemm},~\ref{eqn:enamm} and~\ref{eqn:enpmm} respectively. The
  red points correspond to simulated values for the non-interacting case, in
  approximate agreement with the theoretical behavior. The dashed gray lines
  indicate the $s$-wave temperature limit. The secular part of the average
  kinetic energy is shown on the right. The insets show the difference between
  the solid and dashed blue curves, resembling the micromotion induced heating.
  Green points
  were obtained using a fixed starting/escape sphere, blue with a comoving
  sphere.}
\label{fig:fourionaxial}
\end{figure}

The obtained results for average kinetic energy and secular energy are shown in
Fig.~\ref{fig:fourionaxial}.
Each point was fit by
averaging over at least 30 individual runs. The resulting average kinetic
energies expressed as $T_{\rm{kin}}$ (left) follow approximately the same
quadratic behavior as in the case of a single ion, indicating that the main part
of the kinetic energy is stored in the micromotion.
The quadratic fits (solid blue lines) lead to the increase parameters
$\theta_{E_{\rm{rad}}} = 7.45(3)\,\mu{\rm{K}}\cdot\left({\rm{V/m}}\right)^{-2}$,
$\theta_{E_{\rm{ax}}} = 2705(15)\,\mu{\rm{K}}\cdot\left({\rm{V/m}}\right)^{-2}$ and
$\theta_{\delta \phi_{\rm{rf}}} = 4005(16)\,\mu{\rm{K}}\cdot{\rm{mrad}}^{-2}$, in
almost perfect agreement with the single ion case.  For all three cases the
theoretical approximate energies due to the micromotion is shown as dashed blue
lines.  To verify the validity of these curves, the average kinetic energy for
a crystal without atoms, initialized at zero secular temperature was simulated
as well (red points).  Only for the case of radial excess micromotion
the theoretical prediction (dashed blue)
deviates significantly
from the red points, indicating the approximate nature of the prediction at high
radial micromotion amplitudes.

To quantify the micromotion-induced heating effect on the secular motion, the
secular temperature was extracted as described in section~\ref{sect:crystals1D}.
In all three cases the dependence on the scanned parameter seems to be a bit
weaker than quadratic. The temperature dependence of the individual modes is
discussed in section~\ref{subs:simudecomp}.

In all three cases, the number of collisions required to equilibrate (not shown
in the figures) show a similar behavior as in the single ion case, besides the
fact that they are $N_{\rm{ions}}=4$ times higher because of the reduced
effective density of atoms.
\clearpage
\section*{References}


\begin{thebibliography}{10}
\expandafter\ifx\csname url\endcsname\relax
  \def\url#1{{\tt #1}}\fi
\expandafter\ifx\csname urlprefix\endcsname\relax\def\urlprefix{URL }\fi
\providecommand{\eprint}[2][]{\url{#2}}

\bibitem{Smith:2005}
Smith W~W, Makarov O~P and Lin J 2005 {\em J. Mod. Opt.\/} {\bf 52} 2253--2260
  \urlprefix\url{https://doi.org/10.1080/09500340500275850}

\bibitem{Grier:2009}
Grier A~T, Cetina M, Oru{\v{c}}evi{\'c} F and Vuleti{\'c} V 2009 {\em Phys.
  Rev. Lett.\/} {\bf 102}(22) 223201
  \urlprefix\url{https://link.aps.org/doi/10.1103/PhysRevLett.102.223201}

\bibitem{Zipkes:2010}
Zipkes C, Palzer S, Sias C and K{\"o}hl M 2010 {\em Nature\/} {\bf 464} 388 EP
  -- \urlprefix\url{http://dx.doi.org/10.1038/nature08865}

\bibitem{Schmid:2010}
Schmid S, H\"arter A and Denschlag J~H 2010 {\em Phys. Rev. Lett.\/} {\bf
  105}(13) 133202
  \urlprefix\url{https://link.aps.org/doi/10.1103/PhysRevLett.105.133202}

\bibitem{Zipkes:2010b}
Zipkes C, Palzer S, Ratschbacher L, Sias C and K\"ohl M 2010 {\em Phys. Rev.
  Lett.\/} {\bf 105}(13) 133201
  \urlprefix\url{https://link.aps.org/doi/10.1103/PhysRevLett.105.133201}

\bibitem{Hall:2011}
Hall F~H~J, Aymar M, Bouloufa-Maafa N, Dulieu O and Willitsch S 2011 {\em Phys.
  Rev. Lett.\/} {\bf 107}(24) 243202
  \urlprefix\url{https://link.aps.org/doi/10.1103/PhysRevLett.107.243202}

\bibitem{Hall:2012}
Hall F~H~J and Willitsch S 2012 {\em Phys. Rev. Lett.\/} {\bf 109}(23) 233202
  \urlprefix\url{https://link.aps.org/doi/10.1103/PhysRevLett.109.233202}

\bibitem{Rellergert:2011}
Rellergert W~G, Sullivan S~T, Kotochigova S, Petrov A, Chen K, Schowalter S~J
  and Hudson E~R 2011 {\em Phys. Rev. Lett.\/} {\bf 107}(24) 243201
  \urlprefix\url{https://link.aps.org/doi/10.1103/PhysRevLett.107.243201}

\bibitem{Sullivan:2012}
Sullivan S~T, Rellergert W~G, Kotochigova S and Hudson E~R 2012 {\em Phys. Rev.
  Lett.\/} {\bf 109}(22) 223002
  \urlprefix\url{https://link.aps.org/doi/10.1103/PhysRevLett.109.223002}

\bibitem{Ratschbacher:2012}
Ratschbacher L, Zipkes C, Sias C and K{\"o}hl M 2012 {\em Nature Physics\/}
  {\bf 8} 649 EP -- \urlprefix\url{http://dx.doi.org/10.1038/nphys2373}

\bibitem{Ravi:2012}
Ravi K, Lee S, Sharma A, Werth G and Rangwala S~A 2012 {\em Nature
  Communications\/} {\bf 3} 1126 EP --
  \urlprefix\url{http://dx.doi.org/10.1038/ncomms2131}

\bibitem{Ratschbacher:2013}
Ratschbacher L, Sias C, Carcagni L, Silver J~M, Zipkes C and K\"ohl M 2013 {\em
  Phys. Rev. Lett.\/} {\bf 110}(16) 160402
  \urlprefix\url{https://link.aps.org/doi/10.1103/PhysRevLett.110.160402}

\bibitem{Harter:2013}
H\"arter A and Denschlag J~H 2014 {\em Contemporary Physics\/} {\bf 55} 33--45
  \urlprefix\url{https://doi.org/10.1080/00107514.2013.854618}

\bibitem{Hall:2013}
Hall F~H, Eberle P, Hegi G, Raoult M, Aymar M, Dulieu O and Willitsch S 2013
  {\em Molecular Physics\/} {\bf 111} 2020--2032
  \urlprefix\url{https://doi.org/10.1080/00268976.2013.780107}

\bibitem{Haze:2015}
Haze S, Saito R, Fujinaga M and Mukaiyama T 2015 {\em Phys. Rev. A\/} {\bf
  91}(3) 032709
  \urlprefix\url{https://link.aps.org/doi/10.1103/PhysRevA.91.032709}

\bibitem{Meir:2016}
Meir Z, Sikorsky T, Ben-shlomi R, Akerman N, Dallal Y and Ozeri R 2016 {\em
  Phys. Rev. Lett.\/} {\bf 117}(24) 243401
  \urlprefix\url{https://link.aps.org/doi/10.1103/PhysRevLett.117.243401}

\bibitem{Saito:2017}
Saito R, Haze S, Sasakawa M, Nakai R, Raoult M, Da~Silva H, Dulieu O and
  Mukaiyama T 2017 {\em Phys. Rev. A\/} {\bf 95}(3) 032709
  \urlprefix\url{https://link.aps.org/doi/10.1103/PhysRevA.95.032709}

\bibitem{Tomza:2017cold}
{Tomza} M, {Jachymski} K, {Gerritsma} R, {Negretti} A, {Calarco} T, {Idziaszek}
  Z and {Julienne} P~S 2017 {\em ArXiv e-prints\/} (\textit{Preprint}
  \eprint{1708.07832})

\bibitem{Krych:2010}
Krych M, Skomorowski W, Paw\l{}owski F, Moszynski R and Idziaszek Z 2011 {\em
  Phys. Rev. A\/} {\bf 83}(3) 032723
  \urlprefix\url{https://link.aps.org/doi/10.1103/PhysRevA.83.032723}

\bibitem{Krych:2013}
Krych M and Idziaszek Z 2015 {\em Phys. Rev. A\/} {\bf 91} 023430

\bibitem{Kollath:2007}
Kollath C, K\"ohl M and Giamarchi T 2007 {\em Phys. Rev. A\/} {\bf 76}(6)
  063602 \urlprefix\url{https://link.aps.org/doi/10.1103/PhysRevA.76.063602}

\bibitem{Doerk:2010}
Doerk H, Idziaszek Z and Calarco T 2010 {\em Phys. Rev. A\/} {\bf 81}(1) 012708
  \urlprefix\url{https://link.aps.org/doi/10.1103/PhysRevA.81.012708}

\bibitem{Secker:2016}
Secker T, Gerritsma R, Glaetzle A~W and Negretti A 2016 {\em Phys. Rev. A\/}
  {\bf 94}(1) 013420
  \urlprefix\url{https://link.aps.org/doi/10.1103/PhysRevA.94.013420}

\bibitem{Bissbort:2013}
Bissbort U, Cocks D, Negretti A, Idziaszek Z, Calarco T, Schmidt-Kaler F,
  Hofstetter W and Gerritsma R 2013 {\em Phys. Rev. Lett.\/} {\bf 111}(8)
  080501
  \urlprefix\url{https://link.aps.org/doi/10.1103/PhysRevLett.111.080501}

\bibitem{Idziaszek:2009}
Idziaszek Z, Calarco T, Julienne P~S and Simoni A 2009 {\em Phys. Rev. A\/}
  {\bf 79}(1) 010702
  \urlprefix\url{https://link.aps.org/doi/10.1103/PhysRevA.79.010702}

\bibitem{Idziaszek:2011}
Idziaszek Z, Simoni A, Calarco T and Julienne P~S 2011 {\em New Journal of
  Physics\/} {\bf 13} 083005
  \urlprefix\url{http://stacks.iop.org/1367-2630/13/i=8/a=083005}

\bibitem{Tomza:2015}
Tomza M, Koch C~P and Moszynski R 2015 {\em Phys. Rev. A\/} {\bf 91}(4) 042706
  \urlprefix\url{https://link.aps.org/doi/10.1103/PhysRevA.91.042706}

\bibitem{Gacesa:2017}
Gacesa M and C\^ot\'e R 2017 {\em Phys. Rev. A\/} {\bf 95}(6) 062704
  \urlprefix\url{https://link.aps.org/doi/10.1103/PhysRevA.95.062704}

\bibitem{Furst:2018}
{F{\"u}rst} H, {Feldker} T, {Vincenz Ewald} N, {Joger} J, {Tomza} M and
  {Gerritsma} R 2017 {\em ArXiv e-prints\/} (\textit{Preprint}
  \eprint{1712.07873})

\bibitem{Julienne:2010}
Chin C, Grimm R, Julienne P and Tiesinga E 2010 {\em Rev. Mod. Phys.\/} {\bf
  82}(2) 1225--1286
  \urlprefix\url{https://link.aps.org/doi/10.1103/RevModPhys.82.1225}

\bibitem{Bloch:2012}
Bloch I, Dalibard J and Nascimb{\`e}ne S 2012 {\em Nature Physics\/} {\bf 8}
  267 EP -- \urlprefix\url{http://dx.doi.org/10.1038/nphys2259}

\bibitem{major1968exchange}
Major F~G and Dehmelt H~G 1968 {\em Phys. Rev.\/} {\bf 170}(1) 91--107
  \urlprefix\url{https://link.aps.org/doi/10.1103/PhysRev.170.91}

\bibitem{DeVoe:2009}
DeVoe R~G 2009 {\em Phys. Rev. Lett.\/} {\bf 102}(6) 063001
  \urlprefix\url{https://link.aps.org/doi/10.1103/PhysRevLett.102.063001}

\bibitem{Zipkes:2011}
Zipkes C, Ratschbacher L, Sias C and K\"ohl M 2011 {\em New Journal of
  Physics\/} {\bf 13} 053020
  \urlprefix\url{http://stacks.iop.org/1367-2630/13/i=5/a=053020}

\bibitem{Cetina:2012}
Cetina M, Grier A~T and Vuleti\ifmmode~\acute{c}\else \'{c}\fi{} V 2012 {\em
  Phys. Rev. Lett.\/} {\bf 109}(25) 253201
  \urlprefix\url{https://link.aps.org/doi/10.1103/PhysRevLett.109.253201}

\bibitem{Chen:2014}
Chen K, Sullivan S~T and Hudson E~R 2014 {\em Phys. Rev. Lett.\/} {\bf 112}(14)
  143009
  \urlprefix\url{https://link.aps.org/doi/10.1103/PhysRevLett.112.143009}

\bibitem{Ewald:2015}
Ewald N~V 2015 {\em {Q}uest for an {U}ltracold {H}ybrid {A}tom-{I}on
  {E}xperiment\/} Master's thesis {J}ohannes {G}utenberg-{U}niversit\"at
  {M}ainz

\bibitem{Weckesser:2015}
H\"oltkemeier B, Weckesser P, L\'opez-Carrera H and Weidem\"uller M 2016 {\em
  Phys. Rev. Lett.\/} {\bf 116}(23) 233003
  \urlprefix\url{https://link.aps.org/doi/10.1103/PhysRevLett.116.233003}

\bibitem{Weckesser:2016}
H\"oltkemeier B, Weckesser P, L\'opez-Carrera H and Weidem\"uller M 2016 {\em
  Phys. Rev. A\/} {\bf 94}(6) 062703
  \urlprefix\url{https://link.aps.org/doi/10.1103/PhysRevA.94.062703}

\bibitem{Rouse:2017}
Rouse I and Willitsch S 2017 {\em Phys. Rev. Lett.\/} {\bf 118}(14) 143401
  \urlprefix\url{https://link.aps.org/doi/10.1103/PhysRevLett.118.143401}

\bibitem{Secker:2017}
Secker T, Ewald N, Joger J, F\"urst H, Feldker T and Gerritsma R 2017 {\em
  Phys. Rev. Lett.\/} {\bf 118}(26) 263201
  \urlprefix\url{https://link.aps.org/doi/10.1103/PhysRevLett.118.263201}

\bibitem{Leibfried:2003}
Leibfried D, Blatt R, Monroe C and Wineland D 2003 {\em Rev. Mod. Phys.\/} {\bf
  75}(1) 281--324
  \urlprefix\url{https://link.aps.org/doi/10.1103/RevModPhys.75.281}

\bibitem{Berkeland:1998}
Berkeland D~J, Miller J~D, Bergquist J~C, Itano W~M and Wineland D~J 1998 {\em
  Journal of Applied Physics\/} {\bf 83} 5025--5033
  \urlprefix\url{https://doi.org/10.1063/1.367318}

\bibitem{MeirThesis:2016}
Meir Z 2016 {\em Dynamics of a single, ground-state cooled and trapped ion
  colliding with ultracold atoms: A micromotion tale.\/} Ph.D. thesis Weizmann
  Institute of Science

\bibitem{Meir:2018}
Meir Z, Sikorsky T, Ben-shlomi R, Akerman N, Pinkas M, Dallal Y and Ozeri R
  2018 {\em Journal of Modern Optics\/} {\bf 65} 501--519
  \urlprefix\url{https://doi.org/10.1080/09500340.2017.1397217}

\bibitem{Langevin:1905}
Langevin M~P 1905 {\em Annales de Chimie et de Physique, series\/} {\bf 5}
  245--288 \urlprefix\url{https://ci.nii.ac.jp/naid/10004043377/en/}

\bibitem{weisstein2002sphere}
Weisstein E~W 2002 Sphere point picking
  \urlprefix\url{http://mathworld.wolfram.com/SpherePointPicking.html}

\bibitem{press2007numerical}
Press W~H 2007 {\em Numerical recipes 3rd edition: The art of scientific
  computing\/} (Cambridge university press)

\bibitem{Joger:2017}
Joger J, F\"urst H, Ewald N, Feldker T, Tomza M and Gerritsma R 2017 {\em Phys.
  Rev. A\/} {\bf 96}(3) 030703(R)
  \urlprefix\url{https://link.aps.org/doi/10.1103/PhysRevA.96.030703}

\bibitem{Olmschenk:2007}
Olmschenk S, Younge K~C, Moehring D~L, Matsukevich D~N, Maunz P and Monroe C
  2007 {\em Phys. Rev. A\/} {\bf 76}(5) 052314
  \urlprefix\url{https://link.aps.org/doi/10.1103/PhysRevA.76.052314}

\bibitem{Taylor:1997}
Taylor P, Roberts M, Gateva-Kostova S~V, Clarke R~B~M, Barwood G~P, Rowley
  W~R~C and Gill P 1997 {\em Phys. Rev. A\/} {\bf 56}(4) 2699
  \urlprefix\url{https://link.aps.org/doi/10.1103/PhysRevA.56.2699}

\bibitem{James:1998}
James D 1998 {\em Applied Physics B\/} {\bf 66} 181--190 ISSN 1432-0649
  \urlprefix\url{https://doi.org/10.1007/s003400050373}

\bibitem{cooley1965}
Cooley J~W and Tukey J~W 1965 {\em Mathematics of Computation\/} {\bf 19}
  297--301 ISSN 00255718, 10886842
  \urlprefix\url{http://www.jstor.org/stable/2003354}

\bibitem{Lemmer:2015}
Lemmer A, Cormick C, Schmiegelow C~T, Schmidt-Kaler F and Plenio M~B 2015 {\em
  Phys. Rev. Lett.\/} {\bf 114}(7) 073001
  \urlprefix\url{https://link.aps.org/doi/10.1103/PhysRevLett.114.073001}

\bibitem{Kaufmann:2012}
Kaufmann H, Ulm S, Jacob G, Poschinger U, Landa H, Retzker A, Plenio M~B and
  Schmidt-Kaler F 2012 {\em Phys. Rev. Lett.\/} {\bf 109}(26) 263003
  \urlprefix\url{https://link.aps.org/doi/10.1103/PhysRevLett.109.263003}

\bibitem{Landa:2012}
Landa H, Drewsen M, Reznik B and Retzker A 2012 {\em New Journal of Physics\/}
  {\bf 14} 093023
  \urlprefix\url{http://stacks.iop.org/1367-2630/14/i=9/a=093023}

\bibitem{Shen:2014}
Shen C and Duan L~M 2014 {\em Phys. Rev. A\/} {\bf 90}(2) 022332
  \urlprefix\url{https://link.aps.org/doi/10.1103/PhysRevA.90.022332}

\bibitem{Richerme2016}
Richerme P 2016 {\em Phys. Rev. A\/} {\bf 94}(3) 032320
  \urlprefix\url{https://link.aps.org/doi/10.1103/PhysRevA.94.032320}

\bibitem{Serwane:2011}
Serwane F, Z{\"u}rn G, Lompe T, Ottenstein T~B, Wenz A~N and Jochim S 2011 {\em
  Science\/} {\bf 332} 336--338 ISSN 0036-8075
  \urlprefix\url{http://science.sciencemag.org/content/332/6027/336}

\bibitem{Gross:2016}
Gross C, Gan H~C~J and Dieckmann K 2016 {\em Phys. Rev. A\/} {\bf 93}(5) 053424
  \urlprefix\url{https://link.aps.org/doi/10.1103/PhysRevA.93.053424}

\end{thebibliography}
\providecommand{\newblock}{}

\end{document}